\documentclass[%
 aip,
 amsmath,amssymb,
preprint,%
]{revtex4-1}

\usepackage{graphicx}% Include figure files
\usepackage{dcolumn}% Align table columns on decimal point
\usepackage{bm}% bold math
\usepackage{multirow}% Add multiple rows
\usepackage[utf8]{inputenc}
\usepackage[T1]{fontenc}
\usepackage{mathptmx}
\usepackage{etoolbox}
\usepackage[dvipsnames]{xcolor}
\usepackage{comment}
\renewenvironment{comment}{}{}

\newcommand{\rk}[1]{\textcolor{black}{#1}}

\makeatletter
\def\@email#1#2{%
 \endgroup
 \patchcmd{\titleblock@produce}
  {\frontmatter@RRAPformat}
  {\frontmatter@RRAPformat{\produce@RRAP{*#1\href{mailto:#2}{#2}}}\frontmatter@RRAPformat}
  {}{}
}%
\makeatother
\begin{document}

\preprint{}

\title[Motion of hairpin filaments in the ABL]{On the motion of hairpin filaments in the atmospheric boundary layer}
\author{Abhishek Harikrishnan}
\email{abhishek.harikrishnan@fu-berlin.de}
\author{Marie Rodal}
\author{Rupert Klein}
\affiliation{%
Institute of Mathematics, Freie Universität Berlin, 14195 Berlin, Germany
}%
\author{Daniel Margerit}
\affiliation{Avenue des grands pins, 31660 Toulouse, France}
\author{Nikki Vercauteren}
\affiliation{%
 Department of Geosciences, University of Oslo, 0371 Oslo, Norway
}%

\date{\today}

% Final clean-up

% ----- ABSTRACT ----- %

\begin{abstract}

A recent work of \citet{harikrishnan2021geometry} [arXiv:2110.02253 (2021)] has revealed an abundance of hairpin-like vortex structures, oriented in a similar direction, in the turbulent patches of a stably stratified Ekman flow. The Ekman flow over a smooth wall is a simplified configuration of the Atmospheric Boundary Layer (ABL) where effects of both stratification and rotation are present. In this study, hairpin-like structures are investigated by treating them as slender vortex filaments, i.e., a vortex filament whose diameter $d$ is small when compared to its radius of curvature $R$. The corrected thin-tube model of \citet{KleinKnio1995} [J. Fluid Mech. (1995)] is used to compute the motion of these filaments with the ABL as a background flow. The influence of the mean background flow on the filaments is studied for two stably stratified cases and a neutrally stratified case. Our results suggest that the orientation of the hairpin filament in the spanwise direction is linked to its initial starting height under stable stratification whereas no such dependency can be observed with the neutrally stratified background flow. An improved feature tracking scheme based on volume or spatial overlap for tracking $Q$-criterion vortex structures on the Direct Numerical Simulation (DNS) data is also developed. It overcomes the limitation of using a constant threshold in time by dynamically adjusting the thresholds to accommodate the growth or deterioration of a feature. A comparison between the feature tracking and the filament simulation reveals qualitatively similar temporal developments. Finally, an extension of the asymptotic analysis of \citet{CallegariTing1978} [J. App. Math (1978)] is carried out to include the effect of gravity. The results show that, \rk{in the regime considered here}, a contribution from the gravity term occurs only when the tail of an infinitely long filament is tilted at an angle relative to the wall.

\end{abstract}

\maketitle

% ----- BACKGROUND AND INTRODUCTION ----- %

\section{Background and Introduction}
\label{section: introduction}

\begin{figure*}
    \centering
    \includegraphics[width =  \linewidth]{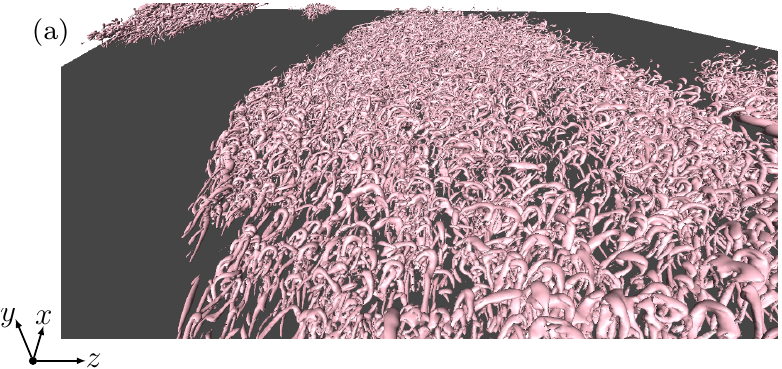}
    \vspace{0.5cm}
    \includegraphics[width = \linewidth]{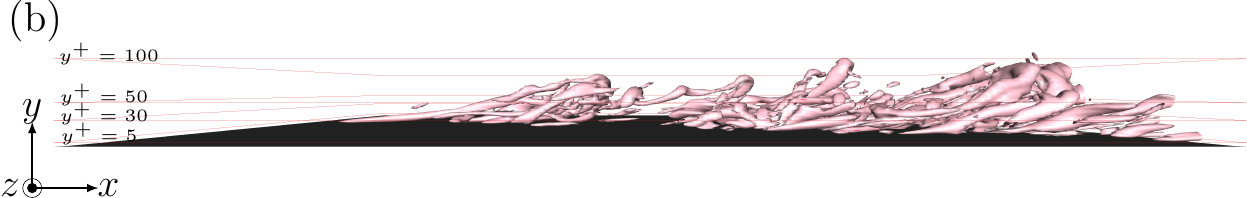}
    \caption{(a) Hairpin-like vortex structures are visualized \rk{based on the} $Q$-criterion for the strongly intermittent case (bulk Richardson number $Ri_B = 2.64$) of \citet{harikrishnan2021geometry}. The visualization is restricted to a wall-normal height of $y^+ = 250$ where the majority of the hairpin-like structures are visible. Regions with no vortex activity correspond to the non-turbulent/inactive regions of the flow. (b) Side view ($x, y$) of a hairpin packet for the \rk{same case.} ($x, y, z$) \rk{denote the streamwise, wall-normal, and spanwise coordinates, respectively.}}
    \label{fig: hairpin_visualization}
\end{figure*}

Understanding the dynamics of three-dimensional vortices is an essential building block to uncover\rk{ing} the mysteries of turbulence. In particular, vortices possessing a unique hairpin-like geometry have been experimentally observed in turbulent boundary layers through the seminal work of \citet{head1981new}. Since then, numerous studies \citep{acarlar1987study, acarlar1987study2, adrian2000vortex, robinson1991kinematics, adrian2007hairpin, adrian2000vortex, zhou1999mechanisms} have been carried out corroborating the existence of these entities. In recent work, \citet{harikrishnan2021geometry} analysed the direct numerical simulations (DNS) of stratified Ekman flows\citep{ansorge2014global, ansorge2016analyses, ansorge2016thesis}, which are simplified representations of the atmospheric boundary layer (ABL). Under very stable conditions, the flow is globally intermittent, i.e., non-turbulent flow regions on scales larger than the coherent motions exist along with the turbulent flow regions close to the wall \citep{mahrt1989intermittency}. For such flows, the authors\citep{harikrishnan2021geometry} observed an abundance of hairpin-like structures within the turbulent regions of the flow. A visualization of this case is shown in figure \ref{fig: hairpin_visualization} for which vortices were detected utilizing the popular $Q$-criterion \citep{hunt1988eddies}, which \rk{utilizes the second invariant of the velocity gradient tensor ($\nabla \mathbf{v}$) to classify} vortices as regions where rotation dominates over the strain, i.e., 
\begin{equation}
\label{eq: qCriterion}
Q = \frac{1}{2} (||\Omega||^2 - ||S||^2) > 0\,.
\end{equation}
In the above equation, $\Omega = \frac{1}{2}[\nabla \mathbf{v} - (\nabla \mathbf{v})^T]$ is the vorticity or spin tensor, $S = \frac{1}{2}[\nabla \mathbf{v} + (\nabla \mathbf{v})^T]$ is the strain-rate tensor and $||\cdot||$ is the Euclidean norm. 

For the same case visualized in figure \ref{fig: hairpin_visualization}, \citet{harikrishnan2020curious} showed that if two hairpin-like vortices \rk{are} extracted randomly from different regions of the flow, the heads of the hairpin-like structures appear to be oriented in a similar direction. In subsequent work\citep{harikrishnan2021lagrangian}, the Lagrangian finite-time Lyapunov exponent (FTLE) also yielded comparable results. While the $Q$-criterion is an Eulerian point-wise characterization of an instantaneous velocity field, FTLE follows fluid particle trajectories to identify regions of maximum material stretching. Initializing and advecting tracers in forward-time yields repelling Lagrangian coherent structures\citep{green2007detection} which have been visualized in \citet{harikrishnan2021lagrangian}. These results suggest that hairpin-like structures oriented in similar directions can be identified using both Eulerian and Lagrangian criteria under very stable conditions of the ABL. 

The identification of hairpin-like structures in the ABL is not new and has been reported in previous works for various stability conditions. For instance, \citet{hommema2003packet} used smoke visualization in their field experiments to identify ramp-like structures (which are conjectured to be hairpin packets) in the first $3$ meters of the ABL under moderately unstable (convective) and neutral conditions. \citet{li2011coherent} found signatures of hairpin vortices under neutral conditions and observed that they change to thermal plumes under highly unstable conditions. Signatures of hairpin vortices \rk{in neutral} to weakly stable conditions were reported \rk{by \citet{heisel2018spatial}}. Under stable (nocturnal) conditions, \citet{oncley2016whirlwinds} noted the presence of counter-rotating vortices which can be interpreted as legs of \rk{hairpin vortex structures}. By studying the DNS of a stably stratified shear layer \rk{akin to those observed} in the atmosphere, \citet{watanabe2019hairpin} not only found a large number of hairpin vortices but noted that they tend to be oriented towards the streamwise and spanwise direction in the middle and top of the shear layer, respectively. 

The motion and behavior of hairpin vortices has been well documented in unstratified fluid flows. By periodically injecting fluid in a subcritical laminar boundary layer, \citet{acarlar1987study} artificially created a low-speed region resembling the low-speed streaks, first experimentally observed by \citet{kline1967structure}. They noted that these regions became unstable and initiated an oscillation which grew to form hairpin vortices. The heads of the vortices which lift-up due to self-induction were then found to be stretched by the wall shear layer as they were carried downstream. With the linear stochastic estimation procedure, \citet{zhou1999mechanisms} were able to isolate and study the evolution of a hairpin structure in the DNS of channel flow. They found that the circulation strength relative to the mean shear can impact the streamwise length of the hairpin vortex, with lower circulation \rk{hairpins} having longer legs due to the dominance of the mean shear over self-induction. In an experimental study, \citet{adrian2000vortex} observed that the inclination angle (denoted by $\gamma$), i.e., the angle of head of the hairpin with respect to the wall, is a function of its location. Near vertical orientation was observed in the outer layer and about $20^{\circ} - 45 ^{\circ}$ inclination was found close to the wall. Furthermore, as the hairpin ages over time, \citet{head1981new, adrian2000vortex} have shown that the hairpins exhibit a characteristic growth angle which ranges between $12^{\circ} - 20^{\circ}$. While these studies give a good picture of the behavior of hairpin vortices in non-stratified turbulent boundary layers, their motion in stratified flows remains unexplored. \rk{Here, we would like to specifically address the following questions in this context}:

\begin{itemize}
    \item [(1)] Starting from an initial perturbation, how do dynamic hairpin characteristics, such as inclination angle, spanwise orientation, wall-normal stretching due to shear, streamwise and spanwise advection, change with respect to stratification? By studying these dynamic characteristics, can the abundance and orientation of hairpin-like structures in the very stable regime of the ABL be explained? 
    \item [(2)] Does the location of the initial perturbation, for example in the buffer or outer layer, have an impact on the dynamics of the hairpin structure? If so, how does it change with respect to stratification?
\end{itemize}

Owing to their abundance, understanding the dynamics of hairpins can be useful to improve parametrizations of the stable boundary layer (SBL) which still remains a challenge \citep{saiki2000large, jimenez2005large}. In particular, the streamwise displacement ($\Delta x$) computed from the inclination angle $\gamma$ of inclined features has been used to improve the wall models of Large Eddy Simulations (LES) \citep{marusic2001experimental, chauhan2013structure}. This motivates the need to study the dynamics of these structures in response to changes in stratification.

By tracking features with spatial overlaps in wall-bounded flows, \citet{lozano2014time} showed that coherent structures such as vortices or quadrant structures such as sweeps and ejections may undergo numerous complex interactions during their lifetimes. While volume tracking of these structures with DNS data can be useful, interactions may complicate understanding of the dynamics of a structure. For instance, once the structure of interest splits into two, there is an ambiguity in following the ``correct'' structure which is exacerbated when there are additional split or merge events during its lifetime. Furthermore, as the method relies on thresholding of scalar fields, it implies that some useful features of the structure may not be adequately captured (see the discussion in subsection \ref{subsec: initial_conditions}). Hence, in this paper, we turn towards a more fundamental approach, similar to the work of \citet{HonWalker1995}, where hairpins are treated as vortex filaments.

In his review paper, \citet{leonard1985computing} identified two methods to compute the motion of vortex filaments, namely the thin-filament and the Local Induction Approximations (LIA). The former is based on slender vortex theory \citep{CallegariTing1978} and on related numerical methods\citep{KleinKnio1995} which assume the vorticity to be highly concentrated along a ``filament centerline'' $\mathcal{L}(t): s \rightarrow \mathbf{X}(s, t)$. An individual vortex core has an averaged diameter $d$ and a characteristic radius of curvature $R$ such that the dimensionless core size parameter $\delta$ satisfies,
\begin{equation}
\label{eq: core_size_parameter}
\delta = \frac{d}{R} \ll 1
\end{equation}
In an unbounded domain, the velocity induced by this filament at a point $\mathbf{P}$ and time $t$ in \rk{a three-dimensional inviscid and irrotational} flow field in the outer flow region, i.e., away from the vortical core, is given by the line-Biot-Savart law as\citep{CallegariTing1978},
\begin{equation}
\label{eq: Biot-Savart Law}
\mathbf{Q}_1 (\textbf{P}, t) = -\frac{\Gamma}{4\pi}\int_{\mathcal{L}} \frac{\left(\mathbf{P}-\mathbf{X}(s',t)\right)\times \mathbf{ds}'}{\left|\mathbf{P}-\mathbf{X}(s',t)\right|^3}\,,
\end{equation}
where $\Gamma$ is the circulation of the filament. As $\mathbf{P}$ moves towards the filament centerline, $\mathbf{Q}_1(\mathbf{P},t)$ becomes singular, and thus the line Biot-Savart integral cannot alone predict the self-induced vortex motion. With matched asymptotic expansions, \citet{CallegariTing1978} showed how this singularity is naturally regularized within the framework of the Navier-Stokes equations, and they provided explicit expressions for the velocity of points on the filament centerline for a closed filament. Adapting this for a non-closed, infinite filament leads to the following equation, 
\begin{equation}\label{eq:FilamentEqnOfMotion}
\frac{\partial}{\partial t}\mathbf{X}(s,t) =
\frac{\Gamma}{4\pi} \kappa \mathbf{b}(s,t) \left(\ln\left(\frac{2}{\delta}\right) + C(t)\right) + \mathbf{Q}_0(s,t)\,,
\end{equation}
where
\begin{equation}
\label{eq:nonlocal_velocity}
\mathbf{Q}_0(s,t) = \mathbf{Q}_{\text{f}}(s,t) + \mathbf{Q}_{2}(\mathbf{X}(s,t))
\end{equation}
is the superposition of the non-singular remainder of the line-Biot-Savart integral $\mathbf{Q}_{f}$, and of a superimposed background flow $\mathbf{Q}_{2}$ (taken on the filament), such as the shear flow in a boundary layer. 
The other method named LIA (for ``Local Induction Approximation'') is a simplification of the thin-filament approximation \citep{hama1962progressive} in which the long-distance induction effects represented by $\mathbf{Q}_0(s,t)$ and the local effects from the core vorticity distribution represented by $C(t)$ in \eqref{eq:FilamentEqnOfMotion} are neglected, so that the filament motion is due to the curvature/binormal term alone. In this paper, it is used mainly for validation purposes and the effect of omission of long-distance effects is exemplified in section \ref{sec: stagnant_flow}. Therefore, hairpin evolution is simulated using the  slender filament approach, implemented numerically by the corrected thin-tube model of \citet{KleinKnio1995}. In this context, in addition to the previous research questions, we also address the following one,
%\DM{This "curvature/binormal" may be strange. Why not say "so that only the logarithm leading order term is kept and is the curvature time the binormal term."?} % I'd say this is still understandable. I've seen from multiple references saying that the contribution is due to the curvature binormal term. So I don't see that this needs to be modified.
%
\begin{itemize}
    \item [(3)] Can a suitable tracking methodology based on volume or spatial overlap be developed which respects the dynamical evolution of features in time? Qualitatively, how much would the results differ with those obtained through simulation with the Biot-Savart law?
\end{itemize}
Finally, the influence of gravity on the self-induced motion of the slender vortex filament is explored. An asymptotic analysis extending the work of \citet{CallegariTing1978} is carried out to include the effect of gravity and is presented in section \ref{sec: mathematical_formulation}. To the author's knowledge, the only attempt at including the effect of gravity in the filament motion equations was presented by \citet{Change2016}. They extended the force balance method of Moore and Saffman\citep{moore1972motion} to include the force of gravity, but \rk{assumed the fluid density in the \emph{a priori} assigned core of the filament to be constant, so that a self-consistent evolution of the fluid density was not considered}. They used this to demonstrate that buoyant vortex rings expand as they rise, as has indeed been observed experimentally by \citet{Turner1957}. In our work, we will \rk{account for the fact that the dominant reason for density variations in the ABL is the transport of air from different heights in the boundary layer, but that there is no diabatic effect, such as combustion, that would specifically affect the vortex core temperature and density. Hence, we} assume that the influence of gravity on the core flow is weak, but that it may have a stronger influence on the external flow. For simplicity, we will refrain from doing the full compressible flow analysis, as was done by  \citet{TingKleinKnio2007} and \citet{Knio2003}, and instead employ the Boussinesq approximation so that the density perturbation only appears in the gravity term. This simplification is justified as our focus of interest is in the ABL which to a large extent can be assumed to be incompressible. This leads us to our final research question:
\begin{itemize}
    \item [(4)] What effect does gravity have on the fluid flow within the vortex filament and hence on the self-induced motion of the filament, given the assumptions listed above? 
\end{itemize}
In section \ref{sec: numerical_methods}, the initial configuration of the hairpin filament is described, along with both numerical methods considered in this paper which are the Local Induction Approximation (LIA) and the corrected thin-tube model of \citet{KleinKnio1995}, henceforth denoted M1 KK. Results of the filament simulation are discussed in section \ref{sec: stagnant_flow} and \ref{sec: ABL_flow}. In the former, comparisons are made between the temporal evolution of hairpin filament with LIA and M1 KK methods which highlight the drawback of LIA \rk{owing to its neglect of nonlocal effects}. In the latter, the evolution of hairpin filaments when subjected to a mean background flow obtained from stratified Ekman flow simulations of \citet{ansorge2016thesis} is studied under different degrees of stratification. A comparison is also made between the filament simulation and tracking of a $Q$-criterion structure in the DNS data with a feature tracking scheme in section \ref{sec: compare_MLP_DNS}. Since our \rk{implementations of the thin-tube filament numerics} do not account for the effect of gravity on the self-induced motion of the \rk{vortex}, a theoretical \rk{justification via matched asymptotics in the regime of weak density variations} is carried out in section \ref{sec: mathematical_formulation}. \rk{Our conclusions are presented in section \ref{sec: conclusion}}.

% ----- NUMERICAL METHODS ----- %

\section{Numerical methods}
\label{sec: numerical_methods}

\begin{figure}
    \centering
    \includegraphics[width =  0.59\linewidth]{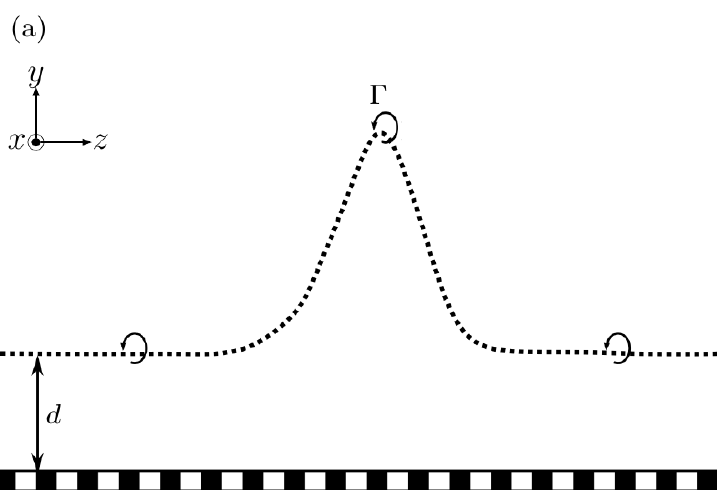}
    \includegraphics[width = 0.39\linewidth]{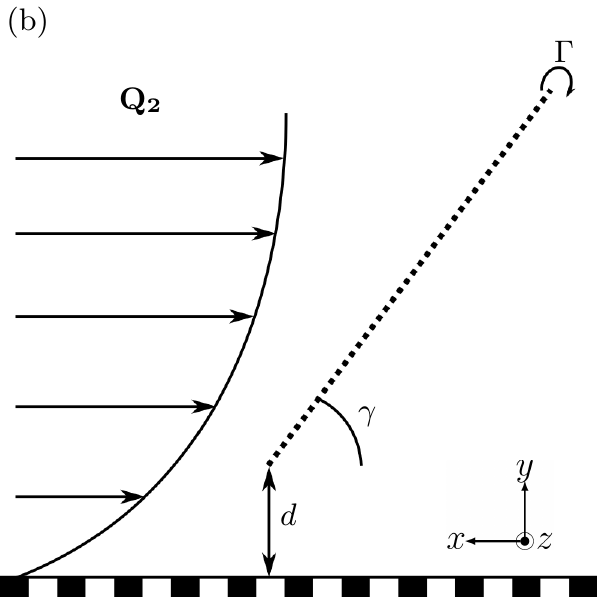}
    \caption{Illustration of (a) hairpin shaped filament (dashed line) when viewed from the front. The filament is placed at a distance $d$ away from the lower boundary. (b) shows the hairpin filament when viewed from the side including the background shear flow, $\mathbf{Q}_2$. Here the filament can be seen tilted at an angle $\gamma$ from the wall.}
    \label{fig:Hairpin_front_side_view}
\end{figure}

For all simulations in this work, the initial configuration of the hairpin is chosen as a small, symmetrical, three-dimensional perturbation as seen in the work of \citet{hon1991evolution} which is given by,
\begin{equation} \label{eq: hairpin_initial}
    \mathbf{X}(s,t) = A\left[(\cos\gamma) \mathbf{\hat{i}} + (\sin\gamma) \mathbf{\hat{j}}\right]e^{-\beta s^2} + \mathbf{\hat{j}} + s \mathbf{\hat{k}}
\end{equation}
where $(\mathbf{\hat{i}},\mathbf{\hat{j}},\mathbf{\hat{k}})$ denote the unit vectors along the streamwise $(x)$, wall-normal $(y)$ and spanwise $(z)$ directions, respectively. The perturbation is symmetric around $s=0$ with an amplitude $A$, and tilted at an angle $\gamma$ with respect to the wall as sketched in figure \ref{fig:Hairpin_front_side_view}. The parameter $\beta$ is simply a large number which controls the initial width of the perturbation.

As stated in the introduction, two methods are implemented to simulate the motion of the centerline, namely (a straightforward discretization of) the LIA and the corrected thin-tube model of \citet{KleinKnio1995}. 

\subsection{Local Induction Approximation} 
\label{subsec: LIA}

Since Local Induction Approximation (LIA) has been discussed extensively in previous works\citep{arms1965localized, zhou1996numerical, margerit2004implementation, batchelor2000introduction}, only a brief overview is presented here. LIA neglects both the long-distance induction effects, $\mathbf{Q}_0 (s, t)$, and the local effects from the core vorticity distribution, $C(t)$. It assumes a small core radius such that the only contribution to the motion of the filament is due to its curvature. This leads to a simplified equation of motion which is written as,
\begin{equation}
\label{eq: LIA_equation_of_motion}
\frac{\partial }{\partial t} \mathbf{X}(s, t) = \frac{\Gamma}{4\pi}\kappa \mathbf{b}(s, t) \ln \bigg(\frac{2}{\delta} \bigg)
\end{equation}
where $\kappa \mathbf{b}(s, t)$ is the curvature in the binormal direction. %The LIA is formally justified in the limit of $\ln(L/\rho) \gg 1$.
\citet{klein1991selfa, klein1991selfb} note that this binormal term alone cannot account for \rk{the self-stretching of vortex filaments, which is due entirely to nonlocal induction}. Therefore, LIA is used only for validation purposes in this paper. In order to make the results of LIA comparable to the corrected thin-tube model presented in the following subsection, we follow the work of \citet{margerit2004implementation} who used the Callegari and Ting equation \citep{CallegariTing1978} without the non-local self-induction term $\mathbf{Q}_0 (s, t)$. For a non-closed, infinite filament, the equation of motion is witten as follows,
\begin{equation}
\label{eq: LIA_motion_DM}
\frac{\partial }{\partial t} \mathbf{X}(s, t) = \frac{\Gamma }{4 \pi} \kappa \mathbf{b}(s, t) \bigg[\ln \bigg(\frac{2}{\delta} \bigg) + C(t) \bigg]
\end{equation}
where $C(t)$ is the core structure coefficient. It is natural to see this local term as an $\mathcal{O}(1)$ correction to the local induction contribution of the self induced velocity and to call it the Local Induction Approximation (or contribution) at $\mathcal{O}$(1). 

\subsection{M1 corrected thin-tube model (M1 KK)} \label{subsec: M1_ttm}

In this method, slender vortices are represented as a chain of overlapping elements satisfying the following overlap condition,
\begin{equation}\label{eq: overlap_condition}
    \underset{i = 1..N} {max} |\delta \bm{\chi}_{i}| < \delta
\end{equation}
Here, $\delta$ is the core radius and $\{\bm{\chi}_i\}_{i = 1}^{N}$ are $N$ vortex elements along the filament centerline. The total vorticity experienced by the filament is given as,
\begin{equation}\label{eq: vorticity_formulation}
    \bm{\omega}(\bm{x}, t) = \sum_{i = 1}^{N} \Gamma \delta \bm{\chi}_{i}(t)f_{\delta}(\bm{x} - \bm{\chi}_{i}^c(t))
\end{equation}
In this formula, $\Gamma$ is a time-independent constant circulation, $\bm{\chi}_i^c(t)$ denote the \rk{centers and $\delta \bm{\chi}_i(t)$ the secant vectors that approximate the filament centerline and are positively aligned with the vorticity}, \emph{i.e.,}
\begin{equation}\label{eq: length_vortex_element}
    \delta \bm{\chi}_i(t) = \bm{\chi}_{i+1}(t) - \bm{\chi}_i(t)\,,
    \quad
    \bm{\chi}_i^c(t) = \frac{\bm{\chi}_{i+1}(t) + \bm{\chi}_i(t)}{2}\,.
\end{equation}
The smoothing function $f_{\delta}$ is related to a \rk{rapidly decaying numerical} core vorticity distribution and is given by,
\begin{equation}\label{eq: core_smoothing_function}
    f_{\delta} = \frac{1}{\delta^3}f \bigg(\frac{|\bm{x}|}{\delta} \bigg)
\end{equation}
The velocity is obtained by inserting equation (\ref{eq: vorticity_formulation}) in the three-dimensional Biot-Savart integral,
\begin{equation}
\label{eq: biot_savart_law_solenoidal}
\bm{v}(\bm{x}, t) = - \frac{1}{4 \pi} \iiint \frac{\bm{x} - \bm{x'}}{|\bm{x} - \bm{x'}|^3} \times \omega(\bm{x'}) d\bm{x'} 
\end{equation}
where $d\bm{x'} = dx_1'dx_2'dx_3'$ is a volume element. The result reads,
\begin{equation}\label{eq: standard_ttm}
    \bm{v}^{\text{ttm}}(\bm{x}, t) = -\frac{\Gamma}{4\pi} \sum_{i = i}^{N} \frac{(\bm{x} - \bm{\chi}_i^c(t)) \times \delta\bm{\chi}_i(t)}{|\bm{x} - \bm{\chi}_i^c(t)|^3} \kappa_{\delta}(\bm{x} - \bm{\chi}_i^c(t))
\end{equation}
and $\kappa_{\delta} \equiv \kappa(|\bm{x}|/\delta)$ is the velocity smoothing function\rk{, which is directly related to the numerical core vorticity distribution $f$ from \eqref{eq: core_smoothing_function}}. Eq. ($\ref{eq: standard_ttm}$) is the \textit{standard} thin-tube model used by \citet{Chorin1980, knio1990numerical}. Since this model assumes that the induced velocity at the nodes \rk{by the \emph{numerical} core vorticity distrbution} is the local filament velocity, it is prone to $\mathcal{O}(1)$ errors. To circumvent this, \citet{KleinKnio1995}, propose three correction strategies based on an asymptotic analysis of the numerical vorticity structure. We choose the third method due to its simplicity. This involves a rescaling of the numerical core radius as follows,
\begin{equation}\label{eq: rescaling_core_radius}
    \delta^{\text{ttm}} = \delta \, \exp \left(C^{\text{ttm}} - C\right) \,.
\end{equation}
Here, $C^{\text{ttm}}$ is the numerical core constant. If we choose the velocity core smoothing function $\kappa(r) = \tanh(r^3)$, then $C^{\text{ttm}} = -0.4202$ as obtained by \citet{knio2000improved}. According to the asymptotic theory, the core structure coefficient $C$ includes contributions from the local swirling and axial velocities, denoted by $C_v$ and $C_w$ respectively,
\begin{equation}\label{eq: core_structure_coefficient}
    C = -1 + C_v + C_w
\end{equation}
Depending on the presence/absence of viscous effects and the initial leading-order velocity profile in the core, different expressions of the core structure coefficients can be obtained. Following \citet{ting1991viscous} for the case of a similar vortex core with the same initial core size and circulation without axial flow, $C_v$ and $C_w$ are given by,
\begin{equation}\label{eq: similarity_solutions_1}
    C_v = \frac{1 + \gamma_E - \ln(2)}{2} - \ln(\delta)
\end{equation}
\begin{equation}\label{eq: similarity_solutions_2}
    C_w = - 2\bigg[\frac{m(0)}{\Gamma \delta}\bigg]^2 \bigg[\frac{S_0}{S(t)}\bigg]^4
\end{equation}
where $S_0$ is the initial length of the filament, $m(0)$ is the initial axial flux of the vortex and $\gamma_E = 0.577$ is Euler's constant. In our case, $m(0) = 0$ implying no contribution from axial velocity. We remark that \eqref{eq: core_structure_coefficient} is usually written as $C = C_v + C_w$ (for instance, see (2) of \citet{knio2000improved}). The inclusion of $-1$ accounts for the difference in the definition of $C_v$ (compare 2.3.73e of \citet{ting1991viscous} and (3) of \citet{knio2000improved}).
\begin{figure*}
    \centering
    \includegraphics[width = 0.9\linewidth]{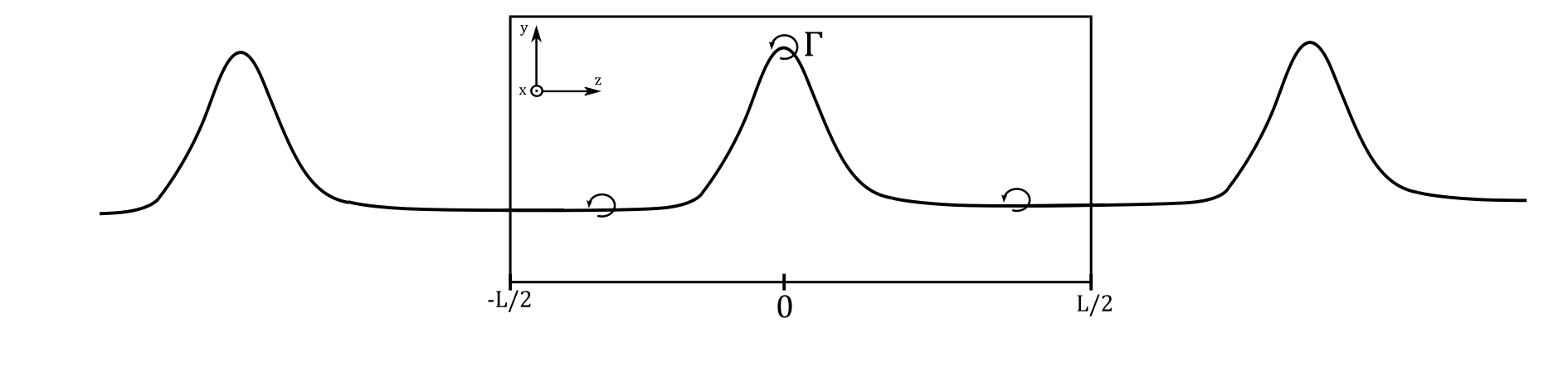}
    \caption{Illustration of the periodic boundary conditions. A long filament, periodic in the axial direction is envisioned. The domain of integration of length $L$ contains one complete period.}
    \label{fig:Hairpin_boundarycond}
\end{figure*}

\noindent \textit{Boundary conditions:} For hairpin filaments which are periodic in the spanwise direction and embedded in an unbounded domain in the other two directions, the total velocity $v^{\text{ttm}}$ requires contributions from an infinite number of images in addition to the elements within the computational domain \citep{knio1991three}. Therefore, (\ref{eq: standard_ttm}) is written as,
\begin{equation}\label{eq: standard_ttm_bc_included}
\bm{v}^{\text{ttm}}(\bm{x}, t) = -\frac{\Gamma}{4\pi} \sum_{k = \pm 1}^{\pm \infty} \sum_{i = i}^{N} \frac{(\bm{x} - \bm{\chi}_i^c(t)) \times \delta\bm{\chi}_i(t)}{|\bm{x} - \bm{\chi}_i^c(t)|^3} \kappa_{\delta}(\bm{x} - \bm{\chi}_i^c(t))
\end{equation}
To overcome the evaluation over an infinite number of images, the equation is split into two components,
\begin{equation}\label{eq: standard_ttm_split}
    \bm{v}^{\text{ttm}} (\bm{x}) = \sum_{i = 1}^{N} (\bm{v}_{\text{center}} (\bm{x}) + \bm{v}_{\text{image}} (\bm{x}))
\end{equation}
where $v_{\text{center}}$ is the contribution from the central part of the domain and $v_{\text{image}}$ from the images on the left and right side. They are computed as follows,
\begin{equation}\label{eq: standard_ttm_center}
    \bm{v}_{\text{center}}(\bm{x}) =  -\frac{\Gamma}{4\pi} \sum_{i = i}^{N} \frac{(\bm{x} - \bm{\chi}_i^c(t)) \times \delta\bm{\chi}_i(t)}{|\bm{x} - \bm{\chi}_i^c(t)|^3} \kappa_{\delta}(\bm{x} - \bm{\chi}_i^c(t))
\end{equation}
\begin{equation}\label{eq: standard_ttm_images}
    \bm{v}_{\text{image}}(\bm{x}) =  -\frac{\Gamma}{4\pi} \sum_{k = \pm 1}^{\pm P} \sum_{i = i}^{N} \frac{(\bm{x} - \bm{\chi}_i^c(t)) \times \delta\bm{\chi}_i(t)}{|\bm{x} - \bm{\chi}_i^c(t)|^3} 
\end{equation}

For the image contribution, the effect of the velocity smoothing function is neglected due to the assumption of $L \gg \delta$ and $\pm P$ is a cutoff number chosen to represent the number of images on either side of the domain. The choice of $P$ is explained in Appendix \ref{appendix: K_alpha}. \\

\noindent \textit{Optimization:} To compensate for the high resolution requirements for thin vortices, \citet{knio2000improved} suggested three optimization techniques. The first method, henceforth referred \rk{to} as M1 technique, is attractive and used in this work since it requires only minor modifications to the existing code and doesn't require computation of the curvature, $\kappa$. The corrected velocity is obtained through a Richardson-type extrapolation in the core size parameter as follows,
\begin{equation}\label{eq: M1_KK_vel}
    \bm{v}^{\text{ttm}}_{\text{corr}} = \bm{v_1} + (\bm{v_1} - \bm{v_2}) \frac{\ln \, (\sigma_1/\delta^{\text{ttm}})}{\ln \, \phi}
\end{equation}
where $\bm{v_1}$ and $\bm{v_2}$ are the velocities corresponding to two large core sizes $\sigma_1$ and $\sigma_2$. If $\sigma_0 (t) = \underset{i = 1..N} {max} |\delta \bm{\chi}_{i}|$ denotes the inter-element separation distance, then 
\begin{equation}\label{eq: fat_mesh_core}
    \sigma_1 = K \, \sigma_0, \ \ \sigma_2 = \phi \, \sigma_1\,.
\end{equation}
The constants $K$ and $\phi$ are chosen as $3$ and $2$, respectively, and the choice of these parameters are described in Appendix \ref{appendix: K_alpha}. The vortex elements move along Lagrangian trajectories with the following equation of motion,
\begin{equation}\label{eq: equation_of_motion}
    \frac{d\bm{\chi}_i(t)}{dt} = v^{\text{ttm}}_{\text{corr}}(\bm{\chi}_i(t), t)
\end{equation}
A summary of the numerical scheme is presented as follows,
\begin{itemize}
    
    \item[(1)] First, the initial configuration of the hairpin is setup.
    
    \item[(2)] For the chosen velocity core smoothing function $\kappa(r) = \tanh(r^3)$, $C^{\text{ttm}}$ is set to $-0.4202$.  The numerical core radius $\delta^{\text{ttm}}$ is computed from (\ref{eq: rescaling_core_radius}). With $\sigma_0 (t) = \underset{i = 1..N} {max} |\delta \bm{\chi}_{i}|$, (\ref{eq: fat_mesh_core}) can be used to compute the two coarse radii. 
    
    \item[(3)] For both core radii, (\ref{eq: standard_ttm}) is evaluated at each node location by applying the periodic boundary condition as illustrated in figure \ref{fig:Hairpin_boundarycond}, i.e., the velocity at a node $x$ inside the computational domain is evaluated by applying a translation along the periodicity direction such that $x$ is at the center of the domain. 
    
    \item [(4)] The corrected velocity is obtained according to (\ref{eq: M1_KK_vel}).
    
    \item [(5)] Finally, the node positions are updated with the equation of motion (\ref{eq: equation_of_motion}) with a fifth order Adams-Bashforth scheme and Runge-Kutta-Fehlberg initialization\citep{butcher2016numerical, hairer1993runge}. 
\end{itemize}
Both LIA and the M1 KK methods are written in python and the filament code is validated against the static test of \citet{KleinKnio1995} which is shown in Appendix \ref{appendix: validation_tests}.

% ----- STAGNANT BG ----- %

\section{Temporal evolution of a hairpin filament in stagnant background flow ($\mathbf{Q_2} = 0$)}
\label{sec: stagnant_flow}

\begin{figure*}
    \centering
    \includegraphics[width = \textwidth]{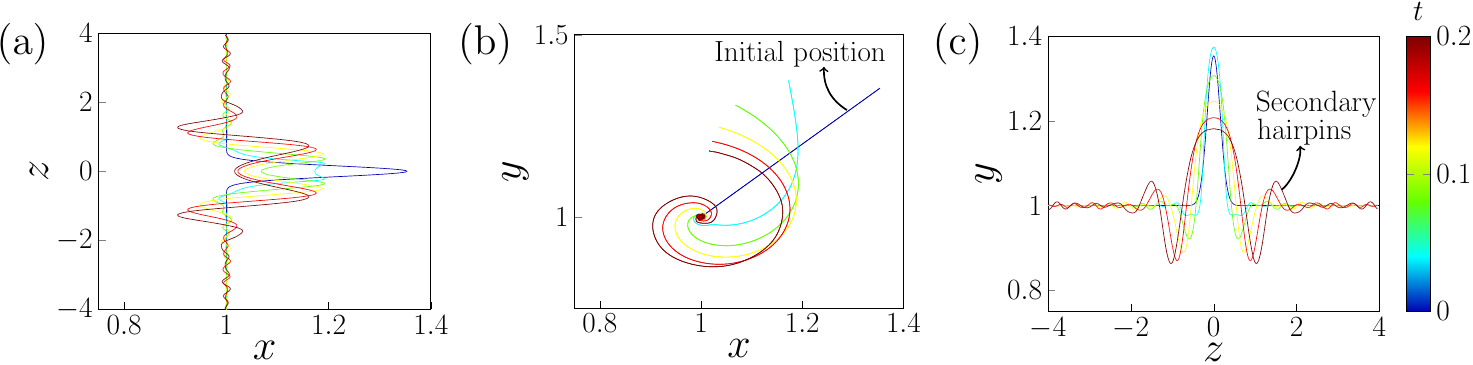}
    \includegraphics[width = \textwidth]{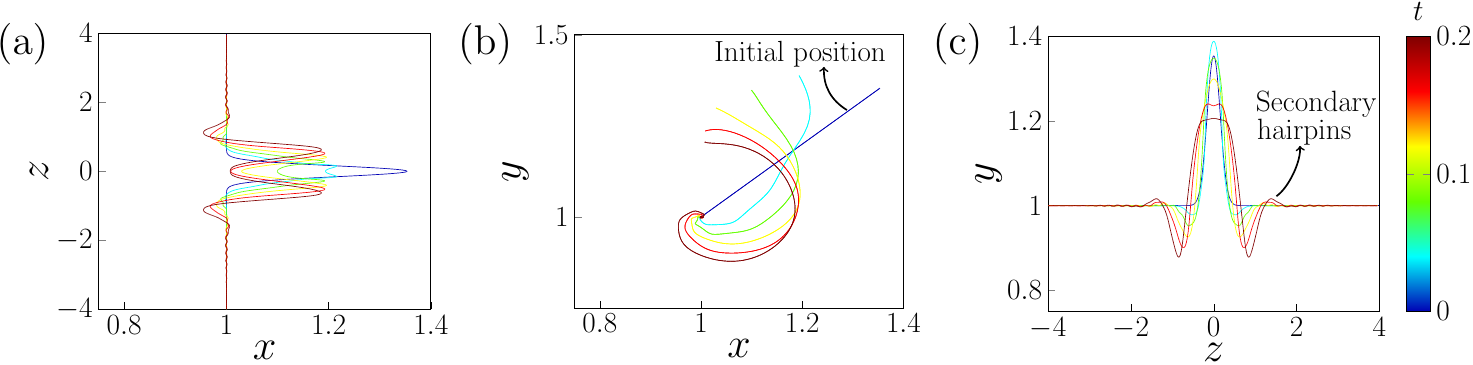}
    \caption{Temporal evolution of a hairpin vortex in a stagnant background flow. (a, b, c) are the top, side and front view of the hairpin, respectively. The simulation performed with LIA is shown in the top panel and M1 KK method in the bottom panel.
    }
    \label{fig: LIA_M1_HW_comparison}
\end{figure*}

In this section, we consider a case \rk{in which} a hairpin filament is allowed to evolve in a stagnant background flow, i.e., $\mathbf{Q_2} = 0$ as shown in the work of \citet{hon1991evolution}. Both \rk{the} LIA and M1 KK methods are used to compute the motion of the filament. The goal of this section is twofold: (1) this case serves as a further validation of the filament code written during the course of this work, and (2) it also highlights the differences between LIA and the M1 KK \rk{approximations}. The hairpin is initially inclined at $\gamma = 45^{\circ}$ with respect to the streamwise direction, having an amplitude $A = 0.5$, spread parameter $\beta = 20$ and a dimensionless core size $\delta = 0.02$. Additionally, circulation is set as $\Gamma = 1$ and a spatial discretization of $700$ nodes over a length of $L = 4$ is used. In all simulations, care is taken to ensure that an adequate number of nodes are used and further refinement does not alter the results and conclusions presented in the paper. 

\begin{figure*}
    \centering
    \includegraphics[width = 0.7\textwidth]{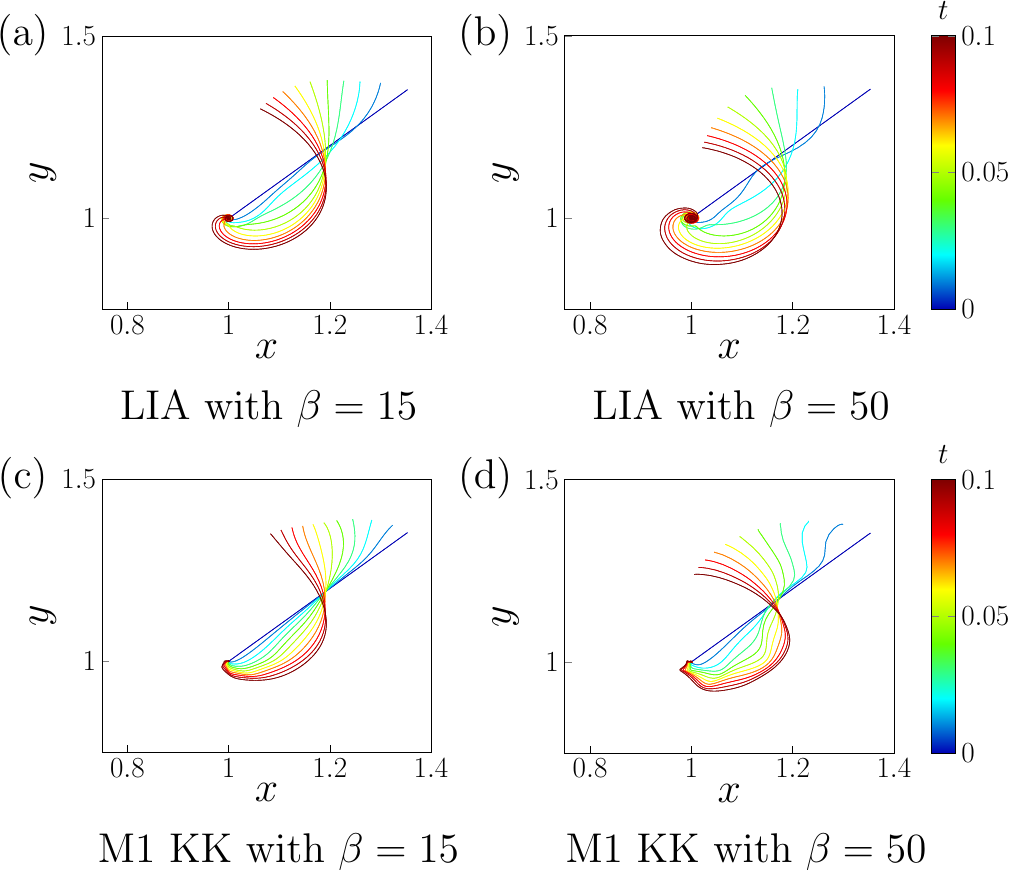}
    \caption{The test shown in figure \ref{fig: LIA_M1_HW_comparison} is repeated with $\beta = 15, 50$ for both methods. (a, b) show the temporal evolution of the hairpin with LIA whereas (c, d) shows the temporal evolution with the M1 KK method. }
    \label{fig: beta_test}
\end{figure*}

The results of the simulation are shown in figure \ref{fig: LIA_M1_HW_comparison}. For LIA, high temporal resolution is necessary to guarantee the smoothness of the filament curve required to allow for the accurate evaluation of its curvature. Accordingly, a time step of $\Delta t = 10^{-5}$ is chosen. The simulation is stopped at $t = 0.2$ when the disturbance or ``wiggles'' have propagated to the ends of the domain. The M1 KK scheme allows for a larger time step $\Delta t = 10^{-3}$ as it \rk{not only} avoids the evaluations of local curvature $\kappa$ where higher-order derivatives need to be computed \citep{knio1990numerical, margerit2004implementation}\rk{, but also is much better conditioned owing to the Richardson-type extrapolation from the artificially enlarged to the actual core size in the correction scheme M1}. \rk{As for} LIA, the simulation is stopped at time $t = 0.2$. Results from both panels in figure \ref{fig: LIA_M1_HW_comparison}(b) show that the methods correctly capture the ``corkscrew'' shape described in \citet{hon1991evolution}, \rk{with the head of the vortex moving} back and towards the wall rapidly due to self-induction in the counter-clockwise direction. Other features such as the formation of hairpin ``legs'' and the secondary hairpins on either side of the main disturbance are also visible. Similar findings were also reported by \citet{moin1986evolution}, who studied a parabolic vortex filament with a cutoff method. See section III A of \citet{moin1986evolution} for details on their setup. 

Keeping everything else constant, two values of spread parameter (which controls the initial width of the perturbation) $\beta = 15, 50$ are tested. The results are shown in figure \ref{fig: beta_test} where the simulation is stopped at $t = 0.1$ once the general trends were evident. It should be noted that the simulation for $\beta = 50$ was run with $1300$ nodes and for LIA, with an even smaller time step $\Delta t = 10^{-6}$ to obtain a stable evolution. It is immediately apparent that there is a difference in the shape of the hairpin legs and how they evolve, particularly at $\beta = 50$. This gives a direct comparison to the hairpin evolution under the absence of nonlocal effects with LIA and when they are correctly represented \rk{by the} M1 KK method. Therefore, all further simulations are carried out only using the M1 KK method. 

\section{Temporal evolution of a hairpin filament in ABL background flow}
\label{sec: ABL_flow}

The evolution of a hairpin filament immersed in an atmospheric boundary layer flow is studied here. In this section, as mentioned in the introduction, we attempt to answer \rk{our questions regarding} the abundance and orientation of hairpin structures in the stably stratified ABL by studying the dynamic hairpin characteristics with respect to changes in stratification. First, the mean background flow profiles are computed from the DNS database and discussed in subsection \ref{subsec: background_flow}. Then, with four initial conditions of the hairpin identified in subsection \ref{subsec: initial_conditions}, the temporal evolution of the hairpin filament is discussed in subsections \ref{subsec: initial_evolution}, \ref{subsec: later_evolution}.

\subsection{Background flow}
\label{subsec: background_flow}

\begin{figure*}
    \centering
    \includegraphics[width = \textwidth]{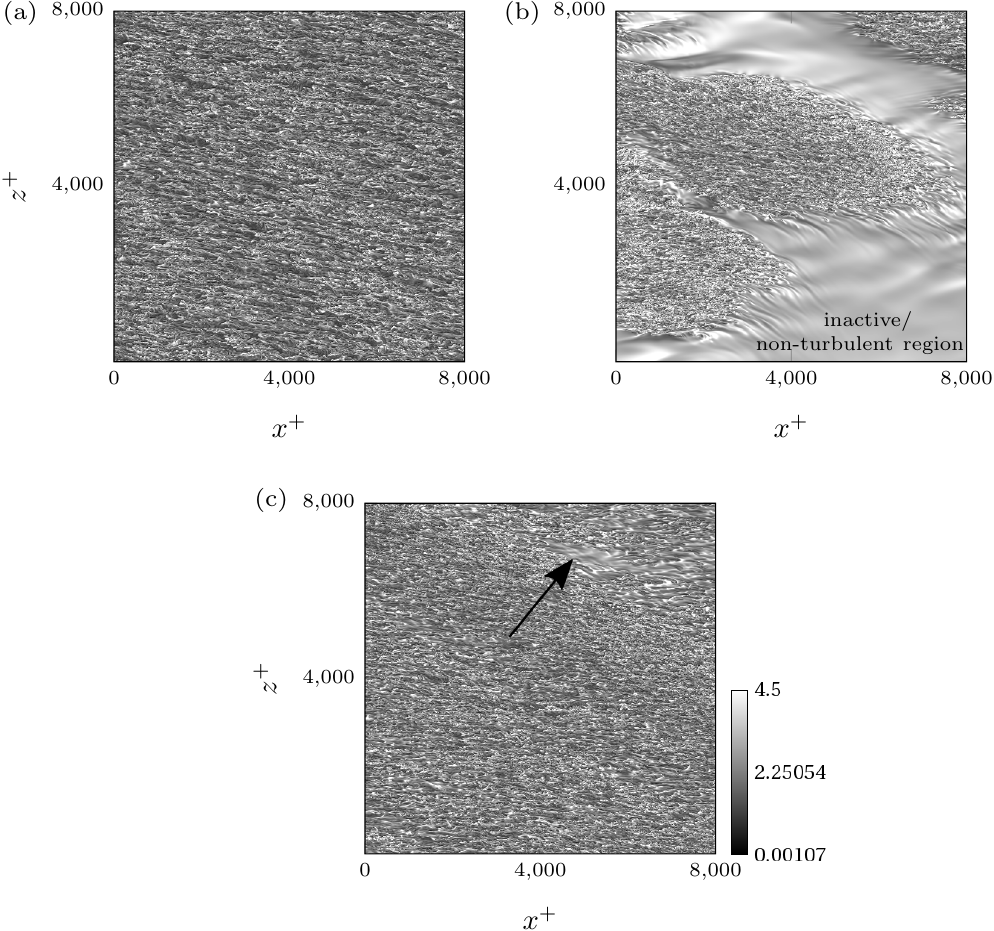}
    \caption{Vorticity magnitude slices are shown at $y^+ \approx 30$. (a, b, c) correspond to case N, S\_1, S\_2, respectively. The inactive/non-turbulent regions are visible in both stably stratified cases. Large patches \rk{of such relatively quiescent air} are clearly visible in (b) whereas an example of \rk{a} patch is indicated by an arrow in~(c).}
    \label{fig: vorticity_magnitude_slice}
\end{figure*}

Before the initial conditions for the hairpin simulations are discussed, a preprocessing step is carried out on the DNS database of \citet{harikrishnan2021geometry} to obtain the mean background flow $\mathbf{Q_2}$. Details regarding the numerical simulation can be found in \citet{ansorge2014global, ansorge2016analyses} and \citet{ansorge2016thesis}. In this paper, the hairpins are simulated in a mean background flow composed of two stably stratified cases (S\_1, S\_2) with different degrees of stratification and a neutrally stratified case (N). By studying the temporal evolution of vertically integrated Turbulent Kinetic Energy (TKE), both S\_1 and S\_2 are classified under the very stable regime \rk{in which} large patches of inactive/non-turbulent regions are visible. This can be seen from figure \ref{fig: vorticity_magnitude_slice}(b, c). The simulation parameters for the three cases are summarized in table \ref{tab: simulation_parameters}. The strength of stratification is quantified with the dimensionless bulk Richardson number ($\text{Ri}_{\text{B}}$) defined as
\begin{table*}[]
    \centering
    \begin{tabular}{| c | c | c | c | c |}
        \hline
        Case & Line specification & Bulk Richardson number  & Froude number & Reynolds number\\
        & & ($\text{Ri}_{\text{B}}$) & ($\text{Fr}$) & ($\text{Re}$)  \\
        \hline
        N & ---------- & $0$ & $\infty$ & \\
        S\_1 & ............. & $2.64$ & $0.02$ & $26\, 450$ \\
        S\_2 & -\,-\,-\,-\,-\,-\,- & $0.58$ & $0.07$ &\\
        \hline
    \end{tabular}
    \caption{The parameters of the DNS simulations used to obtain the background flow $\mathbf{Q_2}$ are listed here. Cases with prefix S indicate stable stratification whereas N is the neutrally stratified case. The Reynolds number is defined here with the boundary layer height $\delta_{\text{h}}$, i.e., $\text{Re} = G\delta_{\text{h}}/\nu$ where $G$ is the geostrophic wind velocity and $\nu$ is the kinematic viscosity.}
    \label{tab: simulation_parameters}
\end{table*}
\begin{figure*}
    \centering
    \includegraphics[width = \textwidth]{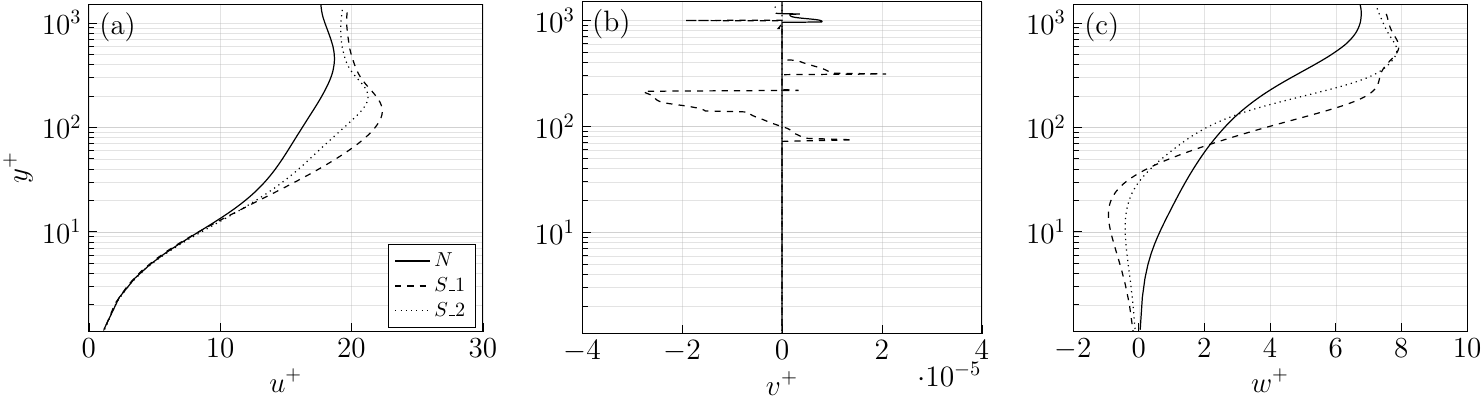}
    \caption{Mean velocity profiles for every wall-normal height until $y^+ = 1500$ \rk{are} plotted for all velocity components. $(\cdot)^+$ \rk{indicates} viscous or wall units. (a, b, c) correspond to \rk{the} streamwise, wall-normal and spanwise \rk{velocities,} respectively.}
    \label{fig: mean_velocity_components}
\end{figure*}
\begin{equation}\label{eq: richardson_number}
\text{Ri}_{\text{B}} \equiv \frac{B_0 \delta_{\text{h}}}{G^2}
\end{equation}
where $B_0$ is the difference in buoyancy between the top and bottom layer, $\delta_{\text{h}}$ is the boundary layer height under neutral conditions and $G$ is the geostrophic wind velocity \rk{magnitude}. This is related to the Froude number ($\text{Fr}$), another parameter for quantifying the strength of stratification, as $\text{Ri}_{\text{B}} = \text{Fr}^{-1}(\delta_{\text{h}}/\Lambda)$ where $\Lambda$ is the Rossby deformation radius. The streamwise ($x$), wall-normal ($y$) and spanwise ($z$) direction along with their corresponding \rk{velocities} are represented in terms of viscous or wall units. For instance,
\begin{equation}\label{eq: wall_units}
    x^+ \equiv \frac{x u_{\tau}}{\nu}, \quad u^+ = \frac{u}{u_{\tau}}
\end{equation}
where $u_{\tau}$ is the friction velocity and $\nu$ is the kinematic viscosity. $\mathbf{Q_2}$ is obtained by computing the horizontal $(x,z)$ mean velocity at every wall-normal height of the flow field. The mean velocity profiles are shown in figure \ref{fig: mean_velocity_components}.

\subsection{Initial conditions}
\label{subsec: initial_conditions}

As described in section \ref{sec: numerical_methods}, the first step of the numerical scheme is to set-up the initial configuration of the hairpin by fixing unknown physical and numerical parameters. We inspect the DNS database to inform our choice on the selection of parameters. 

\begin{figure}
    \centering
    \includegraphics[width = 0.5\textwidth]{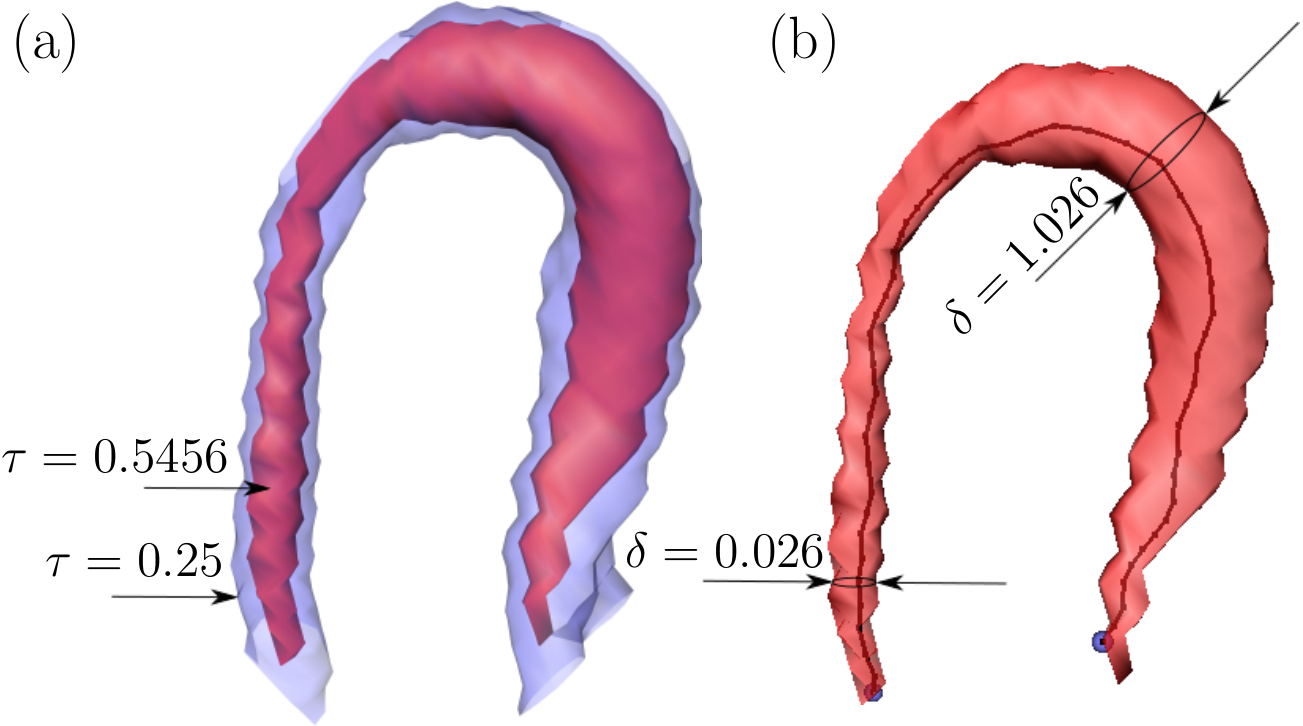}
    \caption{The uncertainty of core size estimation is shown in (a) where isosurfaces of the same structure at two thresholds \rk{of the $Q$-criterion} are visualized. The core sizes are illustrated at two points in (b). Here, $\tau$ is the threshold and $\delta$ is the dimensionless core size parameter.}
    \label{fig: core_size_estimation}
\end{figure}

\begin{table*}[]
	\centering
	\begin{tabular}{| c | l l l | l l l | l l l |}
		\hline
		\multirow{3}{*}{Core size parameter} & \multicolumn{9}{c}{Ekman flow case} \vline \\
		\cline{2-10}
		& \multicolumn{3}{c}{$\text{N}$} \vline & \multicolumn{3}{c}{$\text{S}\_1$} \vline & \multicolumn{3}{c}{$\text{S}\_2$} \vline\\
		\cline{2-10}
		& H1 & H2 & H3  & H1 & H2 & H3  & H1 & H2 & H3 \\
		\hline
		
		$\delta_{\text{min}}$ & $0.024$ & $0.016$ & $0.12$  &  $0.026$ & $0.17$ & $0.13$  & $0.048$ & $0.025$ & $0.026$\\
		
		$\delta_{\text{max}}$ & $1.26$ & $1.89$ & $1.63$  & $1.92$ & $3.45$ & $5.57$  & $5.79$ & $4.18$ & $1.31$ \\
		
		$\delta_{\text{mean}}$ & $0.52$ & $0.49$ & $0.74$  &  $0.52$ & $1.1$ & $1.05$  & $0.92$ & $0.70$ & $0.33$\\
		
		\hline
	\end{tabular}
	\caption{The minimum, maximum and mean values of the dimensionless core size parameter estimated along the centerline of three hairpin-like structures are shown for case N, S\_1 and S\_2. H1, H2, H3 are three randomly chosen hairpin-like structures.}
	\label{tab: estimated_core_sizes}
\end{table*}

Three hairpin-like structures identified using the $Q$-criterion indicator are extracted and examined for case N, S\_1 and S\_2. The \rk{centerlines} for the extracted structures are obtained with the block-wise skeletonization method of \citet{fouard2006blockwise}. At each point on \rk{a} centerline, the diameter $d$ of the vortex core is estimated by fitting the largest sphere within the structure. Additionally, the radius of curvature $R$ at every point of the parametrized centerline $\mathbf{X}$ is given by\citep{abbena2017modern},
\begin{equation}
\label{eq: radius_of_curvature}
R = \frac{1}{\kappa} = \frac{|\mathbf{X}_s|^3}{\sqrt{|\mathbf{X}_s|^2|\mathbf{X}_{ss}|^2 - (\mathbf{X}_s \cdot \mathbf{X}_{ss})^2}}
\end{equation}
where $\mathbf{X}_s, \mathbf{X}_{ss}$ correspond to the first and second derivatives of the space curve along the parameterized centerline. The ratio $d/R$ is the \rk{local} dimensionless core size parameter of the filament \rk{(not to be confused with the asymptotic parameter $\delta$ from \eqref{eq: core_size_parameter}, which is a characteristic value of this quantity for a given filament)}. 

Along the hairpin, the \rk{local} core size parameter estimated from all hairpins and tabulated in \rk{table~\ref{tab: estimated_core_sizes}}, vary between $0.02$ at its thinnest point up to $5.79$ at its thickest. It is important to note that these estimations strongly depend on the threshold, denoted by $\tau$, as illustrated in \ref{fig: core_size_estimation}(a) where smaller thresholds imply thicker cores and vice versa. The thresholds used in this paper were obtained through the work of \citet{harikrishnan2021geometry} where optimum threshold values were computed with percolation analysis. It should also be noted that these estimations are made at a later stage in the lifetime of the hairpin-like structure. Figure \ref{fig: core_size_estimation}(b) shows the variations of the core size along the hairpin. For our simulations, we conservatively choose $\delta = 0.01, 0.05$ to examine the effect of changes in core size.

\begin{table*}[]
	\centering
	\small\begin{tabular}{| c | l l l | l l l | l l l |}
		\hline
		\multirow{3}{*}{Circulation} & \multicolumn{9}{c}{Ekman flow case} \vline \\
		\cline{2-10}
		& \multicolumn{3}{c}{$\text{N}$} \vline & \multicolumn{3}{c}{$\text{S}\_1$} \vline & \multicolumn{3}{c}{$\text{S}\_2$} \vline\\
		\cline{2-10}
		& H1 & H2 & H3  & H1 & H2 & H3  & H1 & H2 & H3 \\
		\hline
		
		\multirow{2}{*}{$\Gamma_{\text{min}}$} & $0.09$ & $0.018$ & $0.015$  &  $0.02$ & $0.049$ & $0.047$  & $0.026$ & $0.0076$ & $0.0014$\\
		& $-0.107$ & $-0.24$ & $-0.217$ & $-0.042$ & $-0.22$ & $-0.137$ & $-0.01$ & $-0.26$ & $-0.035$ \\
		\hline
		\multirow{2}{*}{$\Gamma_{\text{max}}$} & $0.17$ & $0.115$ & $0.318$  & $0.048$ & $0.13$ & $0.11$  & $0.38$ & $0.225$ & $0.049$ \\
		& $-0.009$ & $-0.02$ & $-0.185$ & $-0.048$ & $0.0013$ & $-0.032$ & $-0.0071$ & $-0.0098$ & $0.18$ \\
		\hline
		\multirow{2}{*}{$\Gamma_{\text{mean}}$} & $0.134$ & $0.077$ & $0.124$  &  $0.034$ & $0.094$ & $0.08$  & $0.202$ & $0.113$ & $0.031$\\
		& $-0.045$ & $-0.084$ & $-0.204$ & $-0.025$ & $-0.119$ & $-0.071$ & $-0.0425$ & $-0.122$ & $0.11$ \\
		\hline
	\end{tabular}
	\caption{The minimum, maximum and mean values of the dimensionless core size parameter estimated along the centerline of three hairpin-like structures are shown for case N, S\_1 and S\_2.}
	\label{tab: estimated_circulation}
\end{table*}

For the same hairpin-like structures, circulation is computed on both legs for every wall-normal plane until the head of the structure. The estimated values are listed in table \ref{tab: estimated_circulation} which show positive values on one leg and negative values on the other indicating opposing direction of rotation. Similar to the core size, two values of circulation $\Gamma = 0.01, 0.05$ are chosen to study their impact on the evolution of hairpin.

\begin{table*}[]
	\centering
	\begin{tabular}{| c | c | c | c | c | c | c |}
		\hline
		Case & Core size & Circulation & Spread & Inclination & Amplitude & Length \\
		& ($\delta$) & ($\Gamma$) & ($\beta$) & ($\gamma$) & ($A$) & ($L$) \\
		\hline
		$R_1$ & $0.01$ & $0.01$ & \multirow{4}{*}{$7500$} & \multirow{4}{*}{$0.1^{\circ}$} & \multirow{4}{*}{$10$} & \multirow{4}{*}{$200$}\\
		$R_2$ & $0.05$ & $0.01$ &  &  &  & \\
		$R_3$ & $0.01$ & $0.05$ &  &  &  & \\
		$R_4$ & $0.05$ & $0.05$ &  &  &  & \\
		\hline
	\end{tabular}
	\caption{The four initial configurations of the hairpin chosen for the simulations with ABL background flow.}
	\label{tab: initial_conditions}
\end{table*}

In their experimental study, \citet{acarlar1987study2} showed that low-speed momentum regions (which are also referred to as low-speed streaks) introduced in sub-critical laminar boundary layer oscillate and later break down into hairpin vortices. Therefore, the spread parameter $\beta$ is chosen by computing the average spanwise width over all individual low-speed streaks at a particular height. From figure \ref{fig: hairpin_visualization}(b), numerous hairpin structures can be seen around $y^+ = 50$. Therefore, for this initial height, the mean spanwise width of the low-speed streaks is about $60$ viscous units for which corresponds to $\beta = 7500$.

All other physical parameters are chosen to represent a small perturbation. Therefore, an initial amplitude corresponding to $10$ viscous units and a small angle of inclination $\gamma_{\text{initial}} = 0.1^{\circ}$ are chosen along with a domain length of $200$ viscous units. The initial configurations of the hairpins are listed in table \ref{tab: initial_conditions}.

\subsection{Initial evolution of the hairpin filament}
\label{subsec: initial_evolution}

The effect of stable stratification on the evolution of a hairpin filament is studied with the help of the mean background profiles obtained in subsection \ref{subsec: background_flow}. Before discussing the results, we present suitable arguments to address their validity. 

\begin{figure*}
    \centering
    \includegraphics[width = \textwidth]{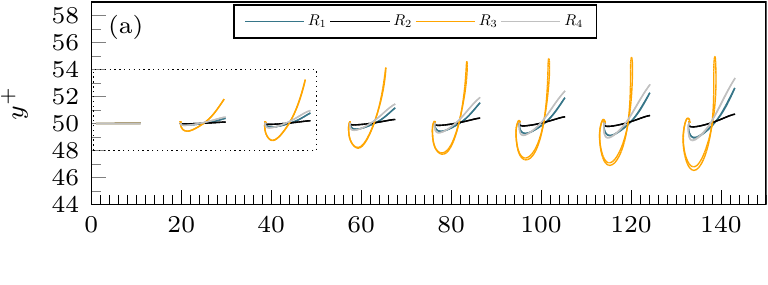}
    \includegraphics[width = \textwidth]{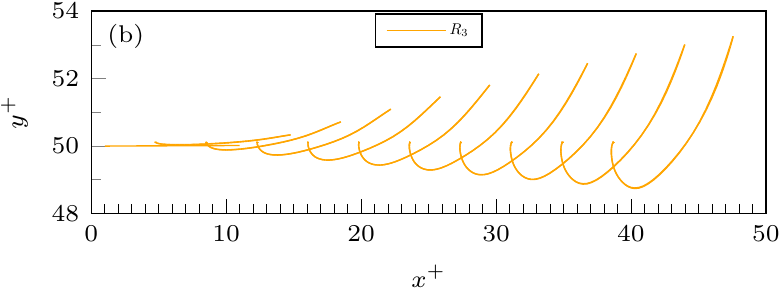}
    \caption{(a) Temporal evolution for all four initial conditions listed in table \ref{tab: initial_conditions} are shown here for case S\_1. The dotted region is zoomed in for initial condition $R_3$ in (b).}
    \label{fig: R1_R2_R3_R4_comparison_FR02}
\end{figure*}

\begin{itemize}
    \item [(1)] As pointed out in the work of \citet{aref1984dynamics}, using a Blasius or boundary-layer background profile for $\mathbf{Q_2}$ requires the implementation of a boundary condition at the wall. This is generally achieved with an image vortex placed behind the wall at the same distance as that of the vortex being simulated. However, \citet{aref1984dynamics} also argue that including the image vortex is only necessary when the filament is approaching the wall very closely. This was examined by \citet{moin1986evolution} who noted that the effect of the image vortex increases as the filament moves closer to the wall. This can be seen in appendix \ref{appendix: image_vortex} below, where the image vortex has a stronger impact on the filament evolution at $y^+ = 15$ than at $y^+ = 30$. Hence, image vortices are used only for these two lower heights and ignored for simulations at higher heights. 
    
    \item [(2)] Although the effect of the background flow \rk{on the motion of the hairpin filament is studied}, it should be noted that this is simply an approximation since there is no feedback mechanism in place, i.e., the action of the filament on the background flow is neglected. \citet{aref1984dynamics} show with an order of magnitude estimate that the neglected \rk{effect} (including vortex stretching) is small provided that \rk{the} core size of the filament is very small compared to the length scale of the background flow field. This condition is fulfilled in our simulations \rk{in which} the core size is several orders of magnitude smaller than the vertical length scale of background flow field. 
    %\RK{Which length scale of the background flow are you referring to here? Is it the boundary layer height or the horizontal extent of the simulation domain? I think it should be the former. Is the ``several orders of magnitude'' assessment still valid under that premise?}
    
    \item [(3)] Viscous effects altering the initial core structure of the filament are also neglected. \citet{moin1986evolution} point out that viscous effects become significant only when two vortex cores approach each other closely. This is often cited as a critical drawback of methods based on Biot-Savart law for vortex reconnection studies, where DNS results show an appreciable deformation of the core even when the core is very small\citep{yao2020singularity, yao2022vortex}. Since reconnecting vortex cores are not the focus of this work, our inviscid core calculations remain valid. Notice, however, that \citet{CallegariTing1978} provide proper evolution equations for the filament vortex core subject to viscous effects, and the M1 KK method would allow us to straightforwardly include them if needed. \rk{Their main impact would be a slow thickening of the vortex cores and an associated reduction of the curvature-binormal term in the filament equation of motion.}
\end{itemize}

For the four initial conditions shown in table \ref{tab: initial_conditions}, the height of the initial perturbation $y^+_{\text{initial}}$ needs to be chosen. From figure \ref{fig: hairpin_visualization}(b), numerous hairpin structures can be seen around $y^+ = 50$. Therefore, at first, $y^+_{\text{initial}} = 50$ is chosen where various hairpin characteristics are studied with respect to stratification. 

\begin{figure}
    \centering
    \includegraphics[width = 0.6\textwidth]{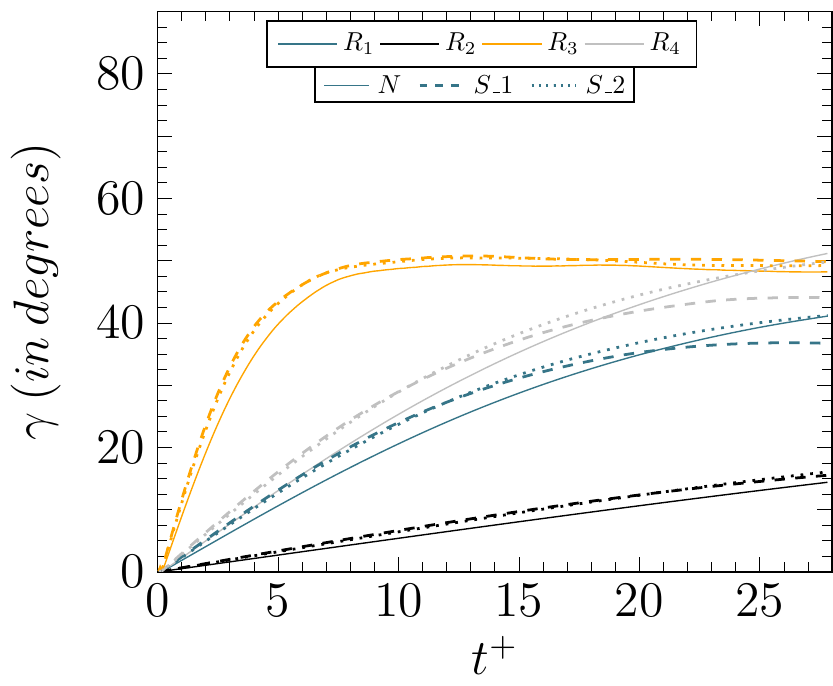}
    \caption{Time history of the inclination angle $\gamma$ computed for all four initial conditions from table \ref{tab: initial_conditions}. The solid, dashed and dotted lines correspond to the different cases N, S\_1, S\_2, respectively.}
    \label{fig: R1_R2_R3_R4_comparison_inclination}
\end{figure}

Unlike the stagnant background flow cases, the inclination angle for all subsequent simulations is set as $\gamma = 0.1^{\circ}$, i.e., a near-planar disturbance. The calculation is carried out for all four initial conditions and three background flow profiles with $500$ nodes. The results for case S\_1 are  visualized in figure \ref{fig: R1_R2_R3_R4_comparison_FR02} for $140$ time steps. This corresponds to a viscous time $t^+ = 7.8$ which is calculated with $t^+ = t \nu/ \delta_{\nu}^2$, where the viscous length scale $\delta_{\nu} = \nu/u_{\tau}$. For cases S\_2 and N, $140$ time steps will correspond to $t^+ = 8.1, 9.75$, respectively. The changes in viscous time units are due to variations in the friction velocity which is not a fixed parameter in Ekman flow simulations\citep{ansorge2014global}. Henceforth, comparisons among the different stratified cases are made at particular viscous time units instead of time steps.

In accordance with previous simulations of hairpin or parabolic vortex filaments in a shear flow (cf. \citet{hon1991evolution, moin1986evolution}), it can be noted from figure \ref{fig: R1_R2_R3_R4_comparison_FR02}(a) that regardless of the changes in core size $\delta$ and circulation $\Gamma$, the head of the filament bends backward and stretches in the wall-normal direction. This curl-up process, as elucidated by \citet{zhou1999mechanisms}, is due to the self-induced velocity of the filament competing with the background mean flow. The formation of the hairpin legs and its movement towards the wall is also observable. 

\begin{figure}
    \centering
    \includegraphics[width = 0.6\textwidth]{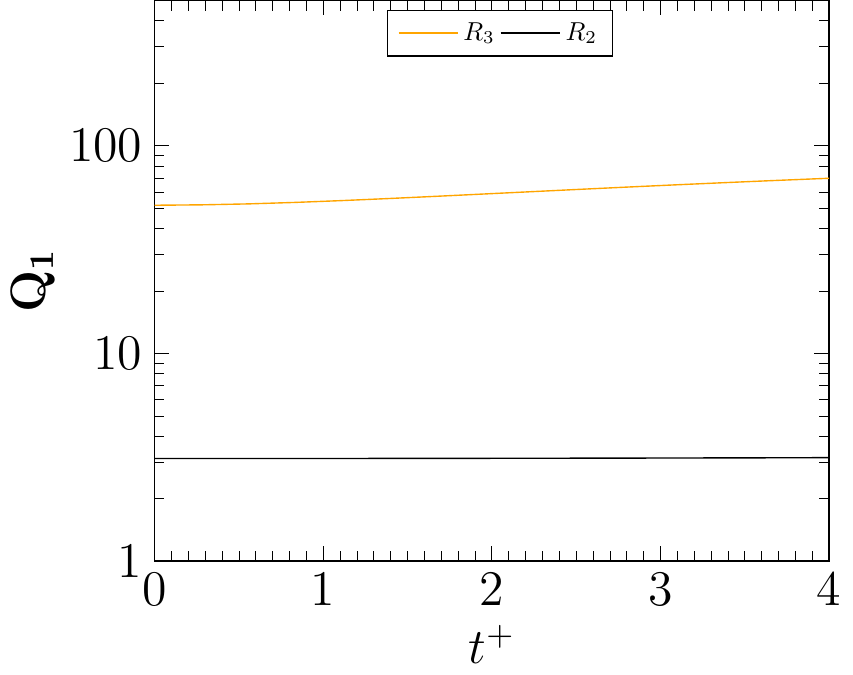}
    \caption{A comparison of self-induced velocity $\mathbf{Q_1}$ summed over the all nodes of the filament is shown for the initial conditions $R_2$ and $R_3$ with the mean background flow of S\_1.}
    \label{fig: R1_R2_R3_R4_comparison_self_induced_velocity}
\end{figure}

First, changes in inclination angle $\gamma = \arctan(\Delta y^+/\Delta x^+)$ are calculated for all simulation cases up to $t^+ = 28$ and shown in figure \ref{fig: R1_R2_R3_R4_comparison_inclination}. Instantly, it can be observed that the initial condition $R_2$, which has a thicker core and a smaller circulation value, exhibits a relatively slower inclination rate than the other three. On the other hand, the initial condition $R_3$ which has a thinner core and stronger circulation shows a faster evolution with a plateauing of the inclination angle around $50^{\circ}$ for all cases S\_1, S\_2 and N. This is close to the results reported by \citet{head1981new} who found hairpins inclined at $45^{\circ}$ with respect to the wall in the outer regions of the boundary layer. 

For initial condition $R_3$, the differences in inclination rate and the maximum inclination angle among the neutrally and stably stratified conditions are negligible due to large values of self-induced velocity which in turn is due to large $\Gamma$ and small $\delta$. As seen from figure \ref{fig: R1_R2_R3_R4_comparison_self_induced_velocity}, the self-induced velocity of $R_3$, computed by summing up the contributions from all nodes along the filament, is at least $16$ times the self-induced velocity of $R_2$. However, if the self-induced velocity takes smaller values, a clear dependence on the strength of stratification can be educed. An increase in the strength of stratification corresponds to a slower inclination rate. Indeed, if the integration of the initial condition $R_2$ is carried out for a longer time (not shown), the maximum inclination angle reached is $\gamma_{max} = 18.4^{\circ}, 22.2^{\circ}, 28.7^{\circ}$ for cases S\_1, S\_2 and N, respectively. This suggests a complex evolution of the hairpin filament strongly hinging on the balance between self-induction and the mean background flow.

\vspace{0.25cm}

\subsection{Further evolution of the hairpin filament}
\label{subsec: later_evolution}

\begin{figure*}
    \centering
    \includegraphics[width = \textwidth]{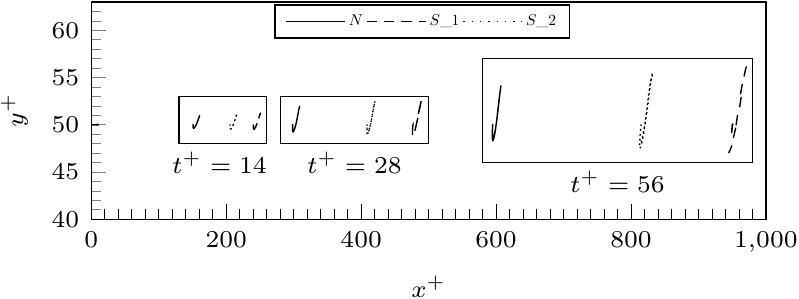}
    \includegraphics[width = \textwidth]{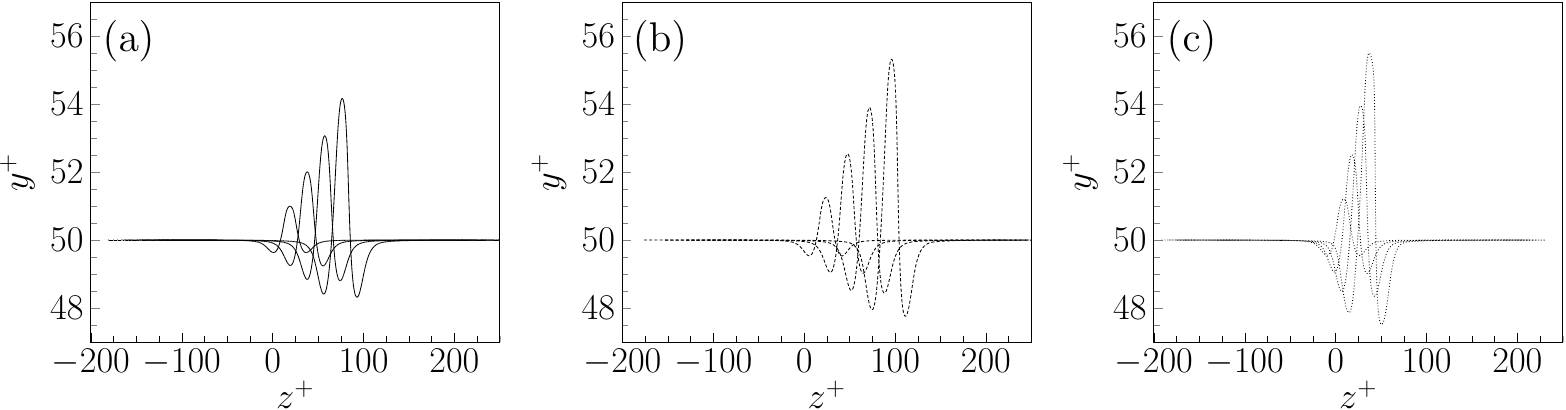}
    \caption{Temporal evolution for $R_2$ is shown for cases N, S\_1, S\_2. The top panel shows a side view of the filaments plotted at $t^+ = 0, 14, 28, 56$ and the bottom panel shows the front view plotted at $t^+ = 0, 14, 28, 42, 56$. In the bottom panel, (a, b, c) correspond to case N, S\_1 and S\_2, respectively.}
    \label{fig: R2_N_S1_S2_comparison}
\end{figure*}

\begin{figure*}
    \centering
    \includegraphics[width = \linewidth]{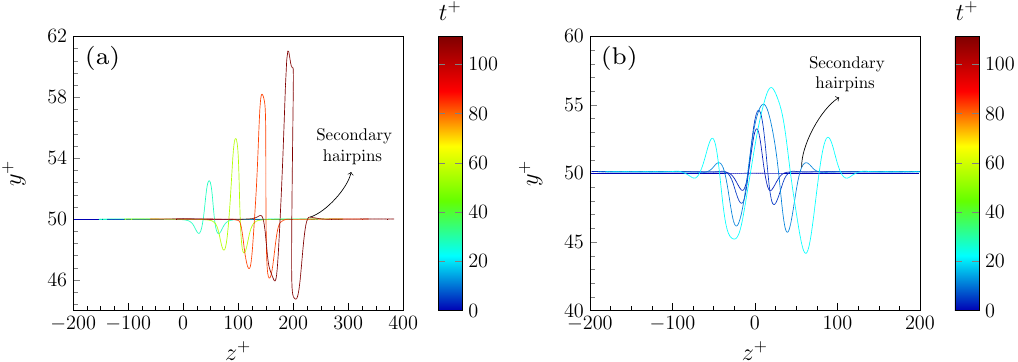}
    \caption{Temporal evolution of a hairpin filament for case S\_1 with (a) initial condition $R_2$ and (b) initial condition $R_3$. The filaments are plotted at $t^+ = 0, 27.8, 55.7, 83.5, 111.4$ in (a) and $t^+ = 0, 2.2, 4.4, 11.1, 22.3$ in (b). }
    \label{fig: R2_R3_FR02_comparison}
\end{figure*}

From the previous subsection, it is clear that $R_2$ has a slower inclination rate than the other three initial conditions. This means that the development of hairpin features such as secondary hairpins (as seen in the stagnant flow case and the shear flow case of \citet{hon1991evolution}) occurs at a much later time. This can be seen in figure \ref{fig: R2_R3_FR02_comparison} where secondary hairpins develop at a viscous time $t^+ = 3.9$ with initial condition $R_3$ and at a much later time $t^+ = 53$ for initial condition $R_2$. Similar trends (not shown) can be observed with case N and S\_2. This hints that the hairpin with initial condition $R_2$, i.e., with a slower inclination rate, may have a longer lifetime and is therefore further investigated in this subsection. The integration on this initial condition is continued with the three background flow profiles and visualized in the top panel of figure \ref{fig: R2_N_S1_S2_comparison} where the side view of hairpin filaments are shown. At time $t^+ = 56$, the filaments have been advected downstream in the streamwise direction for $973, 832, 607$ viscous units for case S\_1, S\_2 and N, respectively. By splitting the flow into turbulent and non-turbulent regions with conditional analysis, \citet{ansorge2016analyses} show that the streamwise velocity in the turbulent partitions for the strongly stratified case are lower than their non-turbulent counterparts, thereby implying a reduction \rk{of the shear intensity} in the turbulent regions of the flow. Numerous differences can also be observed from the front view of these hairpins as shown in the bottom panel of figure \ref{fig: R2_N_S1_S2_comparison}.

\begin{figure*}
    \centering
    \includegraphics[width = \linewidth]{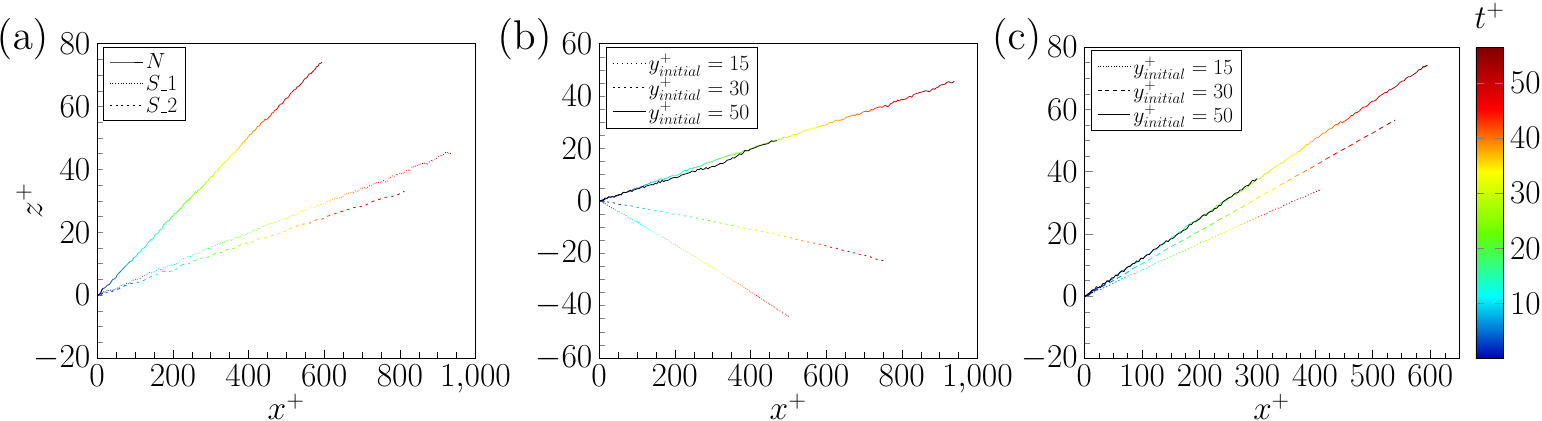}
    \caption{Comparison of spanwise advection among (a) the three cases N, S\_1, S\_2 when the hairpin filament is initialized at a height $y^+ = 50$, (b) at three initial heights $y^+_{initial} = 15, 30, 50$ for the strongly stratified case S\_1, and (c)  at three initial heights $y^+_{initial} = 15, 30, 50$ for the neutrally stratified case N. The solid black line in (b, c) shows the spanwise advection with initial condition $R_3$ wheras all others are shown with initial condition $R_2$.
    }
    \label{fig: R2_N_S1_S2_comparison_xz}
\end{figure*}

Although the initial disturbance is symmetric about $z^+ = 0$, an asymmetry develops during the evolution of the hairpin which is more pronounced for the stably stratified cases. The presence of asymmetric hairpins has been previously reported by \citet{robinson1991kinematics} who found after a detailed probing of the DNS of a channel flow that most hairpin vortices tend to be asymmetric. The development of the asymmetry for all background flow profiles can be attributed to the spanwise velocity gradient as seen in figure \ref{fig: mean_velocity_components}(c). As the hairpin stretches in the wall-normal direction, its head encounters a larger spanwise velocity than its leg which causes the hairpin to tilt in one direction. These results are in line with the presence of an Ekman spiral, \rk{within which} the wind velocity direction rotates with increasing height \citep{ansorge2016thesis}. The degree of tilt appears to increase with the strength of stratification. 

Apart from the asymmetry, the hairpin filament also experiences a strong advection in the spanwise direction. If the midpoint of the hairpin is plotted on a horizontal ($x, z$) plane at every instant in time as shown in figure \ref{fig: R2_N_S1_S2_comparison_xz}(a), the neutrally stratified case exhibits a stronger spanwise advection than the stably stratified cases. The spanwise \rk{drifts} ($\Delta z^+$), calculated as the spanwise distance travelled by the hairpin from $z^+ = 0$, are $74, 46$ and $33$ viscous units for \rk{cases} N, S\_1 and S\_2, respectively, at time $t^+ = 56$. This character remains unchanged for all initial conditions. For instance, a comparison of $R_3$ (solid black line) is made with $R_2$ (solid line with colormap) in figure \ref{fig: R2_N_S1_S2_comparison_xz}(b, c) for \rk{the cases} S\_1 and N, respectively, where the difference in spanwise advection with both initial conditions is negligible. 

If the simulations are repeated for two other initial heights $y^+_{\text{initial}} = 15, 30$ with $R_2$ as an initial condition, it can be seen from figure \ref{fig: R2_N_S1_S2_comparison_xz}(c) that the neutrally stratified case shows only minor changes in the spanwise advection when compared with $y^+_{\text{initial}} = 50$ and the hairpin filament is oriented anticlockwise with respect to the streamwise direction for all heights. However, the spanwise advection for stably stratified case S\_1 shows a change in orientation from clockwise to anticlockwise direction with increase in height (see figure \ref{fig: R2_N_S1_S2_comparison_xz}(b)) which is\rk{, again,} due to the Ekman spiral. This result suggests that the spanwise orientation of \rk{a hairpin filament} is linked to its initial height, i.e., its origin.

\section{Comparison of filament simulation with a feature tracking scheme}
\label{sec: compare_MLP_DNS}

\begin{figure*}
    \centering
    \includegraphics[width=\textwidth]{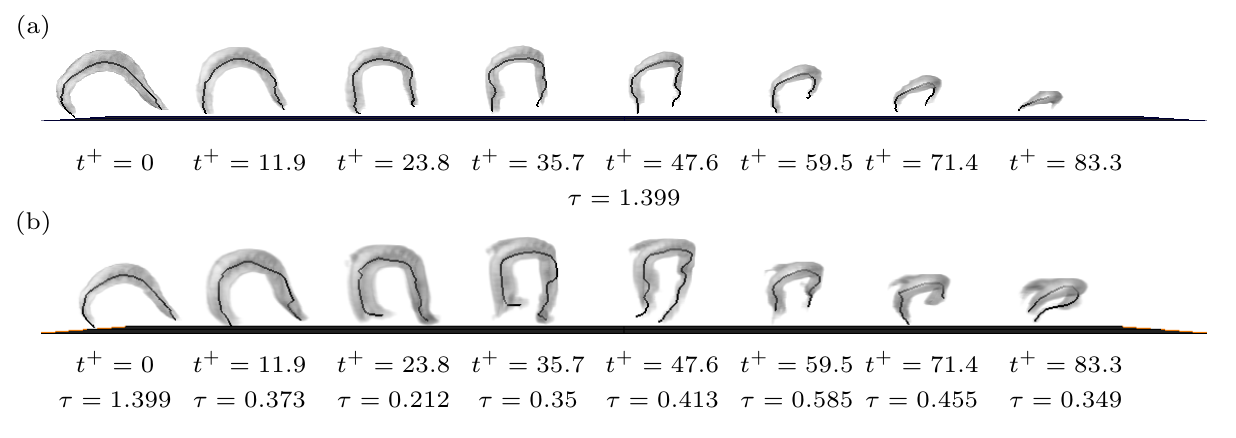}
    \caption{Track of a hairpin-like structure with (a) constant thresholding and (b) MLPT thresholding. The complex geometry of the extracted structures may result in a network of centerlines. Since we are only interested in the main centerline defining the hairpin structure, the remaining are discarded.}
    \label{fig: constant_MLPT_comparison}
\end{figure*}

The availability of DNS data, which is temporally well-resolved, provides an opportunity to track hairpin-like structures in time. The goal of this section is twofold - first, the volume overlap scheme developed in \citet{von2021definition} is extended to automatically choose non-subjective, optimum thresholds \rk{for the $Q$-criterion} in time. Second, initial conditions are carefully chosen to simulate the motion of a hairpin filament and compare the results with feature tracking. The feature tracking is performed on the strongly stratified case S\_1 and the mean velocity \rk{profile} of the same case is used as a background flow for the simulation. 

\subsection{Feature tracking}
\label{subsec: feature_tracking}

Following the work of \citet{moisy2004geometry}, we define a \textit{coherent structure} or \textit{feature} as a connected set of points within a three-dimensional scalar field which exceed a threshold, i.e., 

\begin{equation}
\label{eq: feature_definition}
    \alpha (x) > \tau_p \,\, \overline{\alpha'^2 (y)}^{1/2}
\end{equation}

\noindent where $\alpha$ is a feature indicator, $\overline{\alpha'^2 (y)}^{1/2}$ is its root mean square (RMS) over wall-normal planes and $\tau_p$ is an appropriate threshold. The review paper of \citet{gunther2018state} gives a comprehensive overview of numerous contenders for vortex indicators. In this work, the popular $Q$-criterion is chosen to identify vortical features. 

In a preprocessing step, \rk{scalar field $Q$, normalized with its RMS,} is computed for $1000$ timesteps and stored locally. For the first timestep, a global threshold $\tau_p = 0.0625$ is obtained through percolation analysis\citep{moisy2004geometry, del2006linear}. Individual structures are extracted with the neighbor scanning algorithm with marching cubes correction\citep{harikrishnan2021geometry} (NS+MC). Due to the globally intermittent nature of the flow, \rk{which implies the co-existence of large regions non-turbulent motions with turbulent regions}, the global threshold value tends to identify a large cluster of structures as an individual structure. Hence, the iterative percolation analysis technique called multilevel percolation (MLP) developed by \citet{harikrishnan2021geometry} is used on a small subset of the data to identify and extract individual structures. With manual inspection, one hairpin-like structure is chosen for tracking. 

We use the method of volume overlap to track the feature in time. This is a local tracking technique \rk{by which} the structure extracted at one timestep is matched to\rk{, or identified with,} another \rk{at} a subsequent timestep provided there exists some spatial overlap. The amount of overlap is measured \rk{by} the Dice Similarity Coefficient (DSC)\citep{dice1945measures},
\begin{equation}
\label{eq: similarity_coefficient}
DSC(I_n, I_{n+1}) = \frac{2 |I_n \bigcap I_{n+1}|} {|I_n| + |I_{n+1}|}\,,
\end{equation}
where $I_n$ and $I_{n+1}$ are the structures at time $t_n$ and $t_{n+1}$, respectively, and $|\cdot|$ is the volume of the structure. The value of DSC can range between $0$ and $1$, where $1$ indicates a perfect overlap. The tracking proceeds only when DSC is greater than the user-defined overlap threshold $\tau_{overlap}$. In our case, $\tau_{overlap} = 0.5$. 

Another important parameter that needs to be set is the threshold $\tau$ for subsequent timesteps. In the first run, a constant threshold of $\tau = 1.399$ is used. The results are shown in figure \ref{fig: constant_MLPT_comparison}(a) \rk{in which} both volume rendering and the centerline of the structure are shown. The centerline is again obtained \rk{by} the block-wise skeletonization method of \citet{fouard2006blockwise}. The tracking suggests that the hairpin structure is constantly shrinking in time. This is expected since constant thresholding in time \rk{does not} adapt to the dynamical changes in the size of a structure. This limitation is overcome by combining volume overlap with multilevel percolation analysis. In essence, this technique accommodates for the growth or deterioration of a structure by dynamically adjusting the thresholds in time. Details of this technique are presented in Appendix \ref{appendix: timeMLP}. The result as shown in figure \ref{fig: constant_MLPT_comparison}(b) suggests that the hairpin structure grows in size until $t^+ = 59.5$, after which it starts shrinking. 

\subsection{Comparison with filament simulation}
\label{subsec: simulation_MLP}

\begin{table*}[]
    \centering
    \begin{tabular}{|l | l |}
        \hline
        Parameter & Value\\
        \hline
        Core size parameter, $\delta$ & $0.01$\\
        Circulation, $\Gamma$ & $0.05$\\
        Spread parameter, $\beta$ & $500$\\
        Amplitude, $A$ & $80$\\
        Angle of inclination, $\gamma$ & $0.1^{\circ}$\\
        Initial wall-normal position, $y^+_{\text{initial}}$ & $125$\\
        Domain length, $L$ & $400$\\
        \hline
    \end{tabular}
    \caption{Initial condition for comparison with feature tracking results.}
    \label{tab: filament_comparison_configuration}
\end{table*}

\begin{figure*}
    \centering
    \includegraphics[width=0.59\textwidth]{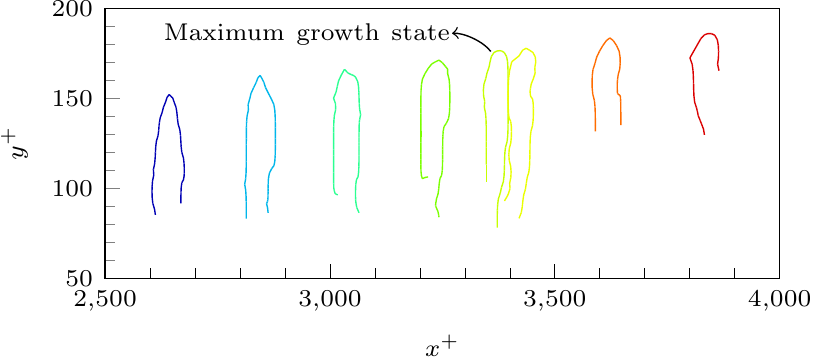}
    \includegraphics[width=0.4\textwidth]{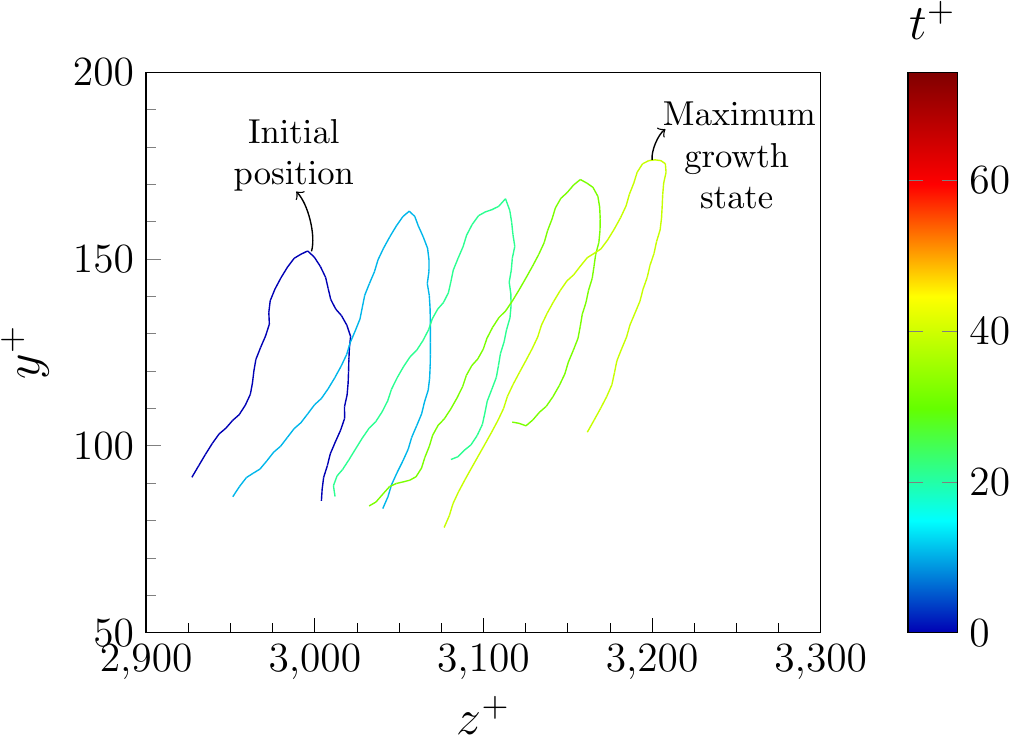}
    \caption{Temporal evolution of the tracked hairpin-like structure with MLPT thresholding as seen in figure \ref{fig: constant_MLPT_comparison}(b). (a, b) show the side view and the front view of the feature tracking results, respectively.}
    \label{fig: MLPT_tikz}
\end{figure*}

The hairpin-like structure tracked with MLPT thresholding is visualized in figure \ref{fig: MLPT_tikz}. Since it is clear that this structure exists in the outer layer in the range $80 < y^+ < 180$, the filament simulation is also initialized at a higher wall-normal height $y^+_{\text{initial}} = 125$ to enable an appropriate comparison. A larger amplitude $A = 80$ and correspondingly, a larger initial width are also chosen. All initial conditions are listed in table \ref{tab: filament_comparison_configuration}. With a spatial discretization of $1500$ nodes, the simulation is carried out until the overlap condition \eqref{eq: overlap_condition_modified} is violated. 

\begin{figure}
    \centering
    \fbox{\includegraphics[width=0.45\textwidth]{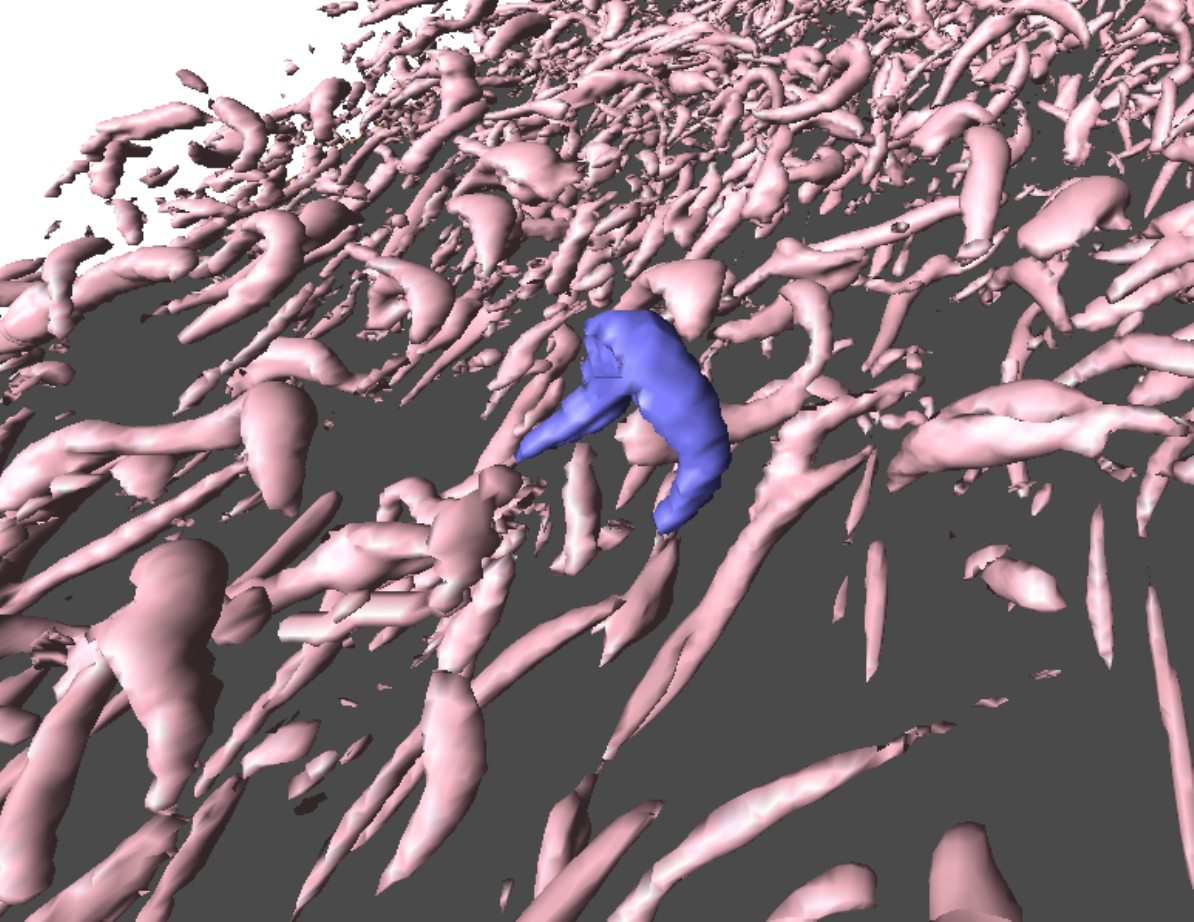}}
    \caption{Isosurfaces of the $Q$-criterion for case S\_1 are shown here for the initial time step. The structure tracked with MLPT is highlighted in blue. }
    \label{fig: mutual_induction}
\end{figure}

Since the initial state of the hairpin structure from the DNS dataset is unknown and the structure is tracked from a later point in its lifetime, it is not possible to establish a direct comparison of the tracking results with the filament simulation. However, some qualitative comparisons can be drawn. The streamwise advection of the hairpin structure, tracked in time for $674$ time steps or $84$ viscous time units, is $1621$ viscous units. The filament simulation shows a streamwise advection of $1682$ viscous units over the same viscous time. This minor discrepancy can be attributed to the choice of the initial height for the filament simulation suggesting that the hairpin structure in the DNS simulation may have originated from a lower height. Differences may also arise due to mutual induction effects of neighboring hairpin structures. It can be seen from figure \ref{fig: mutual_induction} that even at the initial state of the tracked hairpin structure (highlighted blue), at least one other hairpin structure can be seen in close proximity. This may have a strong influence on the later developments of the tracked hairpin structure. 

\begin{figure*}
    \centering
    \includegraphics[width=0.59\textwidth]{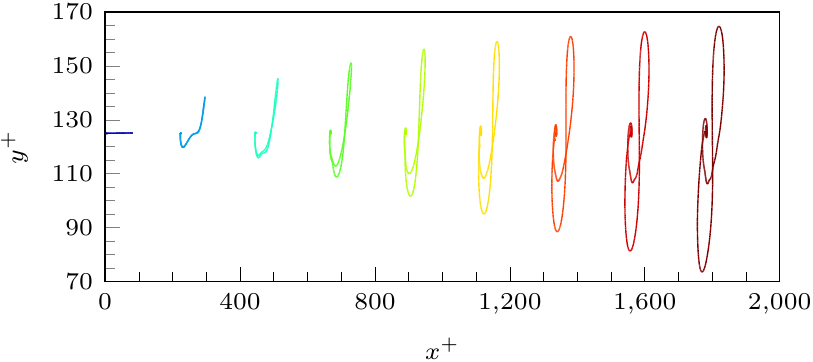}
    \includegraphics[width=0.4\textwidth]{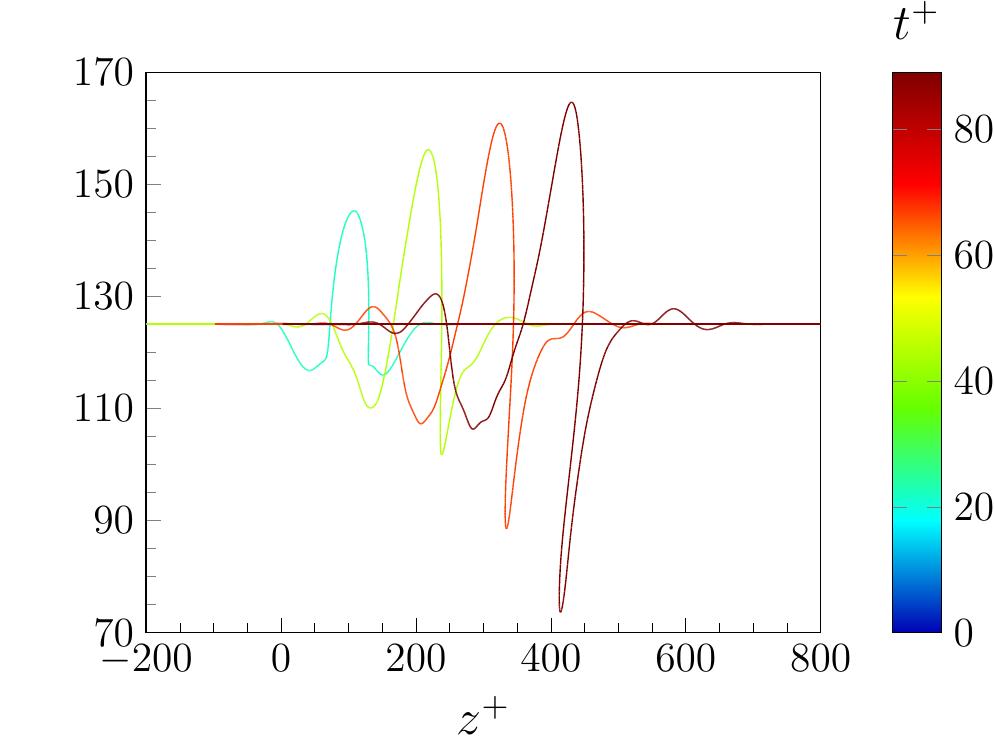}
    \caption{Temporal evolution for a larger hairpin filament at $y^+_{\text{initial}} = 125$. (a, b) show the side and front view of the temporal development, respectively. }
    \label{fig: filament_tikz}
\end{figure*}

The initial state of the DNS hairpin structure as seen from figure \ref{fig: MLPT_tikz} shows that the hairpin, although asymmetric, has both legs at similar wall-normal heights of $91$ and $85$ viscous units. Over time, at its maximum growth state, the asymmetry grows further with one leg at a wall-normal height of $78$ viscous units and the other at $103$ viscous units. A similar effect can be observed in figure \ref{fig: filament_tikz}(b). At time $t^+ = 84$, both legs of the hairpin are at a wall-normal height of $73$ and $106$ viscous units. As described in the further evolution part of section \ref{sec: stagnant_flow}, the persistence of asymmetry over time is due to the influence of spanwise velocity. Unlike the results reported in \citet{zhou1999mechanisms}, who suggested that a sufficiently strong asymmetrical initial configuration is necessary to produce an asymmetrical hairpin vortex, our results indicate that asymmetry develops naturally as a consequence of the mean background flow.

\section{Influence of buoyancy - An asymptotic analysis}
\label{sec: mathematical_formulation}

In all the hairpin filament simulations presented in this paper, the effect of gravity on the self-induced motion of the filament is ignored. In this section, we carry out an asymptotic analysis by extending the work of \citet{CallegariTing1978} to include weak gravitational forces. While we expect the effects of gravity to be predominantly felt through its impact on the external flow, gravity may also have a minor impact on the self-induced velocity of the filament.

We start by reviewing the derivation of the equation for the motion of an incompressible vortex filament found in \citet{CallegariTing1978}; as the influence of gravity is assumed to be weak, the leading and first order solutions will agree with \citet{CallegariTing1978}. The idea upon which the analysis is based is as follows - the self-induced motion of the filament is dictated by an inner solution, describing the flow inside the tube-like structure, and the outer flow, which is simply given by the line Biot-Savart law which is reiterated, 
\begin{equation} \label{BiotSavartLaw}
\mathbf{Q}_1(\mathbf{P},t) = -\frac{\Gamma}{4\pi}\int_{\mathcal{L}} \frac{\left(\mathbf{P}-\mathbf{X}(s',t)\right)\times \mathbf{ds}'}{\left|\mathbf{P}-\mathbf{X}(s',t)\right|^3}
\end{equation}
where $\mathcal{L}$ is the vortex centerline, $\mathbf{X}(s,t)$ is the position vector for any point on $\mathcal{L}$, which is parametrized by the arc length parameter $s$, and $\mathbf{P}$ denotes a point outside of the filament core (see Figure 1). 
Since the Biot-Savart integral diverges on the filament centerline, it cannot be used to determine the motion of the velocity of the vortex core itself. For this reason, through the application of asymptotic techniques, the outer solution described by the Biot-Savart law will be matched with the inner solution derived from the Navier-Stokes equation. We reiterate from the introduction that \citet{KleinKnio1995} provided an alternative derivation of the inner solution based on the vorticity equation, yielding the same result.

\subsection{\label{sec:level5}The outer solution}

\begin{figure}
    \centering
    \includegraphics[width = 0.6\linewidth]{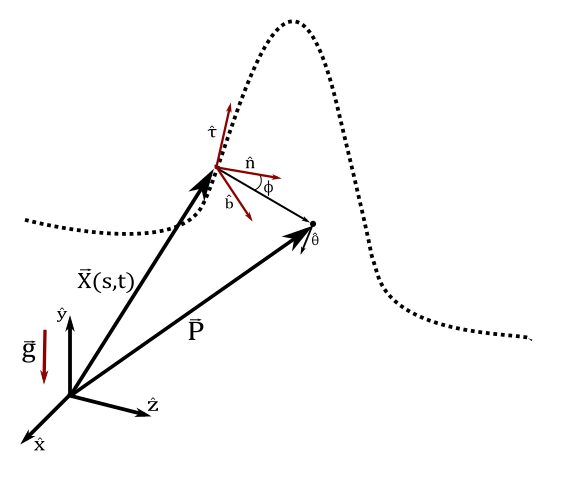}
    \caption{A sketch of the coordinate system with the dashed line denoting the hairpin shaped vortex filament.}
    \label{fig:Coordintaesystem_sketch}
\end{figure}

The velocity at any point is composed of a velocity $\mathbf{Q}_f$ due to the self-induced motion of the vortex filament and a velocity $\mathbf{Q}_2$ given by the background flow field. $\mathbf{Q}_f$, given by (\ref{BiotSavartLaw}), at any point is not on the vortex centerline. The vortex filament is assumed to be given parametrically by $\mathbf{X}(s,t)$, where $s$ is the arc length along the filament, and $t$ is time. The position vector of any point $\mathbf{P}$ in curvilinear coordinates can be associated to $\mathbf{X}(s,t)$ by the equation,

\begin{equation}
    \mathbf{P}(x,y,z) = \mathbf{X}(s,t)+r\hat{\mathbf{r}}
\end{equation}
Here, $r$ denotes the distance from the point $\mathbf{P}$ to the filament $X$, and $\hat{r}$ is the radial unit vector for curvilinear coordinates. The coordinate system is illustrated in figure \ref{fig:Coordintaesystem_sketch}. A careful expansion of the integrand in the Biot-Savart formula leads to an expression for the behaviour of $\mathbf{Q}_f$ as the radial distance to the filament goes to zero,

\begin{equation}
\label{Outer_expansion}
\mathbf{Q}_1(P,t) = \frac{\Gamma}{2\pi r}\hat{\pmb{\theta}} + \frac{\Gamma}{4\pi R}\left[\ln\left(\frac{R}{r}\right)\right]\hat{\mathbf{b}} + \frac{\Gamma}{4\pi R}(\cos\phi)\hat{\pmb{\theta}} + \mathbf{Q}_f
\end{equation}
where $\hat{\theta}$ is the unit circumferential vector and $\hat{b}$ is the unit binormal vector, associated with the point $s$ on the vortex filament $\mathbf{X}(s,t)$. $R=R(s,t)$ is the local radius of curvature, and $\Gamma$ the circulation of the filament. The vector $\mathbf{Q}_f$ is the part of the the Biot-Savart integral which has a limit as $r\rightarrow 0$. Eq. \eqref{Outer_expansion} cannot, however, yield the velocity on the vortex filament as the right hand side diverges as $r \rightarrow 0$. To overcome this problem Callegari and Ting employed a matched asymptotic expansion, where the outer solution, given by equation (\ref{Outer_expansion}), was matched with an inner solution derived from the Navier-Stokes equations with suitable boundary conditions. In the next subsection, we will extend their analysis to include weak gravitational effects. 

\subsection{\label{sec:level6}The inner solution}

The fluid velocity $v(x,t)$ is described relatively to the moving filament by introducing a relative velocity $\mathbf{V}$: \begin{equation}
v(x,t)=\dot{\mathbf{X}}(s,t)+\mathbf{V}(t,r,\theta,s) \end{equation}
The dimensionless incompressible Navier-Stokes equation in curvilinear coordinates is
\begin{eqnarray}
\ddot{\mathbf{X}} + \left(\frac{w}{h_3}-\frac{r}{h_3}\hat{r}_t\cdot\hat{\tau}\right) \dot{\mathbf{X}}_s + \frac{d\mathbf{V}}{dt} \quad\quad\quad\quad\quad\quad\quad\quad\quad\quad\quad \\ = - \frac{\mathbf{\nabla} P}{\rho} + \frac{1}{\text{Re}}\left(\frac{1}{\rho h_3}\left(\frac{1}{h_3}\dot{\mathbf{X}}_s\right)_s + \frac{1}{\rho} \Delta  \mathbf{V}\right) +  \frac{1}{\text{Fr}^2}\hat{\mathbf{g}}\nonumber
\end{eqnarray}
where 
\begin{equation}
    h_ 3 = \sigma[1-\kappa r\cos(\theta + \theta_0)] =\sigma[1-\kappa r\cos\phi]
\end{equation}
is the tangential stretching parameter, Re = Reynolds number, Fr = Froude number, P = Pressure and  $\sigma={|\mathbf{X}}_s|$. The density $\rho$ is assumed to have a fixed part and a variable part which is dependent exclusively on temperature and not on the pressure. The variable part is assumed to be small.
The continuity equation for an incompressible fluid in curvilinear coordinates is,

\begin{eqnarray}
\label{Continutiy_equation_curvilinear}
  \left[\left(\bar{r}h_3 u\right)_{\bar{r}} + \left(h_3 v\right)_\theta + r\left(w_s+\dot{\mathbf{X}}_s\cdot\hat{\pmb{\tau}}\right)\right] = 0 
\end{eqnarray}
We introduce stretched radial coordinates,
\begin{equation}
\label{streteched_coordinates}
\bar{r} = r/\epsilon, \quad
\epsilon = \left( \nu / \Gamma \right)^{1/2}
\end{equation}
with
\begin{equation}
\label{scale_reynoldsnumber}
\left( \nu / \Gamma \right)^{1/2}=\text{Re}^{-1/2}=\bar{\nu}^{1/2}\epsilon
\end{equation}
and expand the dynamical variables in terms of $\epsilon$ as follows: 
\begin{subequations}
\begin{eqnarray}
\label{asymptotic_expansion1}
u(\bar{r}, \theta, s,t ; \epsilon)  &=& u^{(1)}(\bar{r}, \theta, s,t ) + \epsilon u^{(2)}(\bar{r}, \theta, s,t ) + \cdots \\
\label{asymptotic_expansion2}
v(\bar{r}, \theta, s,t ; \epsilon)  &=& \epsilon^{-1}v^{(0)}(\bar{r}, \theta, s,t ) +  v^{(1)}(\bar{r}, \theta, s,t ) + \cdots \\
\label{asymptotic_expansion3}
w(\bar{r}, \theta, s,t ; \epsilon)  &=& \epsilon^{-1}w^{(0)}(\bar{r}, \theta, s,t ) +  w^{(1)}(\bar{r}, \theta, s,t ) + \cdots \\
\label{asymptotic_expansion6}
\mathbf{X}(s,t;\epsilon) &=& \mathbf{X}^{(0)}(s,t) + \epsilon\mathbf{X}^{(1)}(s,t)  + \cdots
\end{eqnarray}
\end{subequations}
Here $u, v$ and $w$ denote the radial, polar and tangential components of the velocity vector $\mathbf{V}$, respectively. In order to obtain non-trivial velocities $v^{(0)}$ and $w^{(0)}$, the following is required,
\begin{equation}
    P(\bar{r}, \theta, s,t ; \epsilon)  =  \epsilon^{-2}P^{(0)}(\bar{r}, \theta, s,t) + \epsilon^{-1} P^{(1)}(\bar{r}, \theta, s,t ) + \cdots \\
\end{equation}
along with
\begin{equation}
    \dot{\mathbf{X}}\cdot\hat{\pmb{\tau}} = 0
\end{equation}
so that the filament core forms a material curve. Through the Serret-Frenet formulas, the geometric parameters $\sigma$, $\kappa$ and $h$ are functions of $\mathbf{X}(s,t)$ and are expanded as follows:
\begin{subequations}
\begin{eqnarray}
\sigma &=& \sigma^{(0)} + \epsilon\sigma^{(1)} + \cdots = |\mathbf{X}^{(0)}_s| +\frac{\epsilon\mathbf{X}^{(0)}_s\cdot\mathbf{X}^{(1)}_s}{|\mathbf{X}^{(0)}_s|} + \cdots \\
\kappa &=& \kappa^{(0)} + \epsilon \kappa^{(1)} +\cdots \\
h_3 &=& h_3^{(0)} + \epsilon h_3^{(1)} + \cdots = \sigma^{(0)} + \epsilon\left[\sigma^{(1)} - \sigma^{(0)}\kappa^{(0)}\bar{r}\cos\varphi^{(0)}\right]
\end{eqnarray}
\end{subequations}

Given our choice of $\epsilon$, the viscosity terms will only enter at higher order. One can show that the term
\begin{equation}
    \left(\frac{1}{\rho h_3}\left(\frac{1}{h_3}\dot{\mathbf{X}}_s\right)_s\right)
\end{equation}
is at least $\mathcal{O}(\epsilon)$, so this term only enters much later in the expansion. 
The Froude number will be assumed to be small such that the gravity term only appears in the second order equations. Specifically, we assume that  $\text{Fr}=\bar{\lambda}\sqrt{\epsilon}$. In the Boussinesq approximation, the density variation only enters in the buoyancy term, $\rho \mathbf{g}$, and can be neglected in the rest of the equation. This yields,
\begin{eqnarray}
\label{Momentum_equation_curvilinear}
\rho_0\left(\ddot{\mathbf{X}} + \frac{1}{h_3}\left(w-r\hat{\mathbf{r}}_t\cdot\hat{\pmb{\tau}}\right) \dot{\mathbf{X}}_s + \frac{d\mathbf{V}}{dt}\right)\quad\quad\quad\quad\quad\quad\quad\\ = - \mathbf{\nabla} P + \frac{1}{\text{Re}}\left(\frac{1}{ h_3}\left(\frac{1}{h_3}\dot{\mathbf{X}}_s\right)_s + \Delta  \mathbf{V}\right) +  \rho\frac{1}{\text{Fr}^2}\hat{\mathbf{g}} \nonumber
\end{eqnarray}
where the temperature and pressure dependent density, $\rho$, has been replaced by a constant density $\rho_0$, except in the buoyancy term which is written as,
\begin{equation}
    (\rho_0 + \Delta\rho)g\,,
\end{equation}
where $\Delta \rho = \rho - \rho_0$ represents the density variation with respect to the reference density $\rho_0$. This can be further rewritten noting that the variation in density will be exclusively due to temperature variations and not pressure variations, which yields
\begin{equation}
    (\rho - \rho_0) g = -\rho_0\beta(T-T_0)g\,,
\end{equation}
where $\beta$ is the coefficient of thermal expansion. For ideal gases, $\beta = 1/T_0$. We expand the perturbation temperature $\Tilde{T} = T-T_0$ in an asymptotic series as follows,
\begin{equation}
    \Tilde{T}(\bar{r},\theta,s,t;\epsilon) = \Tilde{T}^{(0)}(\bar{r},\theta,s,t) + \epsilon \Tilde{T}^{(1)}(\bar{r},\theta,s,t) + \cdots .
\end{equation}

Since we have introduced a density variation to the flow, we also need an energy equation to close the system. The energy equation for a Boussinesq fluid in curvilinear, compressed coordinates, in terms of the temperature, $T$, is 
\begin{equation}
    \frac{\partial T}{\partial t} + \left(\mathbf{V} - \epsilon \bar{r}\frac{\partial \hat{\mathbf{r}}}{\partial t}\right)\cdot\nabla T = \frac{1}{\text{Re}\;\text{Pr}}\Delta T\,,
\end{equation}
where $\text{Pr}$ is the Prandtl number. Here we have used the formula (B.12) for the material derivative provided in \citet{CallegariTing1978} Appendix B. 

In what follows, we shall assume that $\text{Pr} = \mathcal{O}(1) = \bar{\mu}$. By inserting the asymptotic expansion, equations (\ref{asymptotic_expansion1})-(\ref{asymptotic_expansion6}) into equations (\ref{Momentum_equation_curvilinear}) and (\ref{Continutiy_equation_curvilinear}), one gets a hierarchy of equations based on the order of $\epsilon$. From the leading and first order equations, one can derive an expression for the core constant $C(t)$ which appears in the filament velocity equation. Since the gravity term only appears in the second order equations, the derivation follows that of Callegari and Ting exactly. Hence, we will not repeat it in detail here, but merely state the main results. We refer the reader to \citet{CallegariTing1978}, or equivalently, \citet{KleinKnio1995}, for the complete derivation. \\ 

First, the leading order equations imply that the leading order variables, $v^{(0)}$, $w^{(0)}$, $P^{(0)}$ and $\Tilde{T}^{(0)}$, are all $\theta$-independent. Secondly, since  the  leading  order  term  in  the  circumferential  velocity  term  of  the  outer  flow  is  independent  of  the  axial  locations, we will assume that the leading order variables are also independent of the axial coordinate $s$. With these simplifications the first order momentum equations can be solved by introduction of an appropriate stream function and further decomposition into symmetric and anti-symmetric parts, finally yielding the following expression for the core constant,
\begin{eqnarray}
\label{Core_constant}
    C(t) = \frac{1}{8\pi^2}\Big[\frac{1}{2}\lim_{\bar{r}\longrightarrow \infty}\left(\frac{4\pi^2}{\Gamma^2}\int_0^{\bar{r}}\xi\left(v^{(0)}\right)^2\;d\xi -\ln\bar{r}\right) \nonumber \\-\frac{1}{4}-\frac{4\pi^2}{\Gamma^2}\int_0^{\bar{r}}\xi\left(w^{(0)}\right)^2\;d\xi \Big] \quad\quad\quad
\end{eqnarray}
We should note here that, by the \rk{vorticity-based analysis given by} \citet{KleinKnio1995}, the core constant $C(s,t)$ can also be written as,
\begin{equation}
\label{Core_constant_KleinKnio1995}
    C(t) = - \frac{1}{2} - \frac{2\pi}{\Gamma}\int_0^\infty \left[\frac{1}{\kappa}\zeta^{(1)}_{11} +\bar{r}\ln\bar{r}\zeta^{(0)}\right]d\bar{r}\,,
\end{equation}
where $\zeta^{(0)}$ is the leading order vorticity given by
\begin{equation} \label{circulation}
\zeta^{(0)} =  \frac{1}{\bar{r}}\left(\bar{r}v^{(0)}\right)_{\bar{r}}\,,
\end{equation}
and $\zeta_{11}^{(1)}$ is the first cosine Fourier mode of the first order vorticity $\zeta^{(1)}$ with respect to $\theta$,
\begin{equation}
    \zeta_{11}^{(1)} = \frac{1}{2\pi}\int_0^{2\pi}\zeta^{(1)}(\bar{r},\theta,s,t)\cos\theta \;d\theta\,.
\end{equation}

By matching the inner solution with the outer solution, one obtains the following equation for the binormal and normal components of the filament velocity:
\begin{subequations}
\begin{eqnarray}
\label{velocity_filament_binormal}
    \hat{b}\cdot\dot{X}^{(0)} &=& \hat{b}\cdot Q_0 + \kappa^{(0)} \frac{\Gamma}{4\pi} \left(\ln \frac{R^{(0)}}{\epsilon} + C(t)\right) \\
\label{velocity_filament_normal}
    \hat{n}\cdot\dot{X}^{(0)} &=& \hat{n}\cdot Q_0 
\end{eqnarray}
\end{subequations}
where $R^{(0)}= 1/\kappa^{(0)}$ is the leading term of the radius of curvature of the reference line and $C(t)$ is given by equation (\ref{Core_constant}). Thus, we have arrived at an equation for the evolution of the filament, which depends on the leading order velocity components $v^{(0)}$ and $w^{(0)}$. The compatibility conditions, which determine the leading order velocity components, are derived from the second order momentum and continuity equations. It is here that we will see a deviation from the derivation done by Callegari and Ting, as we have the appearance of buoyancy terms in the momentum equations. 

\subsection{\label{sec:level7}The Compatibility Conditions}

The second order momentum equation in the circumferential direction is 
\begin{eqnarray}
v_t^{(0)} + w^{(0)} \hat{\tau}_t\cdot \hat{\theta} +u^{(1)}v^{(1)}_{\bar{r}} + \frac{v^{(1)}v_\theta^{(1)}}{\bar{r}} + \frac{v^{(0)}v_\theta^{(2)}}{\bar{r}} + \frac{v^{(0)}u^{(2)}}{\bar{r}} \quad \nonumber \\ 
+ \frac{v^{(1)}u^{(1)}}{\bar{r}} + \frac{w^{(0)}v_s^{(1)}}{\sigma^{(0)}}  -2w^{(0)}w^{(1)}\kappa^{(0)}\sin\varphi^{(0)} \nonumber \\ - \left(w^{(0)}\right)^{(2)}\left(\frac{\kappa\sigma \sin \varphi}{h_3}\right)^{(1)}   + \frac{\theta^{(0)}\cdot \dot{X}_s^{(0)}w^{(0)}}{\sigma^{(0)}} + u^{(2)}v_{\bar{r}}^{(0)}  \nonumber \\ = -\frac{1}{\bar{r}}\frac{P_\theta^{(2)}}{\rho_0}   + \frac{\bar{\nu}}{\bar{r}}\left(\bar{r}v_{\bar{r}}^{(0)}\right)_{\bar{r}}  - \frac{\bar{\nu} v^{(0)}}{\bar{r^2}} + \alpha g\Tilde{T}^{(0)} \hat{\mathbf{y}}\cdot\hat{\pmb{\theta}} \nonumber
\end{eqnarray}
where $-\hat{\mathbf{y}}g=\hat{\mathbf{g}}$ denotes the acceleration due to gravity. In addition we have defined $\alpha = \beta\bar{\lambda}^{-2}$, where the facor of $\bar{\lambda}^{-2}$ comes from the definition of the Froude number. Upon averaging with respect to $\theta$, we get 
\begin{eqnarray} \label{second_order_circumferential_momentum_theta_averaged}
    v_t^{(0)} - \bar{\nu}\left[\frac{1}{\bar{r}}\left(\bar{r}v_{\bar{r}}^{(0)}\right)_{\bar{r}} - \frac{v^{(0)}}{\bar{r}^2}\right] = - \zeta^{(0)}\left<u^{(2)}\right> - \frac{1}{\bar{r}}\left<u^{(1)}\left(\bar{r}v^{(1)}\right)_{\bar{r}}\right>  \nonumber \\ - \frac{w^{(0)}}{\sigma^{(0)}}\left<v^{(1)}_s\right> + 2\kappa^{(0)}w^{(0)}\left<w^{(1)}\sin\varphi^{(0)}\right>\,, 
    %\quad\quad\quad
\end{eqnarray}
where we have used that $\hat{\pmb{\theta}} = \hat{\mathbf{b}}\cos\varphi^{(0)}-\hat{\mathbf{n}}\sin\varphi^{(0)}$, which in turn implies that 
\begin{eqnarray}
\left<\Tilde{T}^{(0)} \hat{\mathbf{y}}\cdot\hat{\pmb{\theta}}\right> &=& \hat{\mathbf{y}}\cdot\hat{\mathbf{b}}\left<\Tilde{T}^{(0)}\cos\varphi^{(0)}\right> -  \hat{\mathbf{y}}\cdot\hat{\mathbf{n}}\left<\Tilde{T}^{(0)}\sin\varphi^{(0)}\right> 
= 0 
\end{eqnarray}

From this, we anticipate that the evolution equation for $v^{(0)}$, derived from equation (\ref{second_order_circumferential_momentum_theta_averaged}), will not contain any contribution from the gravity term. The second order momentum equation in the tangential direction is
\begin{eqnarray}
w_t^{(0)} + v^{(0)} \hat{\pmb{\theta}}_t^{(0)}\cdot\hat{\pmb{\tau}} +u^{(2)}w^{(0)}_{\bar{r}} + \frac{w^{(1)}_\theta v^{(1)}}{\bar{r}} + u^{(1)}w_{\bar{r}}^{(1)} + \frac{w^{(2)}_\theta v^{(0)}}{\bar{r}} 
\nonumber \\  
+ \frac{w^{(0)}w^{(1)}_s}{\sigma^{(0)}} + w^{(1)}v^{(0)}\kappa^{(0)}\sin\varphi^{(0)}+ w^{(0)}v^{(1)}\kappa^{(0)}\sin\varphi^{(0)} 
\\ 
- w^{(0)}u^{(1)}\kappa^{(0)}\cos\varphi^{(0)} + w^{(0)}v^{(0)}\left(\frac{\kappa\sigma \sin\varphi}{h_3}\right)^{(1)} + \hat{\pmb{\tau}}\cdot \frac{w^{(0)}}{\sigma^{(0)}}\dot{\mathbf{X}}_s^{(0)}  
\nonumber \\  
= -\frac{1}{\sigma^{(0)}}\frac{P_s^{(1)}}{\rho_0} + \frac{\bar{\nu}}{\bar{r}}\left(\bar{r}w_{\bar{r}}^{(0)}\right)_{\bar{r}} + \alpha g\Tilde{T}^{(0)}\hat{\mathbf{y}}\cdot\hat{\pmb{\tau}} 
\nonumber
\end{eqnarray}
which, upon averaging with respect to $\theta$, becomes

\begin{eqnarray} \label{second_order_tangential_momentum_theta_averaged}
w_t^{(0)} - \frac{\bar{\nu}}{\bar{r}}\left(\bar{r}w_{\bar{r}}^{(0)}\right)_{\bar{r}} = -w_{\bar{r}}^{(0)} \left<u^{(2)}\right> - \left<u^{(1)}w_{\bar{r}}^{(1)}\right> - \frac{1}{\bar{r}}\left<v^{(1)}w_\theta^{(1)}\right> \nonumber \\ -\frac{w^{(0)}}{\sigma^{(0)}}\upsilon  - w^{(0)}\kappa^{(0)}\left<v^{(1)}\sin\varphi -u^{(1)}\cos\varphi^{(0)}\right>\quad\quad\quad\quad\quad \\ -v^{(0)}\kappa^{(0)}\left<w^{(1)}\sin\varphi^{(0)}\right> - \frac{\left<P_s^{(1)}\right>}{\rho_0\;\sigma^{(0)}} + \alpha g\Tilde{T}^{(0)}\hat{\mathbf{y}}\cdot\hat{\pmb{\tau}} \nonumber
\end{eqnarray}
where we for the sake of convenience have defined,
\begin{equation}
    \upsilon=\left<w^{(1)}_s\right>+\dot{\mathbf{X}}_s^{(0)}\cdot\hat{\pmb{\tau}^{(0)}}
\end{equation}

Through application of the $\theta$-averaged second order continuity equations and first order momentum equation, this can be rewritten to yield 
\begin{eqnarray} \label{F1}
    \sigma^{(0)} F_1(\bar{r},t) = \frac{1}{\bar{r}}w_{\bar{r}}^{(0)}\int_0^{\bar{r}}\varrho\upsilon\;d\varrho - w^{(0)}\upsilon -\frac{1}{\rho_0}\left<P^{(1)}_s\right>  + \sigma^{(0)}\alpha g\hat{\mathbf{y}}\cdot\hat{\pmb{\tau}}\:\Tilde{T}^{(0)} 
\end{eqnarray}
where
\begin{equation}
  F_1(\bar{r},t)  =     w_t^{(0)} - \frac{\bar{\nu}}{\bar{r}}\left(\bar{r}w^{(0)}_{\bar{r}}\right)_{\bar{r}} 
\end{equation}

Except for the added buoyancy term (the last term on the r.h.s.), equation (\ref{F1}) is the same as in \citet{CallegariTing1978}. Furthermore, we note that $\dot{\mathbf{X}}_s\cdot\hat{\pmb{\tau}}=\dot{\sigma}$ and so $\upsilon=\left<w^{(1)}_s\right>+\dot{\sigma}$. If we define $\Tilde{S}$ such that $\sigma = \Tilde{S}_s$, we have $\Tilde{S}(s,t)=\int_0^{s}\sigma(s',t)ds'$ and $\dot{\sigma}=\Tilde{S}_{st}$. Similarly, equation (\ref{second_order_circumferential_momentum_theta_averaged}) can be reformulated as 
\begin{eqnarray} \label{F_2}
    \sigma^{(0)}F_2(\bar{r},t) = -w^{(0)}\left<v_s^{(1)}\right> +  \frac{\zeta^{(0)}}{\bar{r}}\int_0^{\bar{r}}\varrho\upsilon \; d\varrho 
\end{eqnarray}
where
\begin{equation}
F_2(\bar{r},t) = v_t^{(0)} - \bar{\nu}\left[\frac{1}{\bar{r}}\left(\bar{r}v_{\bar{r}}^{(0)}\right)_{\bar{r}} - \frac{v^{(0)}}{\bar{r}^2}\right]
\end{equation}

Note that the functions $F_1(\bar{r},t)$ and $F_2(\bar{r},t)$, appearing in equations (\ref{F1}) and (\ref{F_2}) are independent of $s$, and that $\sigma^{(0)} = \Tilde{S}_s^{(0)}$, where $\Tilde{S}^{(0)}(s,t)$ is the leading term in the expansion for the arc length at time $t$. Furthermore, 
\begin{equation}
    \dot{\mathbf{X}}_s^{(0)}\cdot\hat{\pmb{\tau}} = \left(\mathbf{X}_s^{(0)}\cdot\hat{\pmb{\tau}}\right)_t = \Tilde{S}_{st}^{(0)}
\end{equation}
Thus, integrating (\ref{F1}) and (\ref{F_2}) over $s$ yields 
\begin{eqnarray}
F_1(\bar{r},t)S^{(0)}(t) &=& \frac{1}{2}\bar{r}^3\left(\frac{w^{(0)}}{\bar{r}^2}\right)_{\bar{r}}\dot{S}^{(0)}(t)  - w^{(0)}\int_{0}^{S_0} \left<w^{(1)}_{s'}\right> ds' - \frac{1}{\rho_0}\int_{0}^{S_0} \left<P^{(1)}_{s'}\right> ds'
 \\ &\quad&\quad\quad + \alpha g \int_0^{S_0}\hat{\mathbf{y}}\cdot\hat{\pmb{\tau}}\:\sigma^{(0)}\:\Tilde{T}^{(0)} ds' \nonumber\\
F_2(\bar{r},t)S^{(0)}(t) &=& \frac{1}{2}\frac{\left(\bar{r}v^{(0)}\right)_{\bar{r}}}{\bar{r}}\dot{S}^{(0)}(t) - w^{(0)}\int_{0}^{S_0} \left<v^{(1)}_{s'}\right> ds' +  \frac{\zeta^{(0)}}{\bar{r}}\int_{0}^{S_0} \rho \Gamma \left<w^{(1)}_{s'}\right> ds'
\end{eqnarray}

For a closed vortex, the integral over $s$ of $\left<w^{(1)}_s\right>$, $\left<v^{(1)}_s\right>$ and $\left<P^{(1)}_s\right>$ all vanish. This also holds true for open filaments, provided it is periodic in some horizontal direction, as was assumed in previous sections, such that $w^{(1)}(s=\text{start})=w^{(1)}(s=\text{end})$. In such cases, we get
\begin{eqnarray} \label{FirstCompatabilityCond_1}
F_1(\bar{r},t)S^{(0)}(t) &=& \frac{1}{2}\bar{r}^3\left(\frac{w^{(0)}}{\bar{r}^2}\right)_{\bar{r}}\dot{S}^{(0)}(t)  
 \\ &\quad&\quad\quad + \alpha g \int_0^{S_0}\hat{\mathbf{y}}\cdot\hat{\pmb{\tau}}\:\sigma^{(0)}\:\Tilde{T}^{(0)} ds \nonumber\\
\label{SecondCompatabilityCond_1}
F_2(\bar{r},t)S^{(0)}(t) &=& \frac{1}{2}\frac{\left(\bar{r}v^{(0)}\right)_{\bar{r}}}{\bar{r}}\dot{S}^{(0)}(t)
\end{eqnarray}

Since we have assumed that all the leading order field components are independent of $s$, we can write,
\begin{equation}
    \int_0^{S_0}\hat{\mathbf{y}}\cdot\hat{\pmb{\tau}}\:\sigma^{(0)}\:\Tilde{T}^{(0)} ds  = \Tilde{T}^{(0)}(\bar{r},t)\int_0^{S_0}\hat{\mathbf{y}}\cdot\hat{\pmb{\tau}}\:\sigma^{(0)} ds
\end{equation}
The integrand can be written as
\begin{equation}
    \hat{\mathbf{y}}\cdot(d\mathbf{X})
\end{equation}
Thus, implying that
\begin{equation}
    \int_0^{S_0} \hat{\mathbf{y}}\cdot\hat{\pmb{\tau}} \sigma^{(0)} ds = \int _0^{S_0} \hat{\mathbf{y}}\cdot(d\mathbf{X}) = \hat{\mathbf{y}}\cdot\left(\mathbf{X}(S_0)-\mathbf{X}(0)\right)
\end{equation}

This is nothing but the height difference between the initial and final points of the filament. Thus, we can conclude that in the case of closed filament, the gravity term vanishes. Furthermore, if the ends of an infinite vortex filament are at the same height, this term will also vanish, indicating that the only time we get a contribution from gravity in the regime outlined here, is when we have an infinitely long filament which is tilted w.r.t. the vertical direction, thus breaking the horizontal symmetry. This rather curious result, is partially due to having assumed that the leading order velocity components are axially symmetric, meaning that one cannot have regions of acceleration (deceleration) of the flow along the center line. 
Thus, in the case of closed or horizontally periodic vortex filaments the influence of gravity on the internal flow vanishes and we can conclude that it was indeed valid to ignore gravity in the numerical simulations in sections II- V. 

For the sake of completeness, we include the derivation of an equation for the temporal evolution of $\Tilde{T}^{(0)}$. This equation is derived from the second order temperature equation,
\begin{eqnarray}
    T_t^{(0)} + \frac{1}{\bar{r}}T^{(2)}_\theta v^{(0)} + \frac{w^{(0)}}{\sigma^{(0)}}T_s^{(1)} + u^{(1)}T^{(1)}_{\bar{r}} + \frac{v^{(1)}}{\bar{r}}T^{(1)}_\theta  \nonumber \\ + u^{(2)}T^{(0)}_{\bar{r}} -\bar{r}\frac{\partial \hat{\mathbf{r}}}{\partial t} \cdot \hat{\mathbf{r}} T^{(0)}_{\bar{r}} = \frac{\bar{\nu}}{\bar{r}\bar{\mu}}\left(\bar{r}T^{(0)}_{\bar{r}}\right)_{\bar{r}}
\end{eqnarray}
Taking the average of this equation w.r.t. $\theta$ yields
\begin{eqnarray}
    T_t^{(0)} + \frac{w^{(0)}}{\sigma^{(0)}}T_s^{(1)} + \left<u^{(1)}T^{(1)}_{\bar{r}}\right> \quad\quad\quad\quad\quad\quad\quad\nonumber \\  + \frac{1}{\bar{r}}\left<v^{(1)}T_\theta^{(1)}\right>  + \left<u^{(2)}\right>T_{\bar{r}}^{(0)}  = \frac{\bar{\nu}}{\bar{r}\bar{\mu}}\left(\bar{r}T^{(0)}_{\bar{r}}\right)_{\bar{r}}
\end{eqnarray}
which, through application of the first order continuity equation, can be written as
\begin{eqnarray} \label{second_order_temperature_theta_averaged}
    T_t^{(0)} + \frac{w^{(0)}}{\sigma^{(0)}}T_s^{(1)} -\frac{T_{\bar{r}}^{(0)}}{\bar{r}\sigma^{(0)}}\int_0^{\bar{r}} \varrho\left[\left<w_s^{(1)}\right>+\left<\dot{\mathbf{X}}_s^{(0)}\cdot\hat{\pmb{\tau}}\right>\right]d\varrho  \nonumber \\ = \frac{\bar{\nu}}{\bar{r}\bar{\mu}}\left(\bar{r}T_{\bar{r}}^{(0)}\right)_{\bar{r}}\,.
\end{eqnarray}
Here, we have used that 
\begin{eqnarray}
    \left<\left(T^{(1)}v^{(1)}\right)_\theta\right> = \left<T^{(1)}u^{(1)}\right> = 0\,,
\end{eqnarray}
which holds for flows in which the leading order field components are axially independent \citep{TingKleinKnio2007}. Integrating equation (\ref{second_order_temperature_theta_averaged}) w.r.t. $s$, using the symmetry condition, yields
\begin{equation} \label{ThirdCompatabilityCond_1}
    T_t^{(0)}S^{(0)} - \frac{1}{2}T_{\bar{r}}^{(0)}\bar{r}\dot{S}^{(0)}(t) - \frac{\bar{\nu}}{\bar{r}\bar{\mu}}\left(\bar{r}T_{\bar{r}}^{(0)}\right)_{\bar{r}}S^{(0)} =0
\end{equation}

Within the Boussinesq approximation, we require that leading order temperature difference, $\Tilde{T}^{(0)}$, decays exponentially as $\bar{r}$ becomes large, or in other words
\begin{equation}
    \Tilde{T}^{(0)} = o(\bar{r}^{-n}) \quad \text{ for all } n \text{ as } \bar{r}\rightarrow \infty
\end{equation}

The exponential decay of $\Tilde{T}^{(0)}$ means that an analytic solution to equation (\ref{ThirdCompatabilityCond_1}) can be obtained, for large time $t$, in terms of Laguerre polynomials. We do not include the calculation here, but refer to \citet{CallegariTing1978} on how to approach this problem. 

Thus, for non closed filaments lacking horizontal symmetry one would need to combine equations (\ref{F1}), (\ref{F_2}) and (\ref{ThirdCompatabilityCond_1}) together with equations for the first order pressure and axial velocity component to arrive at an expression for the core constant. The equations for the first order pressure and axial velocity component are not included here, but can in principle be derived from higher order equations following standard asymptotic techniques. If these equations are solved analytically, they will yield expressions for the leading order velocity field components needed to determine the core constant $C(t)$. This new core constant can be implemented directly in the M1 KK method to observe the effect of gravity on the motion of hairpin filaments. However, this is beyond the current scope of the paper and will be pursued in a future work.

\section{Summary and conclusions}
\label{sec: conclusion}

In this work, we investigated the abundance and orientation of hairpin vortex structures in the strongly stratified Ekman flow reported in \citet{harikrishnan2021geometry}. To study the motion of these structures, we treat them as slender vortex filaments where the diameter of the vortex core $d$ is much smaller than its characteristic radius of curvature $R$.

First, the evolution of the hairpin filament is studied in a stagnant background flow with two methods: the Local Induction Approximation (LIA) and the corrected thin-tube model of \citet{KleinKnio1995} with the M1 optimization technique\citep{knio2000improved} (M1 KK). Results of both methods indicate that the tip region (or head) of the hairpin vortex moves backward and downward rapidly and are able to correctly capture the ``corkscrew" shape (in the side view) as shown in \citet{hon1991evolution}. However, if the initial width of the hairpin filament is reduced, i.e., for $\beta = 50$ as shown in figure \ref{fig: beta_test}(b, d), a marked difference can be observed in the development of the ``legs'' between LIA and the M1 KK scheme, the latter of which can represent the nonlocal effects accurately.

With the M1 KK scheme, the motion of hairpin filaments with an ABL background flow was studied for three cases: two stably stratified (S\_1, S\_2) cases at different degrees of stratification and a neutrally stratified (N) case. The hairpin filament was initialized as a small, nearly two-dimensional perturbation and simulations were run for four different initial conditions with varying core size ($\delta$) and circulation ($\Gamma$). For all four initial conditions, the near-planar perturbation becomes three-dimensional as a result of the self-induced velocity of the filament \rk{which induces} the head of the hairpin vortex \rk{to bend} backward away from the wall, thereby increasing the inclination angle ($\gamma$). In particular, we note the following:
\begin{itemize}
	\item [(1)] For the initial condition $R_3$, which has a relatively thinner core and stronger circulation resulting in a large self-induced velocity, the differences in inclination rate and maximum inclination are negligible for all background flow profiles.
	\item [(2)] For all other initial conditions, it can be seen that an increase in stratification results in a slower inclination rate and a smaller maximum inclination angle. 
	\item [(3)] Since the effect of rotation is present in the Ekman flow, it can be seen that the hairpin filament, which is initially symmetric about $z^+ = 0$, becomes asymmetric over time thereby tilting the vortex. As expected, the degree of tilt increases with an increase in the strength of stratification. Tracking of the $Q$-criterion structure in the DNS dataset shows that the asymmetry becomes stronger as the structure grows in size.
	\item [(4)] Apart from advecting in the streamwise direction, the hairpin filament also experiences strong advection in the spanwise direction. Initializing the filaments at three different heights $y^+ = 15, 30, 50$, we observe an appreciable change in their spanwise orientation, particularly for the strongly stratified case S\_1. The spanwise orientation changes from clockwise to anticlockwise (with respect to the streamwise direction) with increase in height. This hints that, under stable stratification, hairpin filaments initialized at the same height may have a similar spanwise orientation.  
\end{itemize}

Although we present a link between the spanwise orientation of the hairpin filament and its initial starting height for the very stable regime, it is still not clear if this is solely responsible for the similar \rk{orientation} of these structures in different regions of the flow witnessed in \citet{harikrishnan2020curious, harikrishnan2021lagrangian}. Further probing of the DNS dataset is necessary \rk{with hairpin-like structures and their interactions which need} to be tracked in time. Our current efforts are focused in this direction. Further efforts can be directed towards studying the impact of mutual induction of neighboring hairpin filaments (where viscous effects become important) on their orientation. An extension of the M1 KK method incorporating the viscous diffusion has already been presented in \citet{klein1996representation}. This should lead to a straightforward modification of our slender vortex filament code. Additionally, it remains to be seen if the inclusion of gravity as shown in section \ref{sec: mathematical_formulation} will have a noticeable impact on the dynamics of hairpin filaments. We emphasize the need to study the dynamics of hairpin filaments/structures, particularly in the very stable regime, as it is still a poorly understood element of the atmospheric boundary layer\cite{mahrt2014stably}.

% ----- ACKNOWLEDGEMENTS ----- %

\begin{acknowledgments}
The authors would like to thank Professor Omar M. Knio for lending support and \rk{helpful} discussions during the initial stages of the work. They would also like to thank Dr. Cedrick Ansorge who performed the DNS simulations of the stratified Ekman flows and for providing access to the Jülich Supercomputing Center where the data is stored under the project dns2share. This research is funded by the Deutsche Forschungsgemeinschaft (DFG) through grant CRC 1114 ``Scaling Cascades in Complex Systems'', Project Number 235221301, Project B07: Self-similar structures in turbluent flows and the construction of LES closures. 
\end{acknowledgments}

% ----- DATA AVAILABILITY STATEMENT ----- %

\section*{Data Availability Statement}

The code developed for the motion of slender vortex filaments during the course of this work is written in python. This is based on the EZ-vortex code\citep{margerit2004implementation} which was written in C. The python code and the mean velocity profiles from the DNS database are available at the GitHub repository: https://github.com/Phoenixfire1081/SlenderVortexSimulation.

\appendix

% ----- APPENDIX 1 ----- %

\section{Validation of the codes}
\label{appendix: validation_tests}

\begin{figure}
    \centering
    \includegraphics[width = 0.45\textwidth]{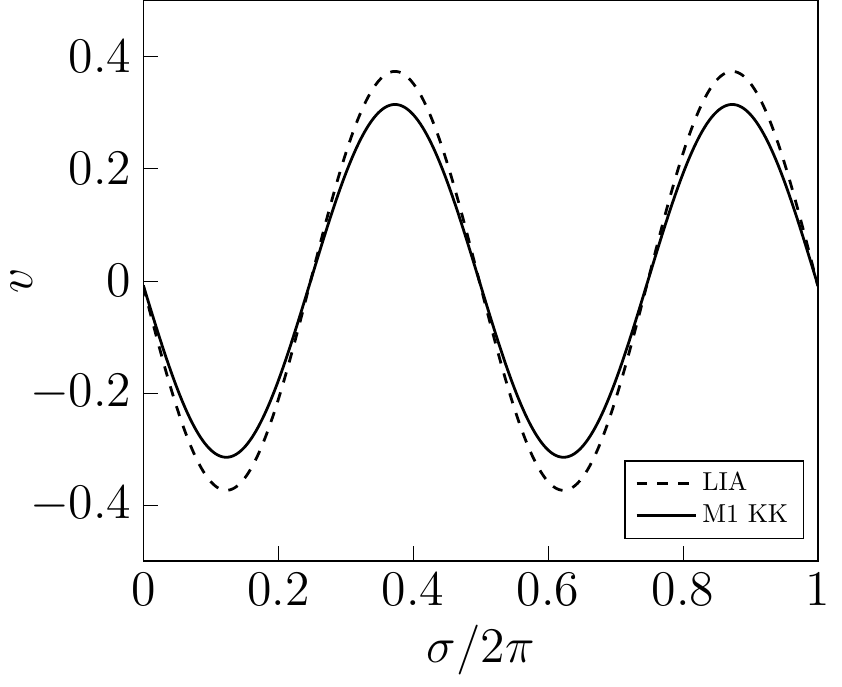}
    \caption{Binormal velocity prediction for the plane curve shown in \ref{eq: sinu_plane_curve} with LIA (dashed line) and the M1 KK (solid line) methods.}
    \label{fig: velocity_prediction}
\end{figure}

Here, the static test presented in section $5.2$ of \citet{KleinKnio1995} is used to validate our slender vortex code. This test which compares the velocity predictions of both LIA and M1 KK schemes eliminates the possibility of errors introduced during temporal discretization due to its static nature. The equation for generating a sinusoidal plane curve is reiterated as,
\begin{equation} 
\label{eq: sinu_plane_curve}
\mathbf{X}(s) = s \mathbf{t} + \epsilon^2 \Tilde{a}\, \sin (2s/\epsilon) \mathbf{n}
\end{equation}
where $\mathbf{t}$ and $\mathbf{n}$ are mutually orthogonal unit vectors, $\Tilde{a}$ is the amplitude and $s$ is a linear spacing of node locations satisfying the overlap condition. The test is performed with the following parameters: $\delta = 0.01$, $\epsilon^2$ = 0.25, $\Gamma = 4\pi$, $\Tilde{a} = 0.01$. It is easy to check that a minimum of $N = 316$ nodes are necessary to satisfy the overlap condition from (\ref{eq: overlap_condition}). However, $1024$ nodes are chosen to eliminate numerical inaccuracies in this test. The choice of mutually orthogonal vectors will induce a velocity normal to the plane i.e., in the direction of $\mathbf{b = t \times n}$. Therefore, the velocity prediction along $\mathbf{b}$ is shown in figure \ref{fig: velocity_prediction}. The maximum deviations in velocity predictions between LIA and M1 KK methods is about 19\%. While this is slightly lower than the 25-30\% reported in \citet{KleinKnio1995} who compared the asymptotic predictions of the Klein-Majda scheme with LIA, the differences clearly highlight the overestimation of velocity by LIA which excludes non-local effects of vortex self-induction. 

% ----- APPENDIX 2 ----- %

\section{Impact of resolution on velocity prediction}

\begin{figure}
    \centering
    \includegraphics[width = 0.45\textwidth]{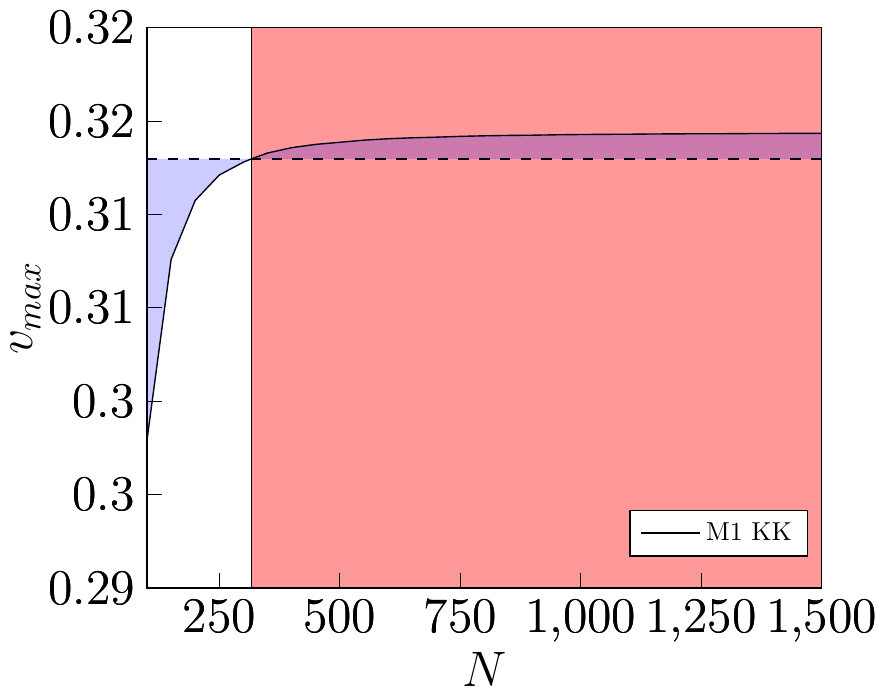}
    \caption{Maximum binormal velocity for the plane curve shown in \ref{eq: sinu_plane_curve} with M1 KK (solid line) for various values of $N$. The dashed line shows the maximum velocity with $N = 317$ which satisfies the overlap condition \ref{eq: overlap_condition} and the blue shaded region shows the deviation. The red shaded region shows all $N$ satisfying the overlap condition.}
    \label{fig: maxVel_vs_N}
\end{figure}

For the test shown in Appendix \ref{appendix: validation_tests}, the number of nodes $N$ is varied and all other parameters are held constant. $N$ is increased in increments of $50$ starting from $100$ nodes and is tested until $1500$. The results in figure \ref{fig: maxVel_vs_N} show that after overlap is satisfied (shaded red region), the deviation between the maximum velocity with $316$ nodes, which is the minumum number of nodes necessary for overlap, and $1500$ nodes is less than $0.5\%$. Therefore, one can surmise that the minumum number of nodes, henceforth referred as \textit{minimum overlap}, gives satisfactory velocity predictions.

% ----- APPENDIX 3 ----- %

\section{Choice of parameters}
\label{appendix: K_alpha}

\begin{figure}
    \centering
    \includegraphics[width = 0.45\textwidth]{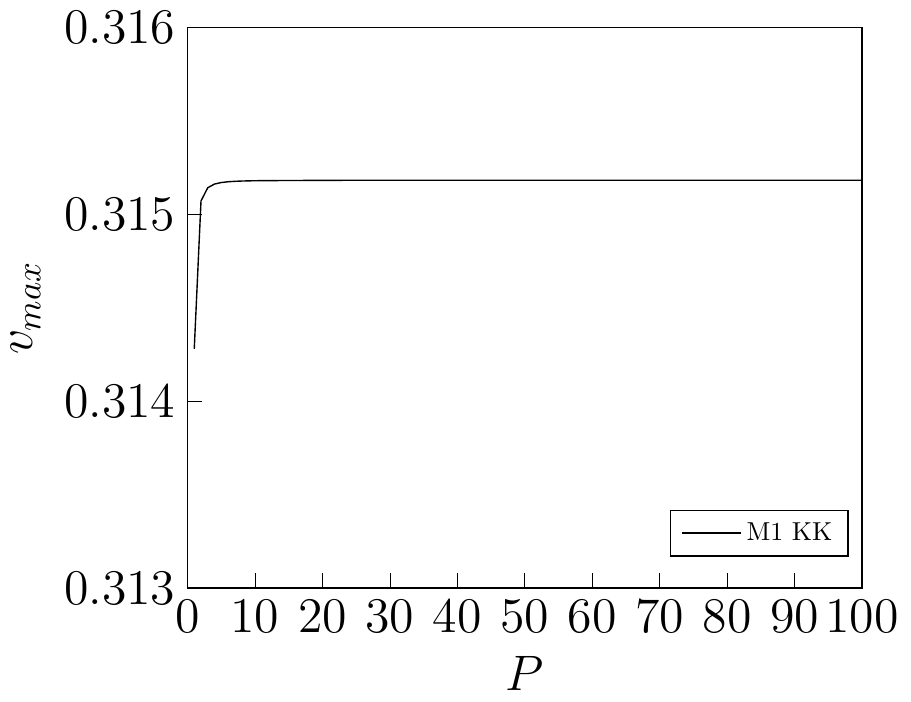}
    \caption{Maximum binormal velocity for the plane curve shown in  \ref{eq: sinu_plane_curve} with M1 KK (solid line) for various values of $P$.}
    \label{fig: maxVel_vs_P}
\end{figure}

With minimum overlap satisfied, the optimum values for three parameters are identified in this section. First, the number of images $P$ is varied from $1$ to $100$ and the maximum binormal velocity is computed in each case. From figure \ref{fig: maxVel_vs_P}, it can be seen that the velocity prediction shows no deviations for $P\geq7$. In all subsequent simulations, we choose $P = 8$. 

\begin{figure}
    \centering
    \includegraphics[width = 0.45\textwidth]{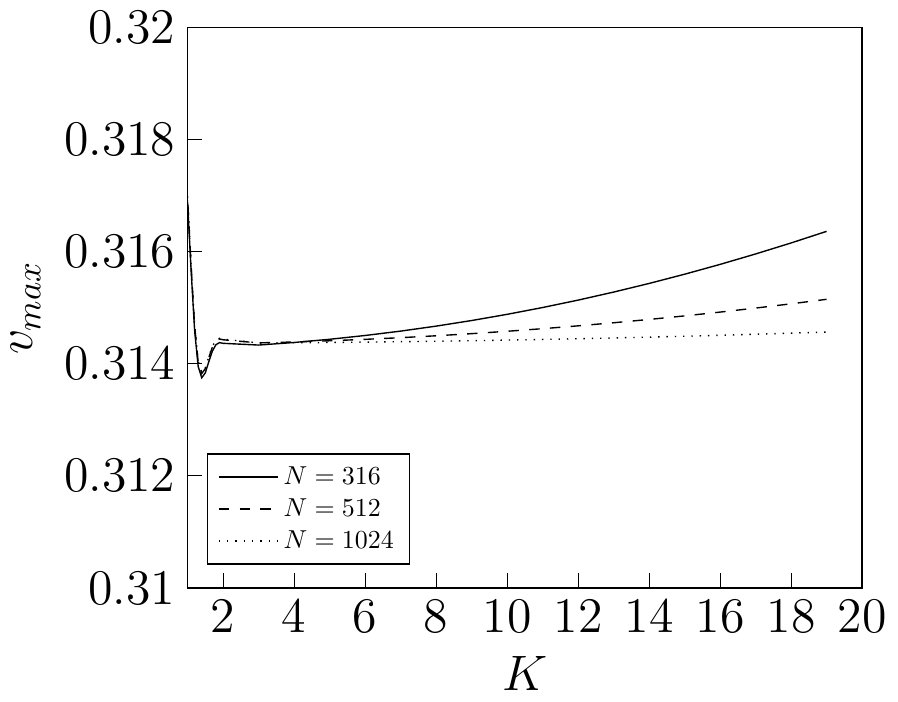}
    \caption{Maximum binormal velocity for the plane curve shown in  \ref{eq: sinu_plane_curve} with M1 KK for various values of $K$. The solid, dashed and dotted lines show the differences with the number of nodes $N$.}
    \label{fig: maxVel_vs_K}
\end{figure}

\begin{figure}
    \centering
    \includegraphics[width = 0.45\textwidth]{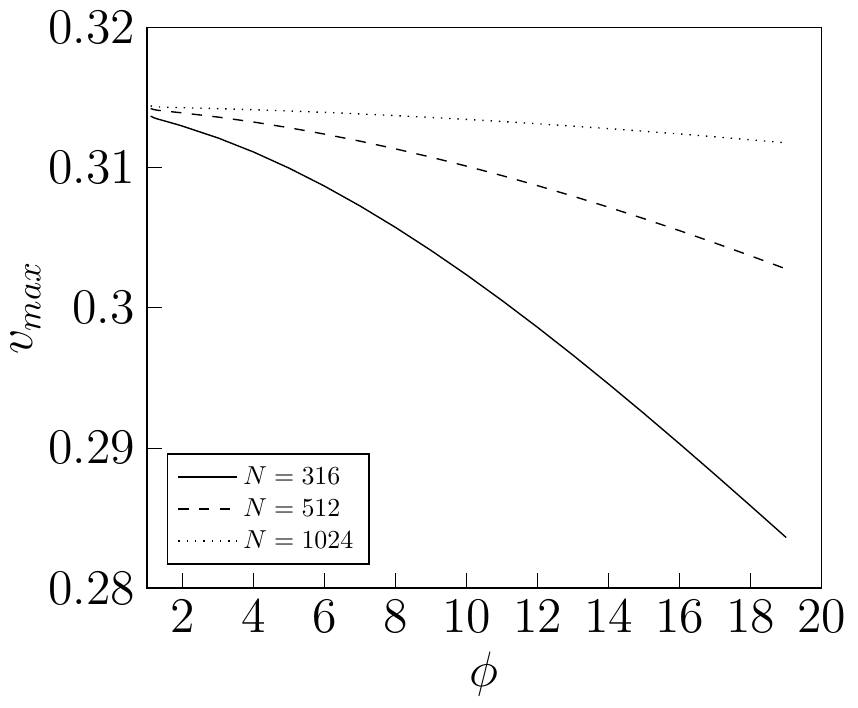}
    \caption{Maximum binormal velocity for the plane curve shown in  \ref{eq: sinu_plane_curve} with M1 KK for various values of $\alpha$. The solid and dashed show the differences with the number of nodes $N$.}
    \label{fig: maxVel_vs_A}
\end{figure}

The parameter $K$ defined in (\ref{eq: fat_mesh_core}) is a ratio between the maximum element length along the filament and the core size can be thought of as an overlap parameter. To estimate this parameter, we use Method 2 velocity correction approach of \citet{knio2000improved}. The corrected velocity is given by

\begin{equation}\label{eq: M2_velocity}
    v^{\text{ttm}}_{\text{corr}} = v_{1} + \frac{\Gamma}{4\pi}\, ln \bigg(\frac{\sigma_1}{\sigma}\bigg)\kappa\mathbf{b}
\end{equation}

\noindent where $\sigma_1$ is described in (\ref{eq: fat_mesh_core}). It can be noted from figure \ref{fig: maxVel_vs_K} that the maximum binormal velocity predictions show less deviations for larger values of $K$ when the nodes are increased. A similar trend can be observed for the parameter $\phi$ in figure \ref{fig: maxVel_vs_A}. Therefore, to keep the numerical errors low, the overlap condition is modified as

\begin{equation}\label{eq: overlap_condition_modified}
    \underset{i = 1..N} {max} |\delta \bm{\chi}_{i}| < \frac{\delta}{3}
\end{equation}

Although this increases the number of nodes necessary for computation threefold, it ensures that the velocity predictions will have sufficient accuracy for each time step. For all simulations in this work, $K = 3$ and $\phi = 2$.

% ----- APPENDIX 4 ----- %

\section{Multilevel percolation thresholding in time}
\label{appendix: timeMLP}

To overcome the limitation of using a constant threshold in time, we introduce a novel thresholding scheme based on multilevel percolation analysis (MLP). 

The MLP scheme, where percolation analysis is applied in an iterative manner, relies on the definition of simple and complex structures to identify an optimum threshold. Given a scalar field, the ratio $V_{max}/V$ (where $V_{max}$ is the volume of the biggest structure in the domain and $V$ is the volume of all structures) is computed over the entire range of threshold values. When $V_{max}/V = 1$ for the entire threshold range, there exists exactly one structure. This type of structure is denoted \textit{simple} because when the threshold is increased, the structure collapses to a single local maximum. A \textit{complex} structure is encountered when $V_{max}/V$ falls below $1$ over the entire threshold range, i.e., two structures may co-exist. In practice, using $V_{max}/V = 1$ to find simple structures is computationally intensive and is therefore relaxed to $V_{max}/V > 0.5$ which implies that at most two structures can co-exist where one structure always has a volume larger than the other. These definitions are used to identify optimum thresholds in time. This method, henceforth called as overlap method with multilevel percolation thresholding in time (MLPT) is discussed below,

\begin{figure*}[h!]
    \centering
    \includegraphics[width=0.75\textwidth]{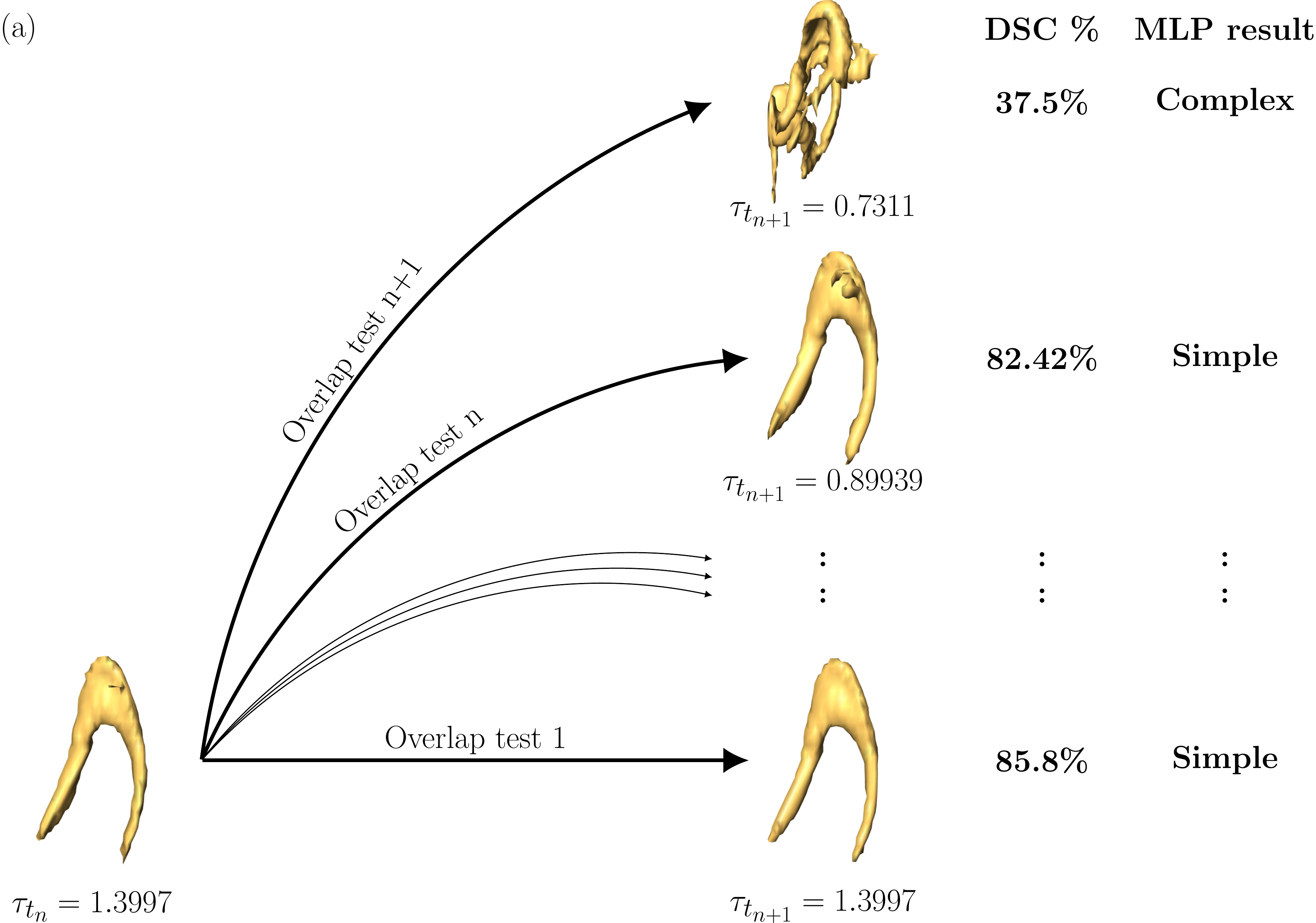}
    \includegraphics[width=0.75\textwidth]{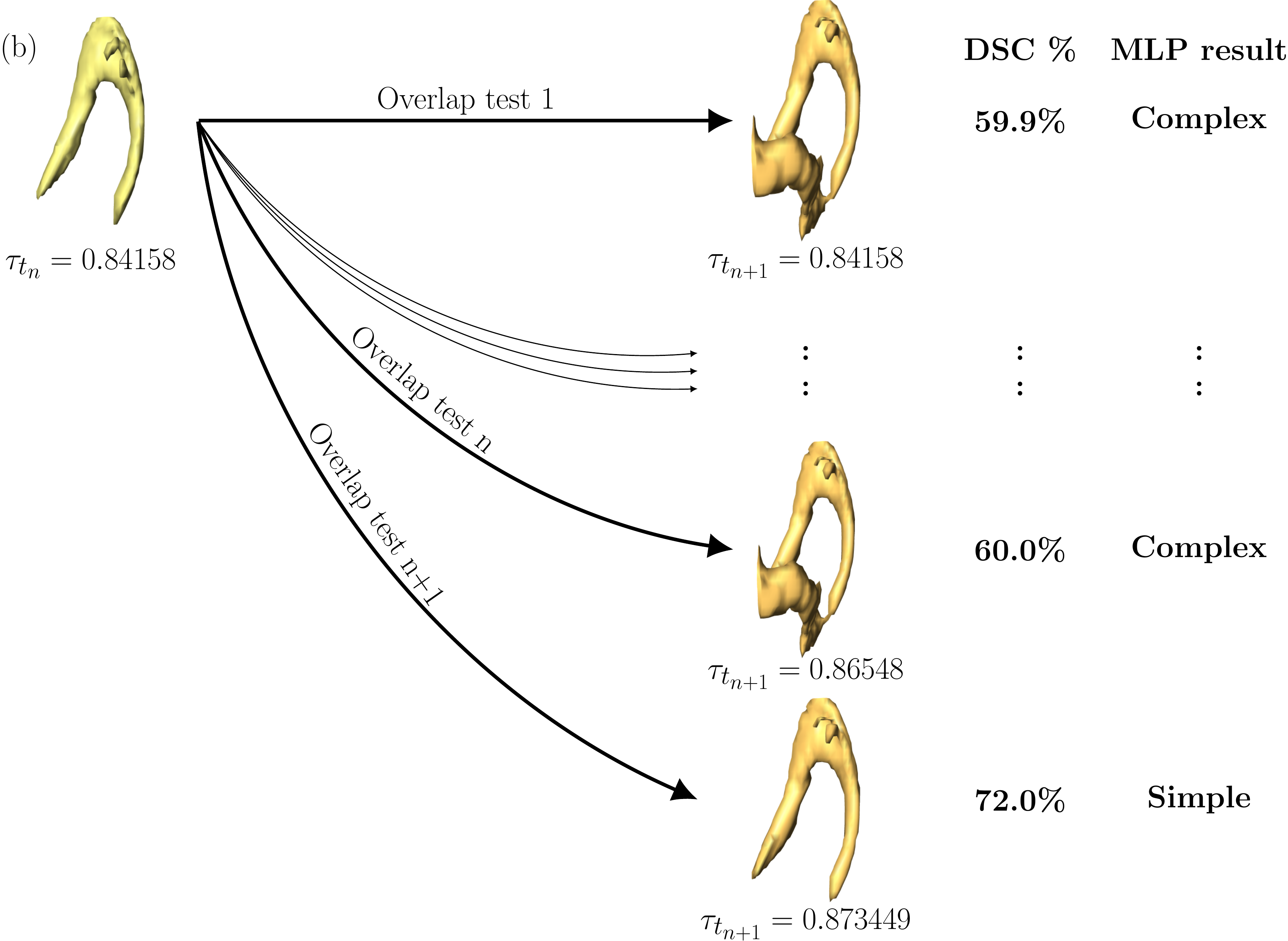}
    \caption{The process of selecting an optimum threshold for a subsequent timestep with MLPT is shown here. The structure is freely allowed to (a) grow and (b) shrink in time with decreasing and increasing the thresholds, respectively. }
    \label{fig: MLPT}
\end{figure*}

\begin{itemize}
    \item [(1)] Once the structure to be tracked is chosen by the user, the NS+MC extraction algorithm is used to extract the structure at the MLP threshold from the scalar field.
    \item [(2)] First, the algorithm checks for structure growth (see \ref{fig: MLPT}(a)) by decreasing the threshold in the subsequent time step. In our case, the minimum threshold to check is taken to be the global percolation threshold, $\tau_p$. However, in practice, the minimum value to check is restricted only by the indicator itself. For instance, with $Q$-criterion, smaller thresholds can be checked as long as $Q > 0$.
    \item [(3)] For every threshold being tested, a structure at timestep $n+1$ is found by overlapping the structure at timestep $n$. Once the $t_{n+1}$ structure is found, it is extracted and subjected to MLP to determine if it is a simple or complex structure. If the structure is found to be simple, the threshold is increased and the process is repeated until a complex structure is found. This signals the algorithm that a previous threshold is the optimum one.
    \item [(4)] If the overlapped structure at $t_{n+1}$ is complex at the MLP threshold itself, then the algorithm checks if the structure is shrinking in time (see \ref{fig: MLPT}(b)) by increasing the threshold.  
    \item [(5)] Steps 2, 3, 4 are repeated for every timestep.
\end{itemize}

\noindent \textit{Optimization: }Since hairpins are small-scale features that need to be tracked in a large domain of size $1024 \times 256 \times 2048$ (approximately $2\, \text{GB}$ for single precision, raw binary format) per timestep, the procedure described above can become computationally expensive. Therefore, two key optimization steps are used: 

\begin{itemize}
    \item [(1)] Since we use temporally well-resolved DNS data for tracking, it can be assumed that the structure is not advected too far in the domain for every time step. Once the initial structure is extracted at the first time step, a smaller domain around the structure of interest can be searched in the subsequent time step which restricts the search space for the extraction algorithm. The smaller computational domain is then moved in time.
    \item [(2)] To reduce RAM overhead, specific bytes of data corresponding to the smaller computational domain is read into memory. 
\end{itemize}

With both optimizations, the structure in figure \ref{fig: constant_MLPT_comparison}(b) was tracked for $674$ time steps (or $1.4$ TB of data) in under $7$ hours.

\section{Impact of image vortices}
\label{appendix: image_vortex}

\begin{figure*}[ht!]
    \centering
    \includegraphics[width=0.75\textwidth]{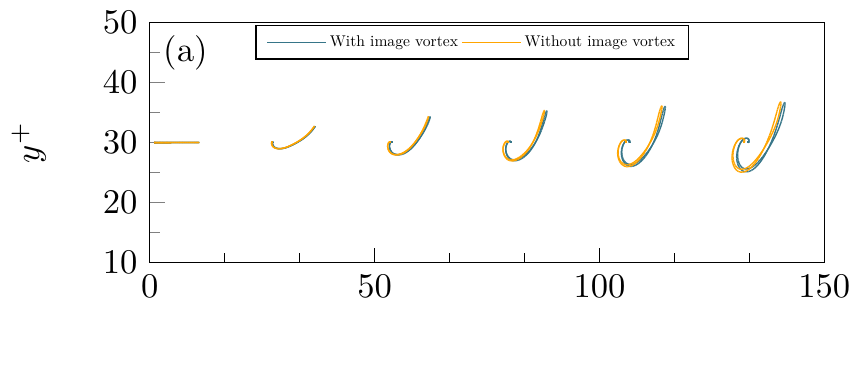}
    \includegraphics[width=0.75\textwidth]{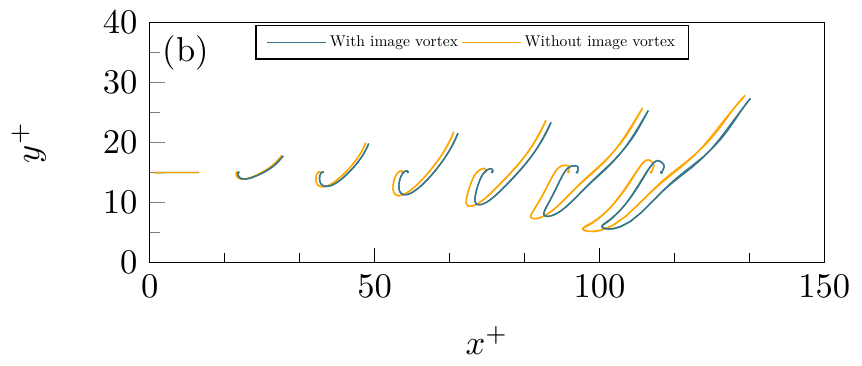}
    \caption{Evolution of a hairpin filament at two different heights (a) $y^+ = 30$ and (b) $y^+ = 15$ are shown both with and without an image vortex.}
    \label{fig: image_vortex}
\end{figure*}

\begin{figure}
    \centering
    \includegraphics[width = 0.45\textwidth]{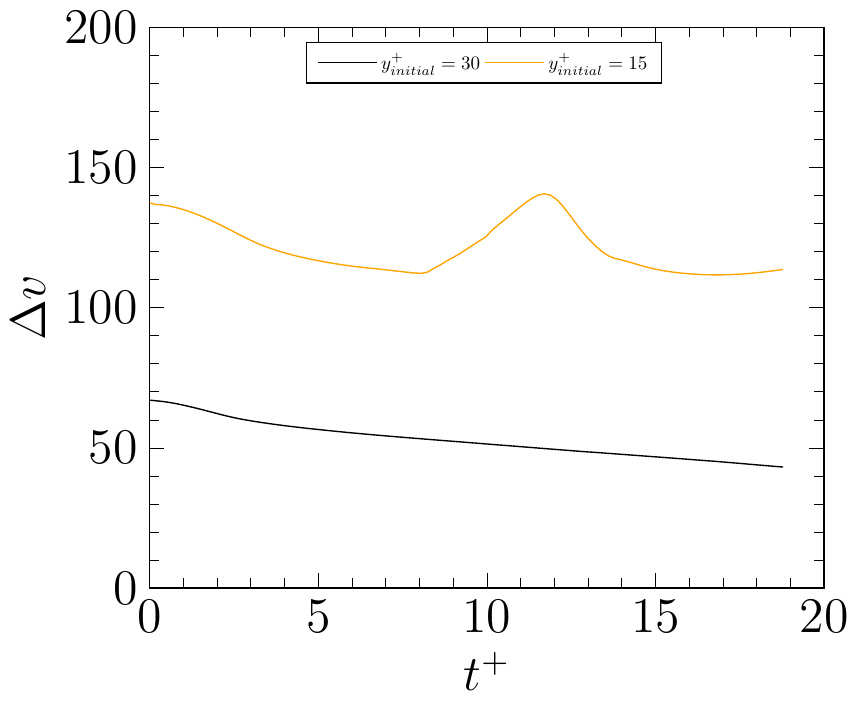}
    \caption{Time history of the difference between the total velocity with and without an image vortex at two different heights as depicted in figure \ref{fig: image_vortex}.}
    \label{fig: velocity_image_vortex}
\end{figure}

The boundary condition of a rigid wall is enforced with the help of an image vortex. In this scenario, an additional step is included in the calculation after step $3$ of the summary presented in section \ref{sec: numerical_methods}. The image vortex also has two contributions as shown in \ref{eq: standard_ttm_split}, i.e., from the central part of the domain and its images on the left and right side to enforce periodicity. It should be noted that the effect of velocity smoothing function is neglected from the central part of the image domain as $L \gg \delta$. Once the image vortex contribution is added, steps $4$ and $5$ are carried out as indicated in section \ref{sec: numerical_methods}. 

We will use initial condition $R_3$ (see table \ref{tab: initial_conditions}) to test the effect of image vortices at two initial heights $y^+_{\text{initial}} = 15, 30$. The image vortices are placed at $y^+_{\text{initial}} = -15, -30$, respectively. Each filament and its image vortex are represented with $600$ nodes. Since twice the number of nodes are used in the calculation when image vortices are present, consequently the calculation time is also doubled. The temporal development of these filaments are shown in figure \ref{fig: image_vortex}. In agreement with the findings reported in \citet{moin1986evolution}, the image vortices appear to enhance the streamwise advection while actively pushing the filament away from the wall. This effect is clearly visible for the filament at $y^+_{\text{initial}} = 15$. In accordance with our expectations, if the total velocity i.e., the sum of self-induced velocity and the background flow velocity is summed up for all nodes along the filament, the difference in velocity when an image vortex is present and absent reduces with increasing height (see figure \ref{fig: velocity_image_vortex}).

% ----- APPENDIX 5 ----- %

\section{Leading and first order temperature equations}
\label{appendix: asymptotics_temp_equation}

\noindent To reiterate, the non-dimensional temperature equation in curvilinear coordinates is 

\begin{equation}
    \frac{\partial T}{\partial t} + \left(V-\epsilon\bar{r}\frac{\partial \hat{r}}{\partial t}\right)\nabla T = \frac{1}{\text{Pr}\:\text{Re}}\Delta T
\end{equation}
where Pr and Re denote the Prandtl and Reynolds numbers, respectively. Given the stretched coordinates in (\ref{streteched_coordinates}), we have Re$^{-1/2} =\bar{\nu}^{1/2}\epsilon$. Furthermore, we have assumed that Pr$=\mathcal{O}(1)=\bar{\mu}$. Hence, the leading order temperature equation is $\mathcal{O}(\epsilon^{-2})$, yielding

\begin{equation}
   T_\theta^{(0)} = 0 
\end{equation}
The first order, or $\mathcal{O}(\epsilon^{-1})$ equation is:

\begin{equation} \label{eq: first order temperature equation}
    \frac{v^{(0)}}{\bar{r}}T_\theta^{(1)} + u^{(1)}T_{\bar{r}}^{(0)} + \frac{w^{(0)}}{\sigma^{(0)}}T_s^{(0)} = 0
\end{equation}
Taking the average of \ref{eq: first order temperature equation} with respect to $\theta$ yields the symmetric first order equation:

\begin{equation}
    u_c^{(1)}T_{\bar{r}}^{(0)}+ \frac{w^{(0)}}{\sigma^{(0)}}T_s^{(0)} = 0
\end{equation}
where the subscript $c$ denotes the symmetric part of $u^{(1)}$. \\
\nocite{*}
\bibliography{aipsamp}% Produces the bibliography via BibTeX.

\providecommand{\noopsort}[1]{}\providecommand{\singleletter}[1]{#1}%
\begin{thebibliography}{64}%
\makeatletter
\providecommand \@ifxundefined [1]{%
 \@ifx{#1\undefined}
}%
\providecommand \@ifnum [1]{%
 \ifnum #1\expandafter \@firstoftwo
 \else \expandafter \@secondoftwo
 \fi
}%
\providecommand \@ifx [1]{%
 \ifx #1\expandafter \@firstoftwo
 \else \expandafter \@secondoftwo
 \fi
}%
\providecommand \natexlab [1]{#1}%
\providecommand \enquote  [1]{``#1''}%
\providecommand \bibnamefont  [1]{#1}%
\providecommand \bibfnamefont [1]{#1}%
\providecommand \citenamefont [1]{#1}%
\providecommand \href@noop [0]{\@secondoftwo}%
\providecommand \href [0]{\begingroup \@sanitize@url \@href}%
\providecommand \@href[1]{\@@startlink{#1}\@@href}%
\providecommand \@@href[1]{\endgroup#1\@@endlink}%
\providecommand \@sanitize@url [0]{\catcode `\\12\catcode `\$12\catcode
  `\&12\catcode `\#12\catcode `\^12\catcode `\_12\catcode `\%12\relax}%
\providecommand \@@startlink[1]{}%
\providecommand \@@endlink[0]{}%
\providecommand \url  [0]{\begingroup\@sanitize@url \@url }%
\providecommand \@url [1]{\endgroup\@href {#1}{\urlprefix }}%
\providecommand \urlprefix  [0]{URL }%
\providecommand \Eprint [0]{\href }%
\providecommand \doibase [0]{http://dx.doi.org/}%
\providecommand \selectlanguage [0]{\@gobble}%
\providecommand \bibinfo  [0]{\@secondoftwo}%
\providecommand \bibfield  [0]{\@secondoftwo}%
\providecommand \translation [1]{[#1]}%
\providecommand \BibitemOpen [0]{}%
\providecommand \bibitemStop [0]{}%
\providecommand \bibitemNoStop [0]{.\EOS\space}%
\providecommand \EOS [0]{\spacefactor3000\relax}%
\providecommand \BibitemShut  [1]{\csname bibitem#1\endcsname}%
\let\auto@bib@innerbib\@empty
%</preamble>
\bibitem [{\citenamefont {Harikrishnan}\ \emph
  {et~al.}(2021{\natexlab{a}})\citenamefont {Harikrishnan}, \citenamefont
  {Ansorge}, \citenamefont {Klein},\ and\ \citenamefont
  {Vercauteren}}]{harikrishnan2021geometry}%
  \BibitemOpen
  \bibfield  {author} {\bibinfo {author} {\bibfnamefont {A.}~\bibnamefont
  {Harikrishnan}}, \bibinfo {author} {\bibfnamefont {C.}~\bibnamefont
  {Ansorge}}, \bibinfo {author} {\bibfnamefont {R.}~\bibnamefont {Klein}}, \
  and\ \bibinfo {author} {\bibfnamefont {N.}~\bibnamefont {Vercauteren}},\
  }\bibfield  {title} {\enquote {\bibinfo {title} {Geometry and organization of
  coherent structures in stably stratified atmospheric boundary layers},}\
  }\href@noop {} {\bibfield  {journal} {\bibinfo  {journal} {arXiv preprint
  arXiv:2110.02253}\ } (\bibinfo {year} {2021}{\natexlab{a}})}\BibitemShut
  {NoStop}%
\bibitem [{\citenamefont {Klein}\ and\ \citenamefont
  {Knio}(1995)}]{KleinKnio1995}%
  \BibitemOpen
  \bibfield  {author} {\bibinfo {author} {\bibfnamefont {R.}~\bibnamefont
  {Klein}}\ and\ \bibinfo {author} {\bibfnamefont {O.~M.}\ \bibnamefont
  {Knio}},\ }\bibfield  {title} {\enquote {\bibinfo {title} {Asymptotic
  vorticity structure and numerical simulation of slender vortex filaments},}\
  }\href@noop {} {\bibfield  {journal} {\bibinfo  {journal} {Journal of Fluid
  Mechanics}\ }\textbf {\bibinfo {volume} {284}} (\bibinfo {year}
  {1995})}\BibitemShut {NoStop}%
\bibitem [{\citenamefont {Callegari}\ and\ \citenamefont
  {Ting}(1978)}]{CallegariTing1978}%
  \BibitemOpen
  \bibfield  {author} {\bibinfo {author} {\bibfnamefont {A.~J.}\ \bibnamefont
  {Callegari}}\ and\ \bibinfo {author} {\bibfnamefont {L.}~\bibnamefont
  {Ting}},\ }\bibfield  {title} {\enquote {\bibinfo {title} {Motion of a curved
  vortex filament with decaying vortical core and axial velocity},}\
  }\href@noop {} {\bibfield  {journal} {\bibinfo  {journal} {Journal of Applied
  Mathematics.}\ }\textbf {\bibinfo {volume} {35}} (\bibinfo {year}
  {1978})}\BibitemShut {NoStop}%
\bibitem [{\citenamefont {Head}\ and\ \citenamefont
  {Bandyopadhyay}(1981)}]{head1981new}%
  \BibitemOpen
  \bibfield  {author} {\bibinfo {author} {\bibfnamefont {M.}~\bibnamefont
  {Head}}\ and\ \bibinfo {author} {\bibfnamefont {P.}~\bibnamefont
  {Bandyopadhyay}},\ }\bibfield  {title} {\enquote {\bibinfo {title} {New
  aspects of turbulent boundary-layer structure},}\ }\href@noop {} {\bibfield
  {journal} {\bibinfo  {journal} {Journal of fluid mechanics}\ }\textbf
  {\bibinfo {volume} {107}},\ \bibinfo {pages} {297--338} (\bibinfo {year}
  {1981})}\BibitemShut {NoStop}%
\bibitem [{\citenamefont {Acarlar}\ and\ \citenamefont
  {Smith}(1987{\natexlab{a}})}]{acarlar1987study}%
  \BibitemOpen
  \bibfield  {author} {\bibinfo {author} {\bibfnamefont {M.}~\bibnamefont
  {Acarlar}}\ and\ \bibinfo {author} {\bibfnamefont {C.}~\bibnamefont
  {Smith}},\ }\bibfield  {title} {\enquote {\bibinfo {title} {A study of
  hairpin vortices in a laminar boundary layer. {P}art 1. {H}airpin vortices
  generated by a hemisphere protuberance},}\ }\href@noop {} {\bibfield
  {journal} {\bibinfo  {journal} {Journal of Fluid Mechanics}\ }\textbf
  {\bibinfo {volume} {175}},\ \bibinfo {pages} {1--41} (\bibinfo {year}
  {1987}{\natexlab{a}})}\BibitemShut {NoStop}%
\bibitem [{\citenamefont {Acarlar}\ and\ \citenamefont
  {Smith}(1987{\natexlab{b}})}]{acarlar1987study2}%
  \BibitemOpen
  \bibfield  {author} {\bibinfo {author} {\bibfnamefont {M.}~\bibnamefont
  {Acarlar}}\ and\ \bibinfo {author} {\bibfnamefont {C.}~\bibnamefont
  {Smith}},\ }\bibfield  {title} {\enquote {\bibinfo {title} {A study of
  hairpin vortices in a laminar boundary layer. {P}art 2. {H}airpin vortices
  generated by fluid injection},}\ }\href@noop {} {\bibfield  {journal}
  {\bibinfo  {journal} {Journal of Fluid Mechanics}\ }\textbf {\bibinfo
  {volume} {175}},\ \bibinfo {pages} {43--83} (\bibinfo {year}
  {1987}{\natexlab{b}})}\BibitemShut {NoStop}%
\bibitem [{\citenamefont {Adrian}, \citenamefont {Meinhart},\ and\
  \citenamefont {Tomkins}(2000)}]{adrian2000vortex}%
  \BibitemOpen
  \bibfield  {author} {\bibinfo {author} {\bibfnamefont {R.~J.}\ \bibnamefont
  {Adrian}}, \bibinfo {author} {\bibfnamefont {C.~D.}\ \bibnamefont
  {Meinhart}}, \ and\ \bibinfo {author} {\bibfnamefont {C.~D.}\ \bibnamefont
  {Tomkins}},\ }\bibfield  {title} {\enquote {\bibinfo {title} {Vortex
  organization in the outer region of the turbulent boundary layer},}\
  }\href@noop {} {\bibfield  {journal} {\bibinfo  {journal} {Journal of fluid
  Mechanics}\ }\textbf {\bibinfo {volume} {422}},\ \bibinfo {pages} {1--54}
  (\bibinfo {year} {2000})}\BibitemShut {NoStop}%
\bibitem [{\citenamefont {Robinson}(1991)}]{robinson1991kinematics}%
  \BibitemOpen
  \bibfield  {author} {\bibinfo {author} {\bibfnamefont {S.~K.}\ \bibnamefont
  {Robinson}},\ }\emph {\bibinfo {title} {The kinematics of turbulent boundary
  layer structure}},\ \href@noop {} {Ph.D. thesis},\ \bibinfo  {school}
  {Stanford University} (\bibinfo {year} {1991})\BibitemShut {NoStop}%
\bibitem [{\citenamefont {Adrian}(2007)}]{adrian2007hairpin}%
  \BibitemOpen
  \bibfield  {author} {\bibinfo {author} {\bibfnamefont {R.~J.}\ \bibnamefont
  {Adrian}},\ }\bibfield  {title} {\enquote {\bibinfo {title} {Hairpin vortex
  organization in wall turbulence},}\ }\href@noop {} {\bibfield  {journal}
  {\bibinfo  {journal} {Physics of fluids}\ }\textbf {\bibinfo {volume} {19}},\
  \bibinfo {pages} {041301} (\bibinfo {year} {2007})}\BibitemShut {NoStop}%
\bibitem [{\citenamefont {Zhou}\ \emph {et~al.}(1999)\citenamefont {Zhou},
  \citenamefont {Adrian}, \citenamefont {Balachandar},\ and\ \citenamefont
  {Kendall}}]{zhou1999mechanisms}%
  \BibitemOpen
  \bibfield  {author} {\bibinfo {author} {\bibfnamefont {J.}~\bibnamefont
  {Zhou}}, \bibinfo {author} {\bibfnamefont {R.~J.}\ \bibnamefont {Adrian}},
  \bibinfo {author} {\bibfnamefont {S.}~\bibnamefont {Balachandar}}, \ and\
  \bibinfo {author} {\bibfnamefont {T.}~\bibnamefont {Kendall}},\ }\bibfield
  {title} {\enquote {\bibinfo {title} {Mechanisms for generating coherent
  packets of hairpin vortices in channel flow},}\ }\href@noop {} {\bibfield
  {journal} {\bibinfo  {journal} {Journal of fluid mechanics}\ }\textbf
  {\bibinfo {volume} {387}},\ \bibinfo {pages} {353--396} (\bibinfo {year}
  {1999})}\BibitemShut {NoStop}%
\bibitem [{\citenamefont {Ansorge}\ and\ \citenamefont
  {Mellado}(2014)}]{ansorge2014global}%
  \BibitemOpen
  \bibfield  {author} {\bibinfo {author} {\bibfnamefont {C.}~\bibnamefont
  {Ansorge}}\ and\ \bibinfo {author} {\bibfnamefont {J.~P.}\ \bibnamefont
  {Mellado}},\ }\bibfield  {title} {\enquote {\bibinfo {title} {Global
  intermittency and collapsing turbulence in the stratified planetary boundary
  layer},}\ }\href@noop {} {\bibfield  {journal} {\bibinfo  {journal}
  {Boundary-layer meteorology}\ }\textbf {\bibinfo {volume} {153}},\ \bibinfo
  {pages} {89--116} (\bibinfo {year} {2014})}\BibitemShut {NoStop}%
\bibitem [{\citenamefont {Ansorge}\ and\ \citenamefont
  {Mellado}(2016)}]{ansorge2016analyses}%
  \BibitemOpen
  \bibfield  {author} {\bibinfo {author} {\bibfnamefont {C.}~\bibnamefont
  {Ansorge}}\ and\ \bibinfo {author} {\bibfnamefont {J.~P.}\ \bibnamefont
  {Mellado}},\ }\bibfield  {title} {\enquote {\bibinfo {title} {Analyses of
  external and global intermittency in the logarithmic layer of {E}kman
  flow},}\ }\href@noop {} {\bibfield  {journal} {\bibinfo  {journal} {Journal
  of Fluid Mechanics}\ }\textbf {\bibinfo {volume} {805}},\ \bibinfo {pages}
  {611--635} (\bibinfo {year} {2016})}\BibitemShut {NoStop}%
\bibitem [{\citenamefont {Ansorge}(2016)}]{ansorge2016thesis}%
  \BibitemOpen
  \bibfield  {author} {\bibinfo {author} {\bibfnamefont {C.}~\bibnamefont
  {Ansorge}},\ }\href@noop {} {\emph {\bibinfo {title} {Analyses of turbulence
  in the neutrally and stably stratified planetary boundary layer}}}\ (\bibinfo
   {publisher} {Springer},\ \bibinfo {year} {2016})\BibitemShut {NoStop}%
\bibitem [{\citenamefont {Mahrt}(1989)}]{mahrt1989intermittency}%
  \BibitemOpen
  \bibfield  {author} {\bibinfo {author} {\bibfnamefont {L.}~\bibnamefont
  {Mahrt}},\ }\bibfield  {title} {\enquote {\bibinfo {title} {Intermittency of
  atmospheric turbulence},}\ }\href@noop {} {\bibfield  {journal} {\bibinfo
  {journal} {Journal of the Atmospheric Sciences}\ }\textbf {\bibinfo {volume}
  {46}},\ \bibinfo {pages} {79--95} (\bibinfo {year} {1989})}\BibitemShut
  {NoStop}%
\bibitem [{\citenamefont {Hunt}, \citenamefont {Wray},\ and\ \citenamefont
  {Moin}(1988)}]{hunt1988eddies}%
  \BibitemOpen
  \bibfield  {author} {\bibinfo {author} {\bibfnamefont {J.~C.}\ \bibnamefont
  {Hunt}}, \bibinfo {author} {\bibfnamefont {A.~A.}\ \bibnamefont {Wray}}, \
  and\ \bibinfo {author} {\bibfnamefont {P.}~\bibnamefont {Moin}},\ }\bibfield
  {title} {\enquote {\bibinfo {title} {Eddies, streams, and convergence zones
  in turbulent flows},}\ }\href@noop {} {\bibfield  {journal} {\bibinfo
  {journal} {Studying turbulence using numerical simulation databases, 2.
  Proceedings of the 1988 summer program}\ } (\bibinfo {year}
  {1988})}\BibitemShut {NoStop}%
\bibitem [{\citenamefont {Harikrishnan}\ \emph {et~al.}(2020)\citenamefont
  {Harikrishnan}, \citenamefont {Ansorge}, \citenamefont {Klein},\ and\
  \citenamefont {Vercauteren}}]{harikrishnan2020curious}%
  \BibitemOpen
  \bibfield  {author} {\bibinfo {author} {\bibfnamefont {A.}~\bibnamefont
  {Harikrishnan}}, \bibinfo {author} {\bibfnamefont {C.}~\bibnamefont
  {Ansorge}}, \bibinfo {author} {\bibfnamefont {R.}~\bibnamefont {Klein}}, \
  and\ \bibinfo {author} {\bibfnamefont {N.}~\bibnamefont {Vercauteren}},\
  }\bibfield  {title} {\enquote {\bibinfo {title} {The curious nature of
  hairpin vortices.}}\ }\href@noop {} {\bibfield  {journal} {\bibinfo
  {journal} {Gallery of Fluid Motion}\ } (\bibinfo {year} {2020})}\BibitemShut
  {NoStop}%
\bibitem [{\citenamefont {Harikrishnan}\ \emph
  {et~al.}(2021{\natexlab{b}})\citenamefont {Harikrishnan}, \citenamefont
  {Ansorge}, \citenamefont {Klein},\ and\ \citenamefont
  {Vercauteren}}]{harikrishnan2021lagrangian}%
  \BibitemOpen
  \bibfield  {author} {\bibinfo {author} {\bibfnamefont {A.}~\bibnamefont
  {Harikrishnan}}, \bibinfo {author} {\bibfnamefont {C.}~\bibnamefont
  {Ansorge}}, \bibinfo {author} {\bibfnamefont {R.}~\bibnamefont {Klein}}, \
  and\ \bibinfo {author} {\bibfnamefont {N.}~\bibnamefont {Vercauteren}},\
  }\bibfield  {title} {\enquote {\bibinfo {title} {Lagrangian hairpins in
  atmospheric boundary layers},}\ }\href@noop {} {\bibfield  {journal}
  {\bibinfo  {journal} {Gallery of Fluid Motion}\ } (\bibinfo {year}
  {2021}{\natexlab{b}})}\BibitemShut {NoStop}%
\bibitem [{\citenamefont {Green}, \citenamefont {Rowley},\ and\ \citenamefont
  {Haller}(2007)}]{green2007detection}%
  \BibitemOpen
  \bibfield  {author} {\bibinfo {author} {\bibfnamefont {M.~A.}\ \bibnamefont
  {Green}}, \bibinfo {author} {\bibfnamefont {C.~W.}\ \bibnamefont {Rowley}}, \
  and\ \bibinfo {author} {\bibfnamefont {G.}~\bibnamefont {Haller}},\
  }\bibfield  {title} {\enquote {\bibinfo {title} {Detection of lagrangian
  coherent structures in three-dimensional turbulence},}\ }\href@noop {}
  {\bibfield  {journal} {\bibinfo  {journal} {Journal of Fluid Mechanics}\
  }\textbf {\bibinfo {volume} {572}},\ \bibinfo {pages} {111--120} (\bibinfo
  {year} {2007})}\BibitemShut {NoStop}%
\bibitem [{\citenamefont {Hommema}\ and\ \citenamefont
  {Adrian}(2003)}]{hommema2003packet}%
  \BibitemOpen
  \bibfield  {author} {\bibinfo {author} {\bibfnamefont {S.~E.}\ \bibnamefont
  {Hommema}}\ and\ \bibinfo {author} {\bibfnamefont {R.~J.}\ \bibnamefont
  {Adrian}},\ }\bibfield  {title} {\enquote {\bibinfo {title} {Packet structure
  of surface eddies in the atmospheric boundary layer},}\ }\href@noop {}
  {\bibfield  {journal} {\bibinfo  {journal} {Boundary-Layer Meteorology}\
  }\textbf {\bibinfo {volume} {106}},\ \bibinfo {pages} {147--170} (\bibinfo
  {year} {2003})}\BibitemShut {NoStop}%
\bibitem [{\citenamefont {Li}\ and\ \citenamefont
  {Bou-Zeid}(2011)}]{li2011coherent}%
  \BibitemOpen
  \bibfield  {author} {\bibinfo {author} {\bibfnamefont {D.}~\bibnamefont
  {Li}}\ and\ \bibinfo {author} {\bibfnamefont {E.}~\bibnamefont {Bou-Zeid}},\
  }\bibfield  {title} {\enquote {\bibinfo {title} {Coherent structures and the
  dissimilarity of turbulent transport of momentum and scalars in the unstable
  atmospheric surface layer},}\ }\href@noop {} {\bibfield  {journal} {\bibinfo
  {journal} {Boundary-Layer Meteorology}\ }\textbf {\bibinfo {volume} {140}},\
  \bibinfo {pages} {243--262} (\bibinfo {year} {2011})}\BibitemShut {NoStop}%
\bibitem [{\citenamefont {Heisel}\ \emph {et~al.}(2018)\citenamefont {Heisel},
  \citenamefont {Dasari}, \citenamefont {Liu}, \citenamefont {Hong},
  \citenamefont {Coletti},\ and\ \citenamefont {Guala}}]{heisel2018spatial}%
  \BibitemOpen
  \bibfield  {author} {\bibinfo {author} {\bibfnamefont {M.}~\bibnamefont
  {Heisel}}, \bibinfo {author} {\bibfnamefont {T.}~\bibnamefont {Dasari}},
  \bibinfo {author} {\bibfnamefont {Y.}~\bibnamefont {Liu}}, \bibinfo {author}
  {\bibfnamefont {J.}~\bibnamefont {Hong}}, \bibinfo {author} {\bibfnamefont
  {F.}~\bibnamefont {Coletti}}, \ and\ \bibinfo {author} {\bibfnamefont
  {M.}~\bibnamefont {Guala}},\ }\bibfield  {title} {\enquote {\bibinfo {title}
  {The spatial structure of the logarithmic region in very-high-reynolds-number
  rough wall turbulent boundary layers},}\ }\href@noop {} {\bibfield  {journal}
  {\bibinfo  {journal} {Journal of Fluid Mechanics}\ }\textbf {\bibinfo
  {volume} {857}},\ \bibinfo {pages} {704--747} (\bibinfo {year}
  {2018})}\BibitemShut {NoStop}%
\bibitem [{\citenamefont {Oncley}, \citenamefont {Hartogensis},\ and\
  \citenamefont {Tong}(2016)}]{oncley2016whirlwinds}%
  \BibitemOpen
  \bibfield  {author} {\bibinfo {author} {\bibfnamefont {S.~P.}\ \bibnamefont
  {Oncley}}, \bibinfo {author} {\bibfnamefont {O.}~\bibnamefont {Hartogensis}},
  \ and\ \bibinfo {author} {\bibfnamefont {C.}~\bibnamefont {Tong}},\
  }\bibfield  {title} {\enquote {\bibinfo {title} {Whirlwinds and hairpins in
  the atmospheric surface layer},}\ }\href@noop {} {\bibfield  {journal}
  {\bibinfo  {journal} {Journal of the Atmospheric Sciences}\ }\textbf
  {\bibinfo {volume} {73}},\ \bibinfo {pages} {4927--4943} (\bibinfo {year}
  {2016})}\BibitemShut {NoStop}%
\bibitem [{\citenamefont {Watanabe}\ \emph {et~al.}(2019)\citenamefont
  {Watanabe}, \citenamefont {Riley}, \citenamefont {Nagata}, \citenamefont
  {Matsuda},\ and\ \citenamefont {Onishi}}]{watanabe2019hairpin}%
  \BibitemOpen
  \bibfield  {author} {\bibinfo {author} {\bibfnamefont {T.}~\bibnamefont
  {Watanabe}}, \bibinfo {author} {\bibfnamefont {J.~J.}\ \bibnamefont {Riley}},
  \bibinfo {author} {\bibfnamefont {K.}~\bibnamefont {Nagata}}, \bibinfo
  {author} {\bibfnamefont {K.}~\bibnamefont {Matsuda}}, \ and\ \bibinfo
  {author} {\bibfnamefont {R.}~\bibnamefont {Onishi}},\ }\bibfield  {title}
  {\enquote {\bibinfo {title} {Hairpin vortices and highly elongated flow
  structures in a stably stratified shear layer},}\ }\href@noop {} {\bibfield
  {journal} {\bibinfo  {journal} {Journal of Fluid Mechanics}\ }\textbf
  {\bibinfo {volume} {878}},\ \bibinfo {pages} {37--61} (\bibinfo {year}
  {2019})}\BibitemShut {NoStop}%
\bibitem [{\citenamefont {Kline}\ \emph {et~al.}(1967)\citenamefont {Kline},
  \citenamefont {Reynolds}, \citenamefont {Schraub},\ and\ \citenamefont
  {Runstadler}}]{kline1967structure}%
  \BibitemOpen
  \bibfield  {author} {\bibinfo {author} {\bibfnamefont {S.~J.}\ \bibnamefont
  {Kline}}, \bibinfo {author} {\bibfnamefont {W.~C.}\ \bibnamefont {Reynolds}},
  \bibinfo {author} {\bibfnamefont {F.}~\bibnamefont {Schraub}}, \ and\
  \bibinfo {author} {\bibfnamefont {P.}~\bibnamefont {Runstadler}},\ }\bibfield
   {title} {\enquote {\bibinfo {title} {The structure of turbulent boundary
  layers},}\ }\href@noop {} {\bibfield  {journal} {\bibinfo  {journal} {Journal
  of Fluid Mechanics}\ }\textbf {\bibinfo {volume} {30}},\ \bibinfo {pages}
  {741--773} (\bibinfo {year} {1967})}\BibitemShut {NoStop}%
\bibitem [{\citenamefont {Saiki}, \citenamefont {Moeng},\ and\ \citenamefont
  {Sullivan}(2000)}]{saiki2000large}%
  \BibitemOpen
  \bibfield  {author} {\bibinfo {author} {\bibfnamefont {E.~M.}\ \bibnamefont
  {Saiki}}, \bibinfo {author} {\bibfnamefont {C.-H.}\ \bibnamefont {Moeng}}, \
  and\ \bibinfo {author} {\bibfnamefont {P.~P.}\ \bibnamefont {Sullivan}},\
  }\bibfield  {title} {\enquote {\bibinfo {title} {Large-eddy simulation of the
  stably stratified planetary boundary layer},}\ }\href@noop {} {\bibfield
  {journal} {\bibinfo  {journal} {Boundary-Layer Meteorology}\ }\textbf
  {\bibinfo {volume} {95}},\ \bibinfo {pages} {1--30} (\bibinfo {year}
  {2000})}\BibitemShut {NoStop}%
\bibitem [{\citenamefont {Jim{\'e}nez}\ and\ \citenamefont
  {Cuxart}(2005)}]{jimenez2005large}%
  \BibitemOpen
  \bibfield  {author} {\bibinfo {author} {\bibfnamefont {M.}~\bibnamefont
  {Jim{\'e}nez}}\ and\ \bibinfo {author} {\bibfnamefont {J.}~\bibnamefont
  {Cuxart}},\ }\bibfield  {title} {\enquote {\bibinfo {title} {Large-eddy
  simulations of the stable boundary layer using the standard kolmogorov
  theory: Range of applicability},}\ }\href@noop {} {\bibfield  {journal}
  {\bibinfo  {journal} {Boundary-Layer Meteorology}\ }\textbf {\bibinfo
  {volume} {115}},\ \bibinfo {pages} {241--261} (\bibinfo {year}
  {2005})}\BibitemShut {NoStop}%
\bibitem [{\citenamefont {Marusic}, \citenamefont {Kunkel},\ and\ \citenamefont
  {Port{\'e}-Agel}(2001)}]{marusic2001experimental}%
  \BibitemOpen
  \bibfield  {author} {\bibinfo {author} {\bibfnamefont {I.}~\bibnamefont
  {Marusic}}, \bibinfo {author} {\bibfnamefont {G.~J.}\ \bibnamefont {Kunkel}},
  \ and\ \bibinfo {author} {\bibfnamefont {F.}~\bibnamefont {Port{\'e}-Agel}},\
  }\bibfield  {title} {\enquote {\bibinfo {title} {Experimental study of wall
  boundary conditions for large-eddy simulation},}\ }\href@noop {} {\bibfield
  {journal} {\bibinfo  {journal} {Journal of Fluid Mechanics}\ }\textbf
  {\bibinfo {volume} {446}},\ \bibinfo {pages} {309--320} (\bibinfo {year}
  {2001})}\BibitemShut {NoStop}%
\bibitem [{\citenamefont {Chauhan}\ \emph {et~al.}(2013)\citenamefont
  {Chauhan}, \citenamefont {Hutchins}, \citenamefont {Monty},\ and\
  \citenamefont {Marusic}}]{chauhan2013structure}%
  \BibitemOpen
  \bibfield  {author} {\bibinfo {author} {\bibfnamefont {K.}~\bibnamefont
  {Chauhan}}, \bibinfo {author} {\bibfnamefont {N.}~\bibnamefont {Hutchins}},
  \bibinfo {author} {\bibfnamefont {J.}~\bibnamefont {Monty}}, \ and\ \bibinfo
  {author} {\bibfnamefont {I.}~\bibnamefont {Marusic}},\ }\bibfield  {title}
  {\enquote {\bibinfo {title} {Structure inclination angles in the convective
  atmospheric surface layer},}\ }\href@noop {} {\bibfield  {journal} {\bibinfo
  {journal} {Boundary-layer meteorology}\ }\textbf {\bibinfo {volume} {147}},\
  \bibinfo {pages} {41--50} (\bibinfo {year} {2013})}\BibitemShut {NoStop}%
\bibitem [{\citenamefont {Lozano-Dur{\'a}n}\ and\ \citenamefont
  {Jim{\'e}nez}(2014)}]{lozano2014time}%
  \BibitemOpen
  \bibfield  {author} {\bibinfo {author} {\bibfnamefont {A.}~\bibnamefont
  {Lozano-Dur{\'a}n}}\ and\ \bibinfo {author} {\bibfnamefont {J.}~\bibnamefont
  {Jim{\'e}nez}},\ }\bibfield  {title} {\enquote {\bibinfo {title}
  {Time-resolved evolution of coherent structures in turbulent channels:
  characterization of eddies and cascades},}\ }\href@noop {} {\bibfield
  {journal} {\bibinfo  {journal} {Journal of fluid mechanics}\ }\textbf
  {\bibinfo {volume} {759}},\ \bibinfo {pages} {432--471} (\bibinfo {year}
  {2014})}\BibitemShut {NoStop}%
\bibitem [{\citenamefont {Hon}\ and\ \citenamefont
  {Walker}(1988)}]{HonWalker1995}%
  \BibitemOpen
  \bibfield  {author} {\bibinfo {author} {\bibfnamefont {T.}~\bibnamefont
  {Hon}}\ and\ \bibinfo {author} {\bibfnamefont {J.}~\bibnamefont {Walker}},\
  }\bibfield  {title} {\enquote {\bibinfo {title} {Evolution of hairpin
  vortices in a shear flow},}\ }\href@noop {} {\bibfield  {journal} {\bibinfo
  {journal} {Nasa Technical Memorandum 100858}\ } (\bibinfo {year}
  {1988})}\BibitemShut {NoStop}%
\bibitem [{\citenamefont {Leonard}(1985)}]{leonard1985computing}%
  \BibitemOpen
  \bibfield  {author} {\bibinfo {author} {\bibfnamefont {A.}~\bibnamefont
  {Leonard}},\ }\bibfield  {title} {\enquote {\bibinfo {title} {Computing
  three-dimensional incompressible flows with vortex elements},}\ }\href@noop
  {} {\bibfield  {journal} {\bibinfo  {journal} {Annual Review of Fluid
  Mechanics}\ }\textbf {\bibinfo {volume} {17}},\ \bibinfo {pages} {523--559}
  (\bibinfo {year} {1985})}\BibitemShut {NoStop}%
\bibitem [{\citenamefont {Hama}(1962)}]{hama1962progressive}%
  \BibitemOpen
  \bibfield  {author} {\bibinfo {author} {\bibfnamefont {F.~R.}\ \bibnamefont
  {Hama}},\ }\bibfield  {title} {\enquote {\bibinfo {title} {Progressive
  deformation of a curved vortex filament by its own induction},}\ }\href@noop
  {} {\bibfield  {journal} {\bibinfo  {journal} {The Physics of Fluids}\
  }\textbf {\bibinfo {volume} {5}},\ \bibinfo {pages} {1156--1162} (\bibinfo
  {year} {1962})}\BibitemShut {NoStop}%
\bibitem [{\citenamefont {Chang}\ and\ \citenamefont
  {Smith}(2016)}]{Change2016}%
  \BibitemOpen
  \bibfield  {author} {\bibinfo {author} {\bibfnamefont {C.}~\bibnamefont
  {Chang}}\ and\ \bibinfo {author} {\bibfnamefont {S.~G.~L.}\ \bibnamefont
  {Smith}},\ }\bibfield  {title} {\enquote {\bibinfo {title} {The motion of a
  buoyant vortex filament},}\ }\href@noop {} {\bibfield  {journal} {\bibinfo
  {journal} {Journal of Fluid Mechanics}\ }\textbf {\bibinfo {volume} {857}}
  (\bibinfo {year} {2016})}\BibitemShut {NoStop}%
\bibitem [{\citenamefont {Moore}\ and\ \citenamefont
  {Saffman}(1972)}]{moore1972motion}%
  \BibitemOpen
  \bibfield  {author} {\bibinfo {author} {\bibfnamefont {D.~W.}\ \bibnamefont
  {Moore}}\ and\ \bibinfo {author} {\bibfnamefont {P.~G.}\ \bibnamefont
  {Saffman}},\ }\bibfield  {title} {\enquote {\bibinfo {title} {The motion of a
  vortex filament with axial flow},}\ }\href@noop {} {\bibfield  {journal}
  {\bibinfo  {journal} {Philosophical Transactions of the Royal Society of
  London. Series A, Mathematical and Physical Sciences}\ }\textbf {\bibinfo
  {volume} {272}},\ \bibinfo {pages} {403--429} (\bibinfo {year}
  {1972})}\BibitemShut {NoStop}%
\bibitem [{\citenamefont {Turner}(1957)}]{Turner1957}%
  \BibitemOpen
  \bibfield  {author} {\bibinfo {author} {\bibfnamefont {J.~S.}\ \bibnamefont
  {Turner}},\ }\bibfield  {title} {\enquote {\bibinfo {title} {Buoyant vortex
  rings},}\ }\href@noop {} {\bibfield  {journal} {\bibinfo  {journal}
  {Philosophical Transactions of the Royal Society of London. Series A,
  Mathematical and Physical Sciences}\ }\textbf {\bibinfo {volume} {239}},\
  \bibinfo {pages} {61–75} (\bibinfo {year} {1957})}\BibitemShut {NoStop}%
\bibitem [{\citenamefont {Ting}, \citenamefont {Klein},\ and\ \citenamefont
  {Knio}(2007)}]{TingKleinKnio2007}%
  \BibitemOpen
  \bibfield  {author} {\bibinfo {author} {\bibfnamefont {L.}~\bibnamefont
  {Ting}}, \bibinfo {author} {\bibfnamefont {R.}~\bibnamefont {Klein}}, \ and\
  \bibinfo {author} {\bibfnamefont {O.~M.}\ \bibnamefont {Knio}},\ }\href@noop
  {} {\emph {\bibinfo {title} {Vortex Dominated Flows: Analysis and Computation
  for Multiple Scale Phenomena}}}\ (\bibinfo  {publisher} {Springer-Verlag},\
  \bibinfo {year} {2007})\BibitemShut {NoStop}%
\bibitem [{\citenamefont {Knio}, \citenamefont {Ting},\ and\ \citenamefont
  {Klein}(2003)}]{Knio2003}%
  \BibitemOpen
  \bibfield  {author} {\bibinfo {author} {\bibfnamefont {O.~M.}\ \bibnamefont
  {Knio}}, \bibinfo {author} {\bibfnamefont {L.}~\bibnamefont {Ting}}, \ and\
  \bibinfo {author} {\bibfnamefont {R.}~\bibnamefont {Klein}},\ }\bibfield
  {title} {\enquote {\bibinfo {title} {Theory of compressible vortex
  filaments},}\ }\href@noop {} {\bibfield  {journal} {\bibinfo  {journal}
  {Proceedings of Second MIT Conference on Computational Fluid Dynamics and
  Solid Mechanics}\ ,\ \bibinfo {pages} {971–973}} (\bibinfo {year}
  {2003})}\BibitemShut {NoStop}%
\bibitem [{\citenamefont {Hon}\ and\ \citenamefont
  {Walker}(1991)}]{hon1991evolution}%
  \BibitemOpen
  \bibfield  {author} {\bibinfo {author} {\bibfnamefont {T.-L.}\ \bibnamefont
  {Hon}}\ and\ \bibinfo {author} {\bibfnamefont {J.~D.~A.}\ \bibnamefont
  {Walker}},\ }\bibfield  {title} {\enquote {\bibinfo {title} {Evolution of
  hairpin vortices in a shear flow},}\ }\href@noop {} {\bibfield  {journal}
  {\bibinfo  {journal} {Computers \& fluids}\ }\textbf {\bibinfo {volume}
  {20}},\ \bibinfo {pages} {343--358} (\bibinfo {year} {1991})}\BibitemShut
  {NoStop}%
\bibitem [{\citenamefont {Arms}\ and\ \citenamefont
  {Hama}(1965)}]{arms1965localized}%
  \BibitemOpen
  \bibfield  {author} {\bibinfo {author} {\bibfnamefont {R.}~\bibnamefont
  {Arms}}\ and\ \bibinfo {author} {\bibfnamefont {F.~R.}\ \bibnamefont
  {Hama}},\ }\bibfield  {title} {\enquote {\bibinfo {title}
  {Localized-induction concept on a curved vortex and motion of an elliptic
  vortex ring},}\ }\href@noop {} {\bibfield  {journal} {\bibinfo  {journal}
  {The Physics of fluids}\ }\textbf {\bibinfo {volume} {8}},\ \bibinfo {pages}
  {553--559} (\bibinfo {year} {1965})}\BibitemShut {NoStop}%
\bibitem [{\citenamefont {Zhou}(1996)}]{zhou1996numerical}%
  \BibitemOpen
  \bibfield  {author} {\bibinfo {author} {\bibfnamefont {H.}~\bibnamefont
  {Zhou}},\ }\href@noop {} {\emph {\bibinfo {title} {Numerical analysis of
  slender vortex motion}}}\ (\bibinfo  {publisher} {University of California,
  Berkeley},\ \bibinfo {year} {1996})\BibitemShut {NoStop}%
\bibitem [{\citenamefont {Margerit}, \citenamefont {Brancher},\ and\
  \citenamefont {Giovannini}(2004)}]{margerit2004implementation}%
  \BibitemOpen
  \bibfield  {author} {\bibinfo {author} {\bibfnamefont {D.}~\bibnamefont
  {Margerit}}, \bibinfo {author} {\bibfnamefont {P.}~\bibnamefont {Brancher}},
  \ and\ \bibinfo {author} {\bibfnamefont {A.}~\bibnamefont {Giovannini}},\
  }\bibfield  {title} {\enquote {\bibinfo {title} {Implementation and
  validation of a slender vortex filament code: Its application to the study of
  a four-vortex wake model},}\ }\href@noop {} {\bibfield  {journal} {\bibinfo
  {journal} {International Journal for Numerical Methods in Fluids}\ }\textbf
  {\bibinfo {volume} {44}},\ \bibinfo {pages} {175--196} (\bibinfo {year}
  {2004})}\BibitemShut {NoStop}%
\bibitem [{\citenamefont {Batchelor}(2000)}]{batchelor2000introduction}%
  \BibitemOpen
  \bibfield  {author} {\bibinfo {author} {\bibfnamefont {G.~K.}\ \bibnamefont
  {Batchelor}},\ }\href@noop {} {\emph {\bibinfo {title} {An introduction to
  fluid dynamics}}}\ (\bibinfo  {publisher} {Cambridge university press},\
  \bibinfo {year} {2000})\BibitemShut {NoStop}%
\bibitem [{\citenamefont {Klein}\ and\ \citenamefont
  {Majda}(1991{\natexlab{a}})}]{klein1991selfa}%
  \BibitemOpen
  \bibfield  {author} {\bibinfo {author} {\bibfnamefont {R.}~\bibnamefont
  {Klein}}\ and\ \bibinfo {author} {\bibfnamefont {A.~J.}\ \bibnamefont
  {Majda}},\ }\bibfield  {title} {\enquote {\bibinfo {title} {Self-stretching
  of a perturbed vortex filament i. the asymptotic equation for deviations from
  a straight line},}\ }\href@noop {} {\bibfield  {journal} {\bibinfo  {journal}
  {Physica D: Nonlinear Phenomena}\ }\textbf {\bibinfo {volume} {49}},\
  \bibinfo {pages} {323--352} (\bibinfo {year}
  {1991}{\natexlab{a}})}\BibitemShut {NoStop}%
\bibitem [{\citenamefont {Klein}\ and\ \citenamefont
  {Majda}(1991{\natexlab{b}})}]{klein1991selfb}%
  \BibitemOpen
  \bibfield  {author} {\bibinfo {author} {\bibfnamefont {R.}~\bibnamefont
  {Klein}}\ and\ \bibinfo {author} {\bibfnamefont {A.~J.}\ \bibnamefont
  {Majda}},\ }\bibfield  {title} {\enquote {\bibinfo {title} {Self-stretching
  of perturbed vortex filaments: Ii. structure of solutions},}\ }\href@noop {}
  {\bibfield  {journal} {\bibinfo  {journal} {Physica D: Nonlinear Phenomena}\
  }\textbf {\bibinfo {volume} {53}},\ \bibinfo {pages} {267--294} (\bibinfo
  {year} {1991}{\natexlab{b}})}\BibitemShut {NoStop}%
\bibitem [{\citenamefont {Chorin}(1980)}]{Chorin1980}%
  \BibitemOpen
  \bibfield  {author} {\bibinfo {author} {\bibfnamefont {A.~J.}\ \bibnamefont
  {Chorin}},\ }\bibfield  {title} {\enquote {\bibinfo {title} {Vortex models
  and boundary layer instability},}\ }\href@noop {} {\bibfield  {journal}
  {\bibinfo  {journal} {SIAM J. Sci. Stat. Comput.}\ }\textbf {\bibinfo
  {volume} {1}} (\bibinfo {year} {1980})}\BibitemShut {NoStop}%
\bibitem [{\citenamefont {Knio}\ and\ \citenamefont
  {Ghoniem}(1990)}]{knio1990numerical}%
  \BibitemOpen
  \bibfield  {author} {\bibinfo {author} {\bibfnamefont {O.~M.}\ \bibnamefont
  {Knio}}\ and\ \bibinfo {author} {\bibfnamefont {A.~F.}\ \bibnamefont
  {Ghoniem}},\ }\bibfield  {title} {\enquote {\bibinfo {title} {Numerical study
  of a three-dimensional vortex method},}\ }\href@noop {} {\bibfield  {journal}
  {\bibinfo  {journal} {Journal of Computational Physics}\ }\textbf {\bibinfo
  {volume} {86}},\ \bibinfo {pages} {75--106} (\bibinfo {year}
  {1990})}\BibitemShut {NoStop}%
\bibitem [{\citenamefont {Knio}\ and\ \citenamefont
  {Klein}(2000)}]{knio2000improved}%
  \BibitemOpen
  \bibfield  {author} {\bibinfo {author} {\bibfnamefont {O.~M.}\ \bibnamefont
  {Knio}}\ and\ \bibinfo {author} {\bibfnamefont {R.}~\bibnamefont {Klein}},\
  }\bibfield  {title} {\enquote {\bibinfo {title} {Improved thin-tube models
  for slender vortex simulations},}\ }\href@noop {} {\bibfield  {journal}
  {\bibinfo  {journal} {Journal of Computational Physics}\ }\textbf {\bibinfo
  {volume} {163}},\ \bibinfo {pages} {68--82} (\bibinfo {year}
  {2000})}\BibitemShut {NoStop}%
\bibitem [{\citenamefont {Ting}\ and\ \citenamefont
  {Klein}(1991)}]{ting1991viscous}%
  \BibitemOpen
  \bibfield  {author} {\bibinfo {author} {\bibfnamefont {L.}~\bibnamefont
  {Ting}}\ and\ \bibinfo {author} {\bibfnamefont {R.}~\bibnamefont {Klein}},\
  }\href@noop {} {\emph {\bibinfo {title} {Viscous vortical flows}}},\ Vol.\
  \bibinfo {volume} {374}\ (\bibinfo  {publisher} {Springer},\ \bibinfo {year}
  {1991})\BibitemShut {NoStop}%
\bibitem [{\citenamefont {Knio}\ and\ \citenamefont
  {Ghoniem}(1991)}]{knio1991three}%
  \BibitemOpen
  \bibfield  {author} {\bibinfo {author} {\bibfnamefont {O.~M.}\ \bibnamefont
  {Knio}}\ and\ \bibinfo {author} {\bibfnamefont {A.~F.}\ \bibnamefont
  {Ghoniem}},\ }\bibfield  {title} {\enquote {\bibinfo {title}
  {Three-dimensional vortex simulation of rollup and entrainment in a shear
  layer},}\ }\href@noop {} {\bibfield  {journal} {\bibinfo  {journal} {Journal
  of Computational Physics}\ }\textbf {\bibinfo {volume} {97}},\ \bibinfo
  {pages} {172--223} (\bibinfo {year} {1991})}\BibitemShut {NoStop}%
\bibitem [{\citenamefont {Butcher}(2016)}]{butcher2016numerical}%
  \BibitemOpen
  \bibfield  {author} {\bibinfo {author} {\bibfnamefont {J.~C.}\ \bibnamefont
  {Butcher}},\ }\href@noop {} {\emph {\bibinfo {title} {Numerical methods for
  ordinary differential equations}}}\ (\bibinfo  {publisher} {John Wiley \&
  Sons},\ \bibinfo {year} {2016})\BibitemShut {NoStop}%
\bibitem [{\citenamefont {Hairer}, \citenamefont {Wanner},\ and\ \citenamefont
  {N{\o}rsett}(1993)}]{hairer1993runge}%
  \BibitemOpen
  \bibfield  {author} {\bibinfo {author} {\bibfnamefont {E.}~\bibnamefont
  {Hairer}}, \bibinfo {author} {\bibfnamefont {G.}~\bibnamefont {Wanner}}, \
  and\ \bibinfo {author} {\bibfnamefont {S.~P.}\ \bibnamefont {N{\o}rsett}},\
  }\bibfield  {title} {\enquote {\bibinfo {title} {Runge-kutta and
  extrapolation methods},}\ }\href@noop {} {\bibfield  {journal} {\bibinfo
  {journal} {Solving Ordinary Differential Equations I: Nonstiff Problems}\ ,\
  \bibinfo {pages} {129--353}} (\bibinfo {year} {1993})}\BibitemShut {NoStop}%
\bibitem [{\citenamefont {Moin}, \citenamefont {Leonard},\ and\ \citenamefont
  {Kim}(1986)}]{moin1986evolution}%
  \BibitemOpen
  \bibfield  {author} {\bibinfo {author} {\bibfnamefont {P.}~\bibnamefont
  {Moin}}, \bibinfo {author} {\bibfnamefont {A.}~\bibnamefont {Leonard}}, \
  and\ \bibinfo {author} {\bibfnamefont {J.}~\bibnamefont {Kim}},\ }\bibfield
  {title} {\enquote {\bibinfo {title} {Evolution of a curved vortex filament
  into a vortex ring},}\ }\href@noop {} {\bibfield  {journal} {\bibinfo
  {journal} {The Physics of fluids}\ }\textbf {\bibinfo {volume} {29}},\
  \bibinfo {pages} {955--963} (\bibinfo {year} {1986})}\BibitemShut {NoStop}%
\bibitem [{\citenamefont {Fouard}\ \emph {et~al.}(2006)\citenamefont {Fouard},
  \citenamefont {Malandain}, \citenamefont {Prohaska},\ and\ \citenamefont
  {Westerhoff}}]{fouard2006blockwise}%
  \BibitemOpen
  \bibfield  {author} {\bibinfo {author} {\bibfnamefont {C.}~\bibnamefont
  {Fouard}}, \bibinfo {author} {\bibfnamefont {G.}~\bibnamefont {Malandain}},
  \bibinfo {author} {\bibfnamefont {S.}~\bibnamefont {Prohaska}}, \ and\
  \bibinfo {author} {\bibfnamefont {M.}~\bibnamefont {Westerhoff}},\ }\bibfield
   {title} {\enquote {\bibinfo {title} {Blockwise processing applied to brain
  microvascular network study},}\ }\href@noop {} {\bibfield  {journal}
  {\bibinfo  {journal} {IEEE Transactions on Medical Imaging}\ }\textbf
  {\bibinfo {volume} {25}},\ \bibinfo {pages} {1319--1328} (\bibinfo {year}
  {2006})}\BibitemShut {NoStop}%
\bibitem [{\citenamefont {Abbena}, \citenamefont {Salamon},\ and\ \citenamefont
  {Gray}(2017)}]{abbena2017modern}%
  \BibitemOpen
  \bibfield  {author} {\bibinfo {author} {\bibfnamefont {E.}~\bibnamefont
  {Abbena}}, \bibinfo {author} {\bibfnamefont {S.}~\bibnamefont {Salamon}}, \
  and\ \bibinfo {author} {\bibfnamefont {A.}~\bibnamefont {Gray}},\ }\href@noop
  {} {\emph {\bibinfo {title} {Modern differential geometry of curves and
  surfaces with Mathematica}}}\ (\bibinfo  {publisher} {Chapman and Hall/CRC},\
  \bibinfo {year} {2017})\BibitemShut {NoStop}%
\bibitem [{\citenamefont {Aref}\ and\ \citenamefont
  {Flinchem}(1984)}]{aref1984dynamics}%
  \BibitemOpen
  \bibfield  {author} {\bibinfo {author} {\bibfnamefont {H.}~\bibnamefont
  {Aref}}\ and\ \bibinfo {author} {\bibfnamefont {E.~P.}\ \bibnamefont
  {Flinchem}},\ }\bibfield  {title} {\enquote {\bibinfo {title} {Dynamics of a
  vortex filament in a shear flow},}\ }\href@noop {} {\bibfield  {journal}
  {\bibinfo  {journal} {Journal of Fluid Mechanics}\ }\textbf {\bibinfo
  {volume} {148}},\ \bibinfo {pages} {477--497} (\bibinfo {year}
  {1984})}\BibitemShut {NoStop}%
\bibitem [{\citenamefont {Yao}\ and\ \citenamefont
  {Hussain}(2020)}]{yao2020singularity}%
  \BibitemOpen
  \bibfield  {author} {\bibinfo {author} {\bibfnamefont {J.}~\bibnamefont
  {Yao}}\ and\ \bibinfo {author} {\bibfnamefont {F.}~\bibnamefont {Hussain}},\
  }\bibfield  {title} {\enquote {\bibinfo {title} {On singularity formation via
  viscous vortex reconnection},}\ }\href@noop {} {\bibfield  {journal}
  {\bibinfo  {journal} {Journal of Fluid Mechanics}\ }\textbf {\bibinfo
  {volume} {888}} (\bibinfo {year} {2020})}\BibitemShut {NoStop}%
\bibitem [{\citenamefont {Yao}\ and\ \citenamefont
  {Hussain}(2022)}]{yao2022vortex}%
  \BibitemOpen
  \bibfield  {author} {\bibinfo {author} {\bibfnamefont {J.}~\bibnamefont
  {Yao}}\ and\ \bibinfo {author} {\bibfnamefont {F.}~\bibnamefont {Hussain}},\
  }\bibfield  {title} {\enquote {\bibinfo {title} {Vortex reconnection and
  turbulence cascade},}\ }\href@noop {} {\bibfield  {journal} {\bibinfo
  {journal} {Annual Review of Fluid Mechanics}\ }\textbf {\bibinfo {volume}
  {54}},\ \bibinfo {pages} {317--347} (\bibinfo {year} {2022})}\BibitemShut
  {NoStop}%
\bibitem [{\citenamefont {von Lindheim}\ \emph {et~al.}(2021)\citenamefont {von
  Lindheim}, \citenamefont {Harikrishnan}, \citenamefont {D{\"o}rffel},
  \citenamefont {Klein}, \citenamefont {Koltai}, \citenamefont {Mikula},
  \citenamefont {M{\"u}ller}, \citenamefont {N{\'e}vir}, \citenamefont {Pacey},
  \citenamefont {Polzin} \emph {et~al.}}]{von2021definition}%
  \BibitemOpen
  \bibfield  {author} {\bibinfo {author} {\bibfnamefont {J.}~\bibnamefont {von
  Lindheim}}, \bibinfo {author} {\bibfnamefont {A.}~\bibnamefont
  {Harikrishnan}}, \bibinfo {author} {\bibfnamefont {T.}~\bibnamefont
  {D{\"o}rffel}}, \bibinfo {author} {\bibfnamefont {R.}~\bibnamefont {Klein}},
  \bibinfo {author} {\bibfnamefont {P.}~\bibnamefont {Koltai}}, \bibinfo
  {author} {\bibfnamefont {N.}~\bibnamefont {Mikula}}, \bibinfo {author}
  {\bibfnamefont {A.}~\bibnamefont {M{\"u}ller}}, \bibinfo {author}
  {\bibfnamefont {P.}~\bibnamefont {N{\'e}vir}}, \bibinfo {author}
  {\bibfnamefont {G.}~\bibnamefont {Pacey}}, \bibinfo {author} {\bibfnamefont
  {R.}~\bibnamefont {Polzin}},  \emph {et~al.},\ }\bibfield  {title} {\enquote
  {\bibinfo {title} {Definition, detection, and tracking of persistent
  structures in atmospheric flows},}\ }\href@noop {} {\bibfield  {journal}
  {\bibinfo  {journal} {arXiv preprint arXiv:2111.13645}\ } (\bibinfo {year}
  {2021})}\BibitemShut {NoStop}%
\bibitem [{\citenamefont {Moisy}\ and\ \citenamefont
  {Jim{\'e}nez}(2004)}]{moisy2004geometry}%
  \BibitemOpen
  \bibfield  {author} {\bibinfo {author} {\bibfnamefont {F.}~\bibnamefont
  {Moisy}}\ and\ \bibinfo {author} {\bibfnamefont {J.}~\bibnamefont
  {Jim{\'e}nez}},\ }\bibfield  {title} {\enquote {\bibinfo {title} {Geometry
  and clustering of intense structures in isotropic turbulence},}\ }\href@noop
  {} {\bibfield  {journal} {\bibinfo  {journal} {Journal of fluid mechanics}\
  }\textbf {\bibinfo {volume} {513}},\ \bibinfo {pages} {111--133} (\bibinfo
  {year} {2004})}\BibitemShut {NoStop}%
\bibitem [{\citenamefont {G{\"u}nther}\ and\ \citenamefont
  {Theisel}(2018)}]{gunther2018state}%
  \BibitemOpen
  \bibfield  {author} {\bibinfo {author} {\bibfnamefont {T.}~\bibnamefont
  {G{\"u}nther}}\ and\ \bibinfo {author} {\bibfnamefont {H.}~\bibnamefont
  {Theisel}},\ }\bibfield  {title} {\enquote {\bibinfo {title} {The state of
  the art in vortex extraction},}\ }in\ \href@noop {} {\emph {\bibinfo
  {booktitle} {Computer Graphics Forum}}},\ Vol.~\bibinfo {volume} {37}\
  (\bibinfo {organization} {Wiley Online Library},\ \bibinfo {year} {2018})\
  pp.\ \bibinfo {pages} {149--173}\BibitemShut {NoStop}%
\bibitem [{\citenamefont {Del~Alamo}\ and\ \citenamefont
  {Jimenez}(2006)}]{del2006linear}%
  \BibitemOpen
  \bibfield  {author} {\bibinfo {author} {\bibfnamefont {J.~C.}\ \bibnamefont
  {Del~Alamo}}\ and\ \bibinfo {author} {\bibfnamefont {J.}~\bibnamefont
  {Jimenez}},\ }\bibfield  {title} {\enquote {\bibinfo {title} {Linear energy
  amplification in turbulent channels},}\ }\href@noop {} {\bibfield  {journal}
  {\bibinfo  {journal} {Journal of Fluid Mechanics}\ }\textbf {\bibinfo
  {volume} {559}},\ \bibinfo {pages} {205--213} (\bibinfo {year}
  {2006})}\BibitemShut {NoStop}%
\bibitem [{\citenamefont {Dice}(1945)}]{dice1945measures}%
  \BibitemOpen
  \bibfield  {author} {\bibinfo {author} {\bibfnamefont {L.~R.}\ \bibnamefont
  {Dice}},\ }\bibfield  {title} {\enquote {\bibinfo {title} {Measures of the
  amount of ecologic association between species},}\ }\href@noop {} {\bibfield
  {journal} {\bibinfo  {journal} {Ecology}\ }\textbf {\bibinfo {volume} {26}},\
  \bibinfo {pages} {297--302} (\bibinfo {year} {1945})}\BibitemShut {NoStop}%
\bibitem [{\citenamefont {Klein}, \citenamefont {Knio},\ and\ \citenamefont
  {Ting}(1996)}]{klein1996representation}%
  \BibitemOpen
  \bibfield  {author} {\bibinfo {author} {\bibfnamefont {R.}~\bibnamefont
  {Klein}}, \bibinfo {author} {\bibfnamefont {O.~M.}\ \bibnamefont {Knio}}, \
  and\ \bibinfo {author} {\bibfnamefont {L.}~\bibnamefont {Ting}},\ }\bibfield
  {title} {\enquote {\bibinfo {title} {Representation of core dynamics in
  slender vortex filament simulations},}\ }\href@noop {} {\bibfield  {journal}
  {\bibinfo  {journal} {Physics of Fluids}\ }\textbf {\bibinfo {volume} {8}},\
  \bibinfo {pages} {2415--2425} (\bibinfo {year} {1996})}\BibitemShut {NoStop}%
\bibitem [{\citenamefont {Mahrt}(2014)}]{mahrt2014stably}%
  \BibitemOpen
  \bibfield  {author} {\bibinfo {author} {\bibfnamefont {L.}~\bibnamefont
  {Mahrt}},\ }\bibfield  {title} {\enquote {\bibinfo {title} {Stably stratified
  atmospheric boundary layers},}\ }\href@noop {} {\bibfield  {journal}
  {\bibinfo  {journal} {Annu. Rev. Fluid Mech}\ }\textbf {\bibinfo {volume}
  {46}},\ \bibinfo {pages} {23--45} (\bibinfo {year} {2014})}\BibitemShut
  {NoStop}%
\end{thebibliography}%

\end{document}